\documentclass[twocolumn]{aastex631}
\usepackage{multirow}
\usepackage[utf8]{inputenc}
\usepackage{pifont,amsmath}
\usepackage{apjfonts}
\usepackage{graphics,graphicx,subfigure,float,color,amsmath,natbib}
\usepackage{hyperref}
\setcitestyle{notesep={ }}
\usepackage{amsmath} 
\usepackage{enumitem}
\usepackage{colortbl}
\usepackage{verbatim} 

\let\oldAA\AA
\renewcommand{\AA}{\text{\normalfont\oldAA}}

\newcommand{\W}{{$\lambda$}}
\newcommand{\CH}[1]{\colhead{#1}}

\setcitestyle{notesep={; },round,aysep={},yysep={;}}

\begin{document}
\shortauthors{Parker et al.}
\shorttitle{CLASSY~XIII}

\author[0000-0002-8809-4608]{Kaelee S. Parker}
\affiliation{Department of Astronomy, University of Texas at Austin, 2515 Speedway, Austin, TX 78712, USA}
\affiliation{Cosmic Frontier Center, The University of Texas at Austin, Austin, TX 78712, USA}

\author[0000-0002-4153-053X]{Danielle A. Berg}
\affiliation{Department of Astronomy, University of Texas at Austin, 2515 Speedway, Austin, TX 78712, USA}
\affiliation{Cosmic Frontier Center, The University of Texas at Austin, Austin, TX 78712, USA}
 
\author[0000-0002-0302-2577]{John Chisholm}
\affiliation{Department of Astronomy, University of Texas at Austin, 2515 Speedway, Austin, TX 78712, USA}
\affiliation{Cosmic Frontier Center, The University of Texas at Austin, Austin, TX 78712, USA}

\author[0000-0002-5659-4974]{Simon Gazagnes}
\affiliation{Department of Astronomy, University of Texas at Austin, 2515 Speedway, Austin, TX 78712, USA}

\author[0000-0002-0159-2613]{Sophia R. Flury}
\affiliation{Institute for Astronomy, University of Edinburgh, Royal Observatory, Edinburgh, EH9 3HJ, UK}

\author[0000-0003-4166-2855]{Cody Carr}
\affiliation{University of Minnesota 316 Church str SE Minneapolis, MN 55455, USA}

\author[0009-0002-9932-4461]{Mason Huberty}
\affiliation{Minnesota Institute for Astrophysics, University of Minnesota, 116 Church Street SE, Minneapolis, MN 55455, USA}

\author[0000-0002-6790-5125]{Anne E. Jaskot}
\affiliation{Department of Astronomy, Williams College, Williamstown, MA 01267, USA}

\author[0000-0001-8587-218X]{Matthew J. Hayes}
\affiliation{Stockholm University, Department of Astronomy and Oskar Klein Centre for Cosmoparticle Physics, AlbaNova University Centre, SE-10691, Stockholm, Sweden}

\author[0000-0001-8419-3062]{Alberto Saldana-Lopez}
\affiliation{Stockholm University, Department of Astronomy and Oskar Klein Centre for Cosmoparticle Physics, AlbaNova University Centre, SE-10691, Stockholm, Sweden}

\author[0000-0003-4857-8699]{Svea Hernandez}
\affiliation{Space Telescope Science Institute, 3700 San Martin Drive, Baltimore, MD 21218, USA}

\author[0000-0003-2804-0648]{Themiya Nanayakkara}
\affiliation{Centre for Astrophysics and Supercomputing, Swinburne University of Technology, PO Box 218, Hawthorn, VIC 3122, Australia}

\author[0000-0003-4372-2006]{Bethan L. James}
\affiliation{AURA for ESA, Space Telescope Science Institute, 3700 San Martin Drive, Baltimore, MD 21218, USA}

\author[0000-0002-2644-3518]{Karla Z. Arellano-C\'{o}rdova}
\affiliation{Department of Astronomy, The University of Texas at Austin, 2515 Speedway, Stop C1400, Austin, TX 78712, USA}

\author[0000-0001-6369-1636]{Allison L. Strom}
\affiliation{Department of Physics and Astronomy and CIERA, Northwestern University, 1800 Sherman Avenue,
Evanston, IL 60201, USA}

\author[0000-0002-9132-6561]{Peter Senchyna}
\affiliation{The Observatories of the Carnegie Institution for Science, 813 Santa Barbara Street, Pasadena, CA 91101, USA}

\author[0000-0003-2589-762X]{Matilde Mingozzi}
\affiliation{AURA for ESA, Space Telescope Science Institute, 3700 San Martin Drive, Baltimore, MD 21218, USA}

\author[0000-0003-1127-7497]{Timothy Heckman}
\affiliation{Center for Astrophysical Sciences, Department of Physics \& Astronomy, Johns Hopkins University, Baltimore, MD 21218, USA}

\author[0000-0002-9217-7051]{Xinfeng Xu}
\affiliation{Center for Astrophysical Sciences, Department of Physics \& Astronomy, Johns Hopkins University, Baltimore, MD 21218, USA}

\author[0000-0002-6586-4446]{Alaina Henry}
\affiliation{Space Telescope Science Institute, 3700 San Martin Drive, Baltimore, MD 21218, USA}

\author[0000-0001-5758-1000]{Ricardo O. Amor\'{i}n} 
\affiliation{Instituto de Astrof\'{i}sica de Andaluc\'{i}a (CSIC), 
Apartado 3004, 18080 Granada, Spain}

\author[0000-0003-0595-9483]{Valentin Mauerhofer}
\affiliation{Kapteyn Astronomical Institute, University of Groningen, PO Box 800 9700 AV Groningen, The Netherlands}

\author[0000-0001-9189-7818]{Crystal L. Martin}
\affiliation{Department of Physics,
University of California Santa Barbara,
Santa Barbara, CA 93106, USA }

\author[0000-0001-9714-2758]{Dawn K. Erb}
\affiliation{Center for Gravitation, Cosmology and Astrophysics, Department of Physics, University of Wisconsin Milwaukee, 3135 N Maryland Ave., Milwaukee, WI 53211, USA}

\author[0000-0003-0605-8732]{Evan D. Skillman}
\affiliation{Minnesota Institute for Astrophysics, University of Minnesota, 116 Church Street SE, Minneapolis, MN 55455, USA}

\author[0000-0001-6248-1864]{Kate H. R. Rubin}
\affiliation{Department of Astronomy, San Diego State University, San Diego, CA 92182, USA}

\author[0000-0002-4957-5468]{John Trevino}
\affiliation{Department of Astronomy, University of Texas at Austin, 2515 Speedway, Austin, TX 78712, USA}

\author[0000-0003-2685-4488]{Claus Leitherer}
\affiliation{Space Telescope Science Institute, 3700 San Martin Drive, Baltimore, MD 21218, USA}

\correspondingauthor{Kaelee S. Parker} 
\email{kaelee.parker@utexas.edu}

\title{CLASSY XIII. Cutting through the Clouds -- Comparing Indirect Tracers of Ionizing Photon Escape}


\begin{abstract}
The Epoch of Reionization (EoR) provides critical insights into the role of early galaxies in shaping the ionization state of the universe. However, because of the opacity of the intergalactic medium, it is often not possible to make direct measurements of the ionizing photon escape fraction ($f_{\mathrm{esc}}^{\: \mathrm{LyC}}$) of high-redshift ($z \gtrsim 4$) galaxies. To explore the agreement and systematics of common indirect approaches, we applied six empirically calibrated diagnostics to predict $f_{\mathrm{esc}}^{\: \mathrm{LyC}}$ for the 45 nearby star-forming galaxies from the COS Legacy Spectroscopic SurveY (CLASSY). These methods-- based on ultraviolet (UV) absorption lines, the UV continuum slope, Ly$\alpha$ kinematics, a multivariate model, radiation-hydrodynamic simulations, and nebular emission line ratios-- enable us to explore systematic differences between predictions and assess how galactic properties influence inferred LyC escape. Despite significant variations in method predictions, there is broad consistency in the resulting weak and strong LyC leaker classifications, with approximately half exhibiting predicted escape fractions $>$1\%. We find evidence for two different pathways of LyC escape in nearby star-forming galaxies: (1) an early escape model driven by very young stellar populations, and (2) a delayed escape model that is consistent with supernova-driven outflows and time-dependent ISM clearing. The early escape model is favored among galaxies with a single, intense burst of recent star formation. In contrast, the delayed escape model is common among galaxies with more extended starburst histories. To interpret ionizing photon escape during the EoR, it will be necessary to recognize and understand this diversity in LyC escape mechanisms.

\vspace{1cm}
\end{abstract}


\section{Introduction} \label{intro}

The Epoch of Reionization (EoR) represents a critical phase in the history of the universe, during which the intergalactic medium (IGM) transitioned from a neutral to an ionized state. This period, which ends roughly around $z \sim 6$, marks a significant phase transition of the universe and is directly related to the formation of the first luminous structures. Understanding the sources responsible for reionizing the IGM and the mechanisms by which ionizing photons were able to escape their host galaxies is crucial for constructing a comprehensive narrative of cosmic evolution. Moreover, characterizing the timeline and topology of reionization provides fundamental insights into the properties of early galaxies and their role in shaping the universe as we observe it today.

Low-mass ($\log(\mathrm{M}_\star/\mathrm{M}_\odot)< 9$) star-forming galaxies are considered to be a primary candidate responsible for driving reionization \citep[e.g.,][]{Wise_2014,Madau_2015,Robertson_2015,Stanway_2016,Finkelstein_2019}. Although these galaxies can produce significant quantities of hydrogen-ionizing radiation \citep[$<$ 912 \AA; Lyman-continuum; LyC; e.g.,][]{Wu_2024,Izotov_2021}, the subset of these photons that actually {\it escape} through the interstellar medium (ISM) and into the host galaxy's surroundings is what contributes to cosmic reionization. However, directly measuring the LyC escape fraction ($f_{\mathrm{esc}}^{\: \mathrm{LyC}}$) of reionization-era galaxies is often not possible, since the IGM becomes opaque to ionizing photons at $z \gtrsim 4$ \citep{Inoue_2014,Bosman_2018,Bosman_2022,Yang_2020,Becker_2021}. Consequently, indirect methods--empirically calibrated at lower redshift--have become essential tools for estimating LyC escape. 

Direct measurements of $f_{\mathrm{esc}}^{\: \mathrm{LyC}}$ from galaxies at $z < 4$ have been used to develop and calibrate a variety of indirect methods to estimate $f_{\mathrm{esc}}^{\: \mathrm{LyC}}$ \citep[e.g.,][]{Bergvall_2006,Leitet_2011,Borthakur_2014,Izotov_2016a,Izotov_2016b,Izotov_2018a,Izotov_2018b,Leitherer_2016,Shapley_2016,Vanzella_2016,deBarros2016,Steidel_2018,Fletcher_2019,Rivera-Thorsen_2019,Wang_2019,Ji_2020,Davis_2021,Marques-Chaves_2021,Xu_2022,Flury_2022a,Flury_2022b,Jaskot2024a,Mascia_2023,Mascia_2024,Choustikov_2024}. These methods are based on empirical trends between $f_{\mathrm{esc}}^{\: \mathrm{LyC}}$ and various observable properties, such as features of the Ly$\alpha$ emission \citep{Verhamme_2015,Verhamme_2017,Rivera-Thorsen_2017,Izotov_2018b,Izotov_2021,Steidel_2018,Gazagnes_2020,Pahl_2021}, the ISM absorption \citep{Reddy_2016,Chisholm_2017,Chisholm_2018,Gazagnes_2018,Steidel_2018,Saldana-Lopez_2022}, or the UV continuum slope \citep{Chisholm_2022}. Although these indirect techniques make it possible to estimate $f_{\mathrm{esc}}^{\: \mathrm{LyC}}$ for galaxies where the ionizing continuum cannot be observed, they can also exhibit significant scatter \citep[e.g.,][]{Roy_2024,Citro_2024}. Understanding the uncertainties and limitations associated with these methods is essential for refining their application to populations of high-redshift galaxies \citep{Mascia_2024}. A large high-resolution sample of local galaxies is an excellent environment for conducting these tests.

In this paper, we indirectly predict $f_{\mathrm{esc}}^{\: \mathrm{LyC}}$ for the CLASSY sample using six techniques, based on the neutral and low-ionization gas covering fraction, $\beta_{\mathrm{UV}}$-slopes, Ly$\alpha$ peak separation, the \texttt{LyCsurv} multivariate code, the best-fits to mock spectra from the radiation hydrodynamic simulation by \citet{Mauerhofer_2021} for CLASSY \citep{Gazagnes_2023}, and the O$_{32}$ ionization parameter ([\ion{O}{3}] \W5007 / [\ion{O}{2}] \W3727). We use a combination of these indirect $f_{\mathrm{esc}}^{\: \mathrm{LyC}}$ predictions to (1) summarize general expectations for ionizing photon escape among CLASSY galaxies, (2) explore systematic differences between these methods, and (3) investigate how global galaxy properties and star formation histories can influence the amount of ionizing radiation that escapes. 

The remainder of this paper is organized as follows.
Section~\ref{data} introduces the FUV spectra utilized in this work and our \texttt{Starburst99} \citep{Leitherer_1999} stellar population fits.
Section~\ref{fesc} discusses the methodology for each of the techniques we use to indirectly predict $f_{\mathrm{esc}}^{\: \mathrm{LyC}}$.
In Section~\ref{results}, we investigate systematic differences between the predictions of $f_{\mathrm{esc}}^{\: \mathrm{LyC}}$ from these methods.
Section~\ref{discuss} explores trends between the recent star formation episodes and other galaxy characteristics among the CLASSY sample, particularly concerning properties of their ISM and its potential impact on the LyC escape fraction. Finally, Section~\ref{summary} is a summary of our findings.


\section{Data} \label{data}

In this section, we discuss characteristics of the CLASSY galaxy sample (\S~\ref{sub:classy}) and the CLASSY Treasury's high-resolution FUV spectra (\S~\ref{sub:FUV}), with a discussion of the properties from our \texttt{Starburst99} \citep{Leitherer_1999} stellar population fits in \S~\ref{subsub:sps} and the absorption line fits from \cite{Parker2024} in \S~\ref{subsub:fits}. We use measurements from these observations and fits to indirectly estimate $f_{\mathrm{esc}}^{\: \mathrm{LyC}}$ in Section \ref{fesc}. 


\subsection{The CLASSY sample} \label{sub:classy}

The COS Legacy Spectroscopic SurveY\footnote{The data is available at MAST: \dataset[doi: 10.17909/m3fq-jj25]{http://dx.doi.org/10.17909/m3fq-jj25}} \citep[CLASSY;][]{Berg_2022,James_2022} consists of high-S/N (S/N$_{1500} \gtrsim$ 5 per resolution element), high-resolution ($R \sim 15,000$), FUV ($\sim1100-2000$ \AA) spectra from the Cosmic Origins Spectrograph (COS) on the Hubble Space Telescope (HST) for 45 low redshift ($z<0.18$) star-forming galaxies, many of which have 21-cm \ion{H}{1} observations from the Green Bank Telescope \citep{Parker2024}. These galaxies span a diverse range of properties, with a relatively large range of stellar masses ($6.2 < \log \mathrm{M}_\star/\mathrm{M}_\odot < 10.1$), star formation rate (SFR; $-2.0 < \log$ SFR (M$_\odot$ yr$^{-1}) < 1.6$), direct gas-phase metallicity (7.0 $<$ 12+log(O/H) $<$ 8.8), ionization (0.5 $<$ O$_{32} <$ 38.0), reddening ($0.02 < E(B-V)_{\mathrm{neb}} < 0.67$), and electron density ($10 < n_e$ (cm$^{-3}$) $< 1120$). The CLASSY galaxies were selected for their high UV luminosities, resulting in a sample of galaxies that exhibit enhanced star formation that is more comparable to galaxies at $z\sim2-3$ than those at $z<0.18$ \citep[see Figure 8 in][]{Berg_2022}. 


\subsection{FUV spectra} \label{sub:FUV}

We used the high-level science product spectra of the CLASSY survey,\footnote{accessible via \url{https://archive.stsci.edu/hlsp/classy} and \url{https://mast.stsci.edu/search/ui/\#/classy}} specifically the high-resolution (HR) coadded spectra (G130M+G160M gratings), for multiple of the measurements used to indirectly estimate $f_{\mathrm{esc}}^{\: \mathrm{LyC}}$. For technical details of the coaddition process, see \citet{Berg_2022} and \citet{James_2022}. The HR coadded spectra cover wavelengths between 1200 and 1750 $\AA$ and have an average spectral resolution of 0.073 $\AA$ resel$^{-1}$ \citep{Berg_2022}, where one resolution element = six native pixels for COS \citep{Soderblom_2021}.


\subsubsection{Stellar population fits} \label{subsub:sps}
The $f_{\mathrm{esc}}^{\: \mathrm{LyC}}$ calibrations used in this work require spectral measurements from continuum-normalized rest-frame UV spectra. We note that UV stellar continuum fits were utilized in previous CLASSY papers \citep[e..g,][]{Xu_2022,Parker2024}, but we present the resulting massive star population properties here for the first time. To fit the UV stellar continuum of the CLASSY spectra, we follow the approach of \citet{Chisholm_2019}. We model the total UV stellar continuum, which is dominated by the light from young, massive stars, as a linear combination of 50 single-age simple stellar population (SSP) continuum models to match the stellar wind features (i.e., \ion{N}{5} \W\W1238,1240, \ion{Si}{4} \W\W1393,1402, \ion{C}{4} \W\W1548,1550, and \ion{N}{4} \W1720) and photospheric features (i.e., \ion{Fe}{5} \W1430, \ion{C}{3} \W1478, \ion{S}{5} \W1502, \ion{C}{3} \W1527, \ion{Si}{2} \W1533, \ion{C}{3} \W1620, and \ion{Fe}{5} \W1662) in the CLASSY spectra, which are distinctly sensitive to either age or metallicity of the stellar population \citep[e.g.,][]{deMello_2000,Vidal_Garcia_2017}. The models, taken from the \texttt{Starburst99} library \citep[\texttt{SB99};][]{Leitherer_1999}, span five metallicities (0.05, 0.2, 0.4, 1, and 2 Z$_\odot$) and 10 distinct population ages (1, 2, 3, 4, 5, 8, 10, 15, 20, and 40 Myr), and were generated using a Kroupa initial mass function with a high and low mass exponent of 2.3 and 1.3, respectively, a high-mass cutoff of 100~M$_\odot$, and the stellar evolution tracks with high mass loss from \citet{Meynet_1994}. The \texttt{SB99} models have a fixed 0.4~\AA\ resolution, and we convolve the data to match this resolution before fitting.

To reduce the risk of unphysical discrepancies between the fitted stellar population metallicities and the optically measured gas-phase abundances \cite[from][]{Berg_2022}, we constrain the grid of stellar metallicities in these fits to only the three values closest to the observed gas-phase metallicity (out of the five total metallicity options). However, for four galaxies (J0036-3333, J1157+3220, J1525+0757, J1428+1653), the reduced metallicity fits were unable to reproduce features in the spectra that are associated with older stars (such as the broad wings of \ion{Si}{4} \W1400). In these cases, using the full grid of metallicities could fit these features associated with older populations but only with higher metallicity templates. We, therefore, chose to use the ages returned by the full-grid metallicity fit but the stellar metallicities from the reduced-grid fit for these four galaxies.

As detailed in \citet{Chisholm_2019}, we include a nebular continuum in each single-age and metallicity stellar population model. These nebular continua are generated with \texttt{Cloudy} \citep{Ferland_2013, Ferland_2017}, using consistent gas-phase and stellar metallicities, a volume hydrogen density of 100 cm$^{-3}$, and an ionization parameter $\log U = -2.5$. We note that allowing the ionization parameter to vary would change the nebular-to-stellar continuum ratio; however, the close match of the fixed $\log U$ models to the observed stellar features and overall continuum shape of the CLASSY Sample suggests that these simpler models are sufficient for our work. 

Finally, to account for dust, we used a uniform dust screen model and assumed the far-UV extinction curve from \citet{Reddy_2016}.  We manually mask out strong absorption and emission features in both the observed and rest-frame spectra, and fit for any damped Ly$\alpha$ absorbers using a custom fitting routine. Altogether, we use \texttt{mpfit} \citep{Markwardt_2009} to fit for the 50 linear coefficients -- "light fractions"-- multiplied to each single age and single metallicity model, as well as the single dust extinction parameter. This corresponds to a star formation history represented by a combination of discrete bursts. We used the best-fit light fractions to estimate the light-weighted age and metallicity using Equations 2 and 3 in \citet{Chisholm_2019}. Uncertainties on the stellar population properties are estimated using a Monte Carlo Markov Chain method that varies the observed flux density in each pixel by the flux-density uncertainty, refitting the model, tabulating the stellar population properties, and repeating the process 1,000 times to establish a distribution of the light-weighted ages and metallicities. 

\begin{figure*}
    \centering
    \includegraphics[width=\linewidth]{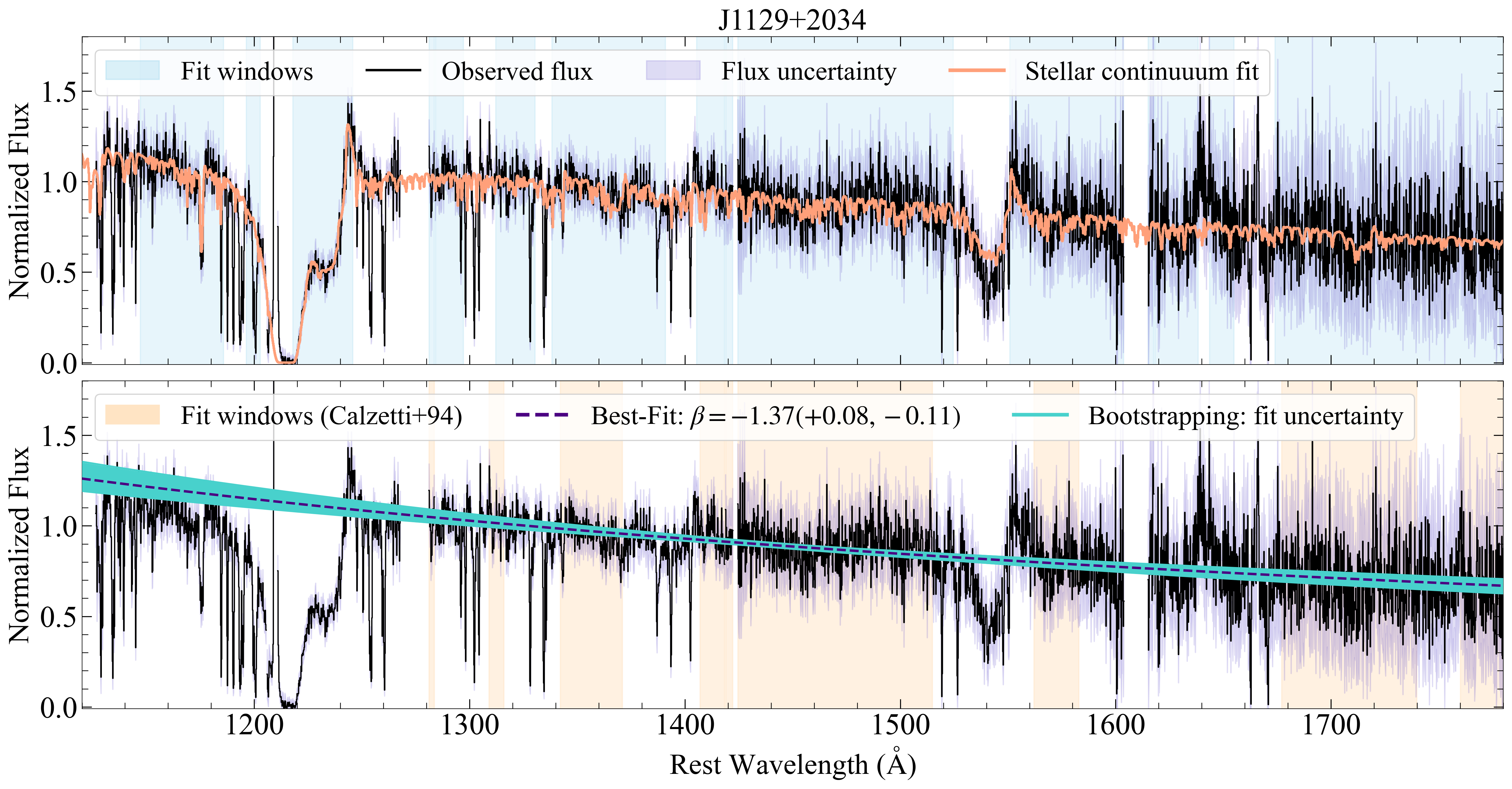}
     \caption{Example fits for one CLASSY galaxy, J1129+2034. \textit{Top:} the \texttt{Starburst99} \citep{Leitherer_1999} stellar population fit and \ion{H}{1} fit (orange), which results in a light-weighted metallicity of Z/Z$_\odot$ = 0.21 and light-weighted age of 2.56 Myr for this galaxy. \textit{Bottom:} the $\beta_{\mathrm{UV}}$-slope fit (dashed black line with cyan MC uncertainties). In both panels, the observed spectrum is in black with purple uncertainties. The windows used in each fit are shaded in light blue and light orange in the top and bottom panels, respectively. The windows for the SPS fits were chosen to exclude regions with strong nebular features (such as the ISM absorption component of \ion{C}{4} \W\W1549,1550). In addition, in galaxies with broad \ion{He}{2} \W1640 emission, we masked out this feature since \texttt{SB99} has does not include sufficient Wolf-Rayet star templates to reproduce this broad emission, especially at low-metallicities and/or young ages \citep{Leitherer_2018,Chisholm_2019}.} The regions used to fit the $\beta_{\mathrm{UV}}$-slopes are from \citet{Calzetti_1994}, also selected to span regions dominated largely by stellar features.
    \label{fig:SPS}
\end{figure*}

Table \ref{tab:fit_props} contains the parameters of the SPS fits: metallicity $Z/Z_\odot$, population ages, and dust attenuation $E(B-V)_\star$. The upper panel of Figure \ref{fig:SPS} shows an example of the SPS fit for one CLASSY galaxy (J1129+2034), including both the stellar \texttt{SB99} model and a Voigt profile fit to the Ly$\alpha$ transition. The windows used to perform this fit are shaded in light blue, where regions with prominent nebular features have been masked.


\startlongtable
\begin{deluxetable}{l|ccc}
	 \tablewidth{0pt}
	 \setlength{\tabcolsep}{5pt}
	 \tabletypesize{\footnotesize}
	 \tablecaption{Light-weighted population properties from \texttt{Starburst99} fits to the stellar continuum}
	 \tablehead{\CH{} & \CH{} & \CH{Age} & \CH{$E(B-V)_\star$} \\ [-2ex]
	 \CH{Galaxy} & \CH{$Z/Z_\odot$} & \CH{(Myr)} & \CH{(mag)}}
	 \startdata
	 J0021+0052 & 0.20 $\pm$ 0.05 & 4.12 $\pm$ 0.37 & 0.133 $\pm$ 0.004 \\ 
	 J0036-3333 & 0.25 $\pm$ 0.02 & 19.52 $\pm$ 0.83 & 0.123 $\pm$ 0.002 \\ 
	 J0127-0619 & 0.50 $\pm$ 0.05 & 2.28 $\pm$ 0.08 & 0.307 $\pm$ 0.006 \\ 
	 J0144+0453 & 0.05 $\pm$ 0.01 & 7.15 $\pm$ 0.81 & 0.070 $\pm$ 0.005 \\ 
	 J0337-0502 & 0.06 $\pm$ 0.00 & 5.27 $\pm$ 1.13 & 0.000 $\pm$ 0.004 \\ 
	 J0405-3648 & 0.05 $\pm$ 0.01 & 21.46 $\pm$ 3.15 & 0.071 $\pm$ 0.009 \\ 
	 J0808+3948 & 1.52 $\pm$ 0.03 & 3.37 $\pm$ 0.08 & 0.163 $\pm$ 0.004 \\ 
	 J0823+2806 & 0.20 $\pm$ 0.06 & 4.34 $\pm$ 2.74 & 0.255 $\pm$ 0.008 \\ 
	 J0926+4427 & 0.08 $\pm$ 0.04 & 4.80 $\pm$ 0.37 & 0.095 $\pm$ 0.004 \\ 
	 J0934+5514 & 0.05 $\pm$ 0.00 & 13.48 $\pm$ 1.43 & 0.000 $\pm$ 0.000 \\ 
	 J0938+5428 & 0.20 $\pm$ 0.04 & 4.79 $\pm$ 0.61 & 0.073 $\pm$ 0.005 \\ 
	 J0940+2935 & 0.08 $\pm$ 0.03 & 10.31 $\pm$ 2.70 & 0.110 $\pm$ 0.008 \\ 
	 J0942+3547 & 0.10 $\pm$ 0.01 & 4.86 $\pm$ 0.10 & 0.071 $\pm$ 0.002 \\ 
	 J0944-0038 & 0.05 $\pm$ 0.02 & 13.51 $\pm$ 2.68 & 0.250 $\pm$ 0.011 \\ 
	 J0944+3442 & 0.11 $\pm$ 0.05 & 14.06 $\pm$ 4.71 & 0.195 $\pm$ 0.016 \\ 
	 J1016+3754 & 0.06 $\pm$ 0.02 & 11.61 $\pm$ 1.57 & 0.052 $\pm$ 0.006 \\ 
	 J1024+0524 & 0.06 $\pm$ 0.03 & 5.16 $\pm$ 0.24 & 0.071 $\pm$ 0.005 \\ 
	 J1025+3622 & 0.20 $\pm$ 0.04 & 5.36 $\pm$ 1.17 & 0.119 $\pm$ 0.005 \\ 
	 J1044+0353 & 0.05 $\pm$ 0.01 & 10.62 $\pm$ 2.05 & 0.229 $\pm$ 0.007 \\ 
	 J1105+4444 & 0.10 $\pm$ 0.03 & 11.31 $\pm$ 1.85 & 0.310 $\pm$ 0.005 \\ 
	 J1112+5503 & 0.40 $\pm$ 0.05 & 3.81 $\pm$ 0.46 & 0.256 $\pm$ 0.006 \\ 
	 J1119+5130 & 0.05 $\pm$ 0.00 & 12.64 $\pm$ 2.23 & 0.080 $\pm$ 0.005 \\ 
	 J1129+2034 & 0.21 $\pm$ 0.04 & 3.67 $\pm$ 0.79 & 0.189 $\pm$ 0.005 \\ 
	 J1132+1411 & 0.18 $\pm$ 0.05 & 11.23 $\pm$ 1.94 & 0.201 $\pm$ 0.006 \\ 
	 J1132+5722 & 0.05 $\pm$ 0.04 & 13.85 $\pm$ 4.46 & 0.000 $\pm$ 0.000 \\ 
	 J1144+4012 & 0.20 $\pm$ 0.07 & 6.10 $\pm$ 2.20 & 0.216 $\pm$ 0.008 \\ 
	 J1148+2546 & 0.05 $\pm$ 0.03 & 5.67 $\pm$ 0.41 & 0.232 $\pm$ 0.006 \\ 
	 J1150+1501 & 0.19 $\pm$ 0.03 & 7.55 $\pm$ 1.58 & 0.161 $\pm$ 0.004 \\ 
	 J1157+3220 & 0.21 $\pm$ 0.02 & 11.31 $\pm$ 1.21 & 0.106 $\pm$ 0.002 \\ 
	 J1200+1343 & 0.17 $\pm$ 0.03 & 3.85 $\pm$ 0.38 & 0.197 $\pm$ 0.005 \\ 
	 J1225+6109 & 0.18 $\pm$ 0.02 & 8.92 $\pm$ 1.18 & 0.185 $\pm$ 0.003 \\ 
	 J1253-0312 & 0.13 $\pm$ 0.02 & 3.81 $\pm$ 0.71 & 0.195 $\pm$ 0.004 \\ 
	 J1314+3452 & 0.22 $\pm$ 0.03 & 2.36 $\pm$ 0.04 & 0.217 $\pm$ 0.003 \\ 
	 J1323-0132 & 0.13 $\pm$ 0.01 & 1.52 $\pm$ 0.11 & 0.072 $\pm$ 0.004 \\ 
	 J1359+5726 & 0.09 $\pm$ 0.02 & 15.70 $\pm$ 1.53 & 0.106 $\pm$ 0.006 \\ 
	 J1416+1223 & 0.46 $\pm$ 0.05 & 4.48 $\pm$ 0.64 & 0.157 $\pm$ 0.005 \\ 
	 J1418+2102 & 0.06 $\pm$ 0.03 & 1.64 $\pm$ 0.80 & 0.181 $\pm$ 0.006 \\ 
	 J1428+1653 & 0.25 $\pm$ 0.04 & 12.88 $\pm$ 2.29 & 0.081 $\pm$ 0.009 \\ 
	 J1429+0643 & 0.22 $\pm$ 0.05 & 3.14 $\pm$ 1.36 & 0.129 $\pm$ 0.007 \\ 
	 J1444+4237 & 0.05 $\pm$ 0.01 & 15.32 $\pm$ 1.67 & 0.056 $\pm$ 0.007 \\ 
	 J1448-0110 & 0.16 $\pm$ 0.05 & 3.15 $\pm$ 0.31 & 0.215 $\pm$ 0.008 \\ 
	 J1521+0759 & 0.20 $\pm$ 0.03 & 5.84 $\pm$ 0.56 & 0.016 $\pm$ 0.004 \\ 
	 J1525+0757 & 0.40 $\pm$ 0.07 & 16.10 $\pm$ 3.16 & 0.225 $\pm$ 0.007 \\ 
	 J1545+0858 & 0.07 $\pm$ 0.03 & 4.39 $\pm$ 0.83 & 0.123 $\pm$ 0.009 \\ 
	 J1612+0817 & 0.40 $\pm$ 0.06 & 4.45 $\pm$ 0.88 & 0.202 $\pm$ 0.006 \\ 
	 \enddata
     \tablecomments{Column 1: Galaxy name. Based on the \texttt{Starburst99} \citep{Leitherer_1999} stellar continuum fits, this table lists the light-weighted metallicities ($Z$; Column 2), population ages (Column 3), and dust attenuations ($E(B-V)_\star$; Column 4).}
	 \label{tab:fit_props}
\end{deluxetable}


\subsubsection{Absorption-line properties} \label{subsub:fits}

\cite{Parker2024} performed a detailed analysis of the neutral ISM properties of the CLASSY sample using a combination of the FUV interstellar metal absorption lines measured from the continuum-subtracted spectra and the 21-cm \ion{H}{1} emission. 
In this paper, we use the covering fractions from \cite{Parker2024} to indirectly predict $f_{\mathrm{esc}}^{\: \mathrm{LyC}}$. To estimate these properties, they simultaneously fit a series of Voigt profiles to a subset of 15 interstellar low-ionization state (LIS) and neutral absorption lines (\ion{O}{1}, \ion{C}{2}, \ion{Si}{2}, \ion{S}{2}, \ion{Al}{2}) using \texttt{lmfit} minimization and a Monte Carlo implementation, similar to the procedures in \cite{Gazagnes_2018} and \cite{Chisholm_2018}. Through various tests, \cite{Parker2024} demonstrated that this approach is capable of breaking the degeneracies between column density and Doppler broadening, which arise from fitting saturated lines, at least for high-resolution spectra \citep{Jennings_2025}. 
The covering fractions determined from these fits are ion-specific (e.g., $C_f$(\ion{O}{1}), $C_f$(\ion{C}{2})) and were determined from fits to multiple transitions of each ion, when possible. We describe how we use these covering fraction measurements to predict $f_{\mathrm{esc}}^{\: \mathrm{LyC}}$ in $\S$~\ref{sub:method1}. 


\section{Predicting the LyC escape fraction} \label{fesc}

We use six methods to predict the escape fraction of ionizing LyC photons for CLASSY galaxies, which are described individually below: the \ion{H}{1}/LIS covering fraction method is discussed in $\S$~\ref{sub:method1}, the $\beta_{\mathrm{UV}}$-slope method in $\S$~\ref{sub:method2}, the Ly$\alpha$ peak separations method in $\S$~\ref{sub:method3}, the multivariate approach using the \texttt{LyCsurv} software package from \citet{Jaskot2024a,Jaskot2024b} in $\S$~\ref{sub:method4}, mock spectra fits from a radiation hydrodynamic simulation by \cite{Gazagnes_2023} in $\S$~\ref{sub:method5}, and the O$_{32}$ ionization ratio method in $\S$~\ref{sub:method6}. We utilize multiple methods to estimate $f_{\mathrm{esc}}^{\: \mathrm{LyC}}$ because each of these indirect methods is primarily based on differing assumptions and observables.

Several of the methods we use to predict the LyC escape fraction rely on measurements from, or empirical trends that were fit using, the Low-Redshift Lyman Continuum Survey (LzLCS; \citealt{Flury_2022a,Flury_2022b}). The LzLCS+ is the largest and most comprehensive sample of spectroscopically confirmed LyC leaking galaxies in the nearby universe ($z \sim 0.2-0.4$), consisting of the 66 galaxies from the LzLCS \citep{Flury_2022a} and an additional 23 $z \sim 0.3$ galaxies with measured LyC, or stringent upper limits, from archival datasets \citep{Izotov_2016a,Izotov_2016b,Izotov_2018a,Izotov_2018b,Wang_2019,Izotov_2021}. This sample has the unique advantage of providing homogeneously processed-- although low resolution (R$\sim$1000)-- spectra, ensuring direct detection of LyC photons while simultaneously allowing the measurement of indirect diagnostics such as Ly\(\alpha\) emission properties, ISM absorption features, and nebular emission lines.

We classify galaxies with $f_{\mathrm{esc}}^{\: \mathrm{LyC}} < 5\%$ as LyC non-leakers, those with $5\% \leq f_{\mathrm{esc}}^{\: \mathrm{LyC}} < 20\%$ as weak LyC leakers, and those with $f_{\mathrm{esc}}^{\: \mathrm{LyC}} \geq 20\%$ as strong LyC leakers.\footnote{Note that these thresholds differ from those used by \citet{Flury_2022a, Flury_2024} for the LzLCS, where weak LyC leakers were defined as those with $1\% \leq f_{\mathrm{esc}}^{\: \mathrm{LyC}} < 5\%$ and strong leakers had $f_{\mathrm{esc}}^{\: \mathrm{LyC}} \geq 5\%$.} These thresholds were chosen to distinguish between different pictures of the EoR, based on the types of galaxies that could have been the most significant contributors to reionization. Galaxies with $f_{\mathrm{esc}}^{\: \mathrm{LyC}} < 5\%$ are unlikely to contribute significantly to the ionizing photon budget unless they are particularly numerous and have high intrinsic ionizing photon production rates \citep[e.g.,][]{Wise_2014,Finkelstein_2019,Yung_2020,Dayal_2020}. This means that a universe of predominantly non-leaking galaxies at the time of the EoR would correspond to a slow reionization of the universe and an earlier start \citep{Robertson_2015,Finkelstein_2019,Rosdahl_2022,Munoz_2024}. 

It is important to note that the contribution of a galaxy to reionization depends not only on $f_{\mathrm{esc}}^{\: \mathrm{LyC}}$ but also on its UV magnitude ($M_\mathrm{1500}$) and ionizing photon production efficiency ($\xi_{\mathrm{ion}}$). Faint galaxies with low escape fractions can still contribute substantially to reionization if they are sufficiently numerous. Thus, our classification should be interpreted within the context of the M$_\mathrm{UV}$ range probed by CLASSY.


\subsection{$f_{\mathrm{esc}}^{\: C_f}$: from \ion{H}{1} covering fraction and E(B-V)} \label{sub:method1}

The covering fraction of neutral gas provides a valuable approach for indirectly estimating the escape fraction of ionizing photons from galaxies \citep[e.g.,][]{Reddy_2016,Chisholm_2018,Steidel_2018}. When the covering fraction is high, the line-of-sight opacity is high, and fewer LyC photons can escape; a lower covering fraction, in contrast, implies greater porosity and higher escape fractions. This makes the covering fraction, when considered alongside dust attenuation, a useful proxy for estimating $f_{\mathrm{esc}}^{\: \mathrm{LyC}}$. 

Strong, saturated LIS absorption indicates a high covering fraction of metal-enriched neutral gas, while shallow or partial absorption implies optically thin regions along the line-of-sight (with lower column densities), and/or lower covering fractions of absorbing gas. However, as discussed by \cite{Flury_2025}, this interpretation can be complicated by line saturation effects: saturated lines can appear shallow if the gas geometry or instrument resolution leads to partial infilling or dilution of absorption features. In such cases, the observed line depth or equivalent width may not reliably reflect the true covering fraction or column density. These effects are particularly relevant in the context of LyC escape and highlight the importance of carefully disentangling covering fraction from saturation and radiative transfer effects.

Using a sample of nine low-redshift ($z < 0.3$) LyC leakers, \citet{Chisholm_2018} established an empirical relationship between the \ion{H}{1} covering fraction and $f_{\mathrm{esc}}^{\: \mathrm{LyC}}$ (Equation \ref{eq:cf_fesc} below). This formulation assumes that LyC photons are absorbed by optically thick \ion{H}{1} gas with $N_{\mathrm{HI}} \gtrsim 10^{17.2}$ cm$^{-2}$ and that dust attenuation can be modeled using $k_{912} = 12.87$ from \citet{Reddy_2016b} and the $E(B-V)_\star$ values derived from our \texttt{Starburst99} fits (listed in Table~\ref{tab:fit_props}).
\begin{equation} \label{eq:cf_fesc}
    f_{\mathrm{esc}}^{\: C_f} = 10^{-0.4 \: k_{912} \: E(B-V)_\star}\times \left[1 - C_f\mathrm{(H \: I)}\right]
\end{equation}

Direct measurements of $C_f$(\ion{H}{1}) based on Ly$\beta$ absorption are not possible for most CLASSY galaxies due to the blue limit of the COS G130M grating. We therefore estimate $C_f$(\ion{H}{1}) using the empirical relation established with the LzLCS+ galaxies by \citet{Saldana-Lopez_2022} and the LIS covering fractions derived by \citet{Parker2024} from low-ionization metal absorption lines (e.g., \ion{C}{2}, \ion{Si}{2}).
 \begin{equation} \label{eq:cf}
     C_f\mathrm{(H \: I)} = (0.63 \pm 0.19) \times C_f\mathrm{(LIS)} \: + \: (0.54 \pm 0.09)
 \end{equation}

In this work, $C_f$(LIS) represents the fraction of the UV continuum obscured by metal-enriched LIS gas along the line-of-sight. In contrast, $C_f$(\ion{H}{1}) traces the covering of optically thick hydrogen (i.e., $N_{\text{H\,\textsc{i}}} \gtrsim 10^{17.2} \, \text{cm}^{-2}$), often inferred from the damping wings of Lyman-series absorption lines. Although both are derived from down-the-barrel absorption, these two covering fractions trace different column density regimes and can diverge when the neutral ISM is clumpy or partially ionized.

There is an ongoing debate on which covering fraction-- $C_f$(LIS) or $C_f$(\ion{H}{1})-- is more appropriate for predicting LyC escape. While only optically thick \ion{H}{1} can directly block LyC photons, such gas may fully cover the sightline while still allowing LyC escape if the column is below the Lyman limit threshold or the gas is partially ionized. Conversely, the LIS absorption lines trace higher metallicity material and are often associated with the denser ISM components, where LyC photons are most efficiently absorbed. As shown in \citet{Flury_2025}, $C_f(\text{LIS})$ provides a more reliable tracer of LyC escape than $C_f$(\ion{H}{1}). These findings align with the conclusions of the \citet{Saldana-Lopez_2022} that LIS lines may better reflect the geometry and clumpiness of optically thick gas than the \ion{H}{1} lines themselves.

Overall, however, there are several caveats to our estimates of $f_{\mathrm{esc}}^{\: \mathrm{LyC}}$ from the $C_f$(LIS) measurements. The $C_f$(LIS) values from \citet{Saldana-Lopez_2022} were derived using COS G140L spectra, which have significantly lower spectral resolution than the G130M observations in CLASSY.
The broader absorption profiles from the lower resolution spectra used to establish Equation \ref{eq:cf} would cause the measured residual flux ($R_f$; the normalized flux at the line's minimum) to be systematically biased towards higher values than the CLASSY measurements, resulting in lower inferred covering fractions when approximated as $C_f \sim 1 - R_f$. This resolution-induced bias in the covering fraction has also been quantitatively demonstrated by \citet[][using a Gaussian line spread function]{Jennings_2025} and \citet[][using the COS line spread function]{Flury_2024}, who showed that higher spectral resolution can reveal substructure in absorption troughs, leading to higher and more accurate $C_f$ values, particularly for partially covered sightlines. To assess the extent of this bias, we compare $C_f$(Ly$\beta$) for the 10 CLASSY galaxies with coverage of Ly$\beta$ with the estimated $C_f$(\ion{H}{1}) from Equation \ref{eq:cf} using the LIS lines in these galaxies. For galaxies with $C_f$(Ly$\beta$) $>$ 0.9, we find excellent agreement between the direct and estimated measurements. However, more variation may exist at lower covering fractions, though we cannot make a strong conclusion since there is only one galaxy for which this is the case. For this single galaxy (J0926+4427), $C_f$(Ly$\beta$) is $\sim0.1$ dex higher than the estimate of $C_f$(\ion{H}{1}) from the LIS lines (0.86 versus 0.76, respectively), corresponding to a $\sim$2.5\% lower prediction of $f_{\mathrm{esc}}^{\: C_f}$. There are 12 CLASSY galaxies for which we estimate $C_f$(\ion{H}{1}) to be less than 0.9 and so, at least for this subset, our derived values of $f_{\mathrm{esc}}^{\: C_f}$ may be systematically underestimated due to spectral resolution effects.

Finally, we note that the different LIS ions trace slightly different gas phases. \ion{O}{1} has an ionization potential nearly identical to \ion{H}{1} and is less affected by dust depletion than \ion{C}{2}, \ion{Si}{2}, or \ion{Al}{2}. Although LIS ions cover a slightly higher range of ionization energies, \citet{Parker2024} demonstrates that they are still closely tied to \ion{O}{1} in the ISM. Therefore, we adopt an error-weighted average of the available LIS lines and \ion{O}{1} from \citet{Parker2024}.


\subsection{$f_{\mathrm{esc}}^\beta$: from $\beta_{\mathrm{UV}}$-slope} \label{sub:method2}

The $\beta_{\mathrm{UV}}$-slope, defined as the slope of a galaxy's spectrum in the UV range, is expressed in the form \(F_\lambda \propto \lambda^\beta\), where \(F_\lambda\) represents the flux density at wavelength \W. Young, massive stars are the primary contributors to the UV continuum in star-forming galaxies; however, the observed $\beta_{\mathrm{UV}}$-slope is influenced by interstellar dust within the galaxy, which absorbs and scatters UV light, making the spectrum appear redder. Thus, the $\beta_{\mathrm{UV}}$-slope serves as an indicator of the properties of the ionizing stellar population and the amount of dust attenuation.

Using galaxies from LzLCS+, \citet{Chisholm_2022} found an empirical relationship between the LyC escape fraction and the $\beta_{\mathrm{UV}}$-slope:
\begin{equation} \label{eq:beta}
    f_{\mathrm{esc}}^\beta = (1.3 \pm 0.6) \times 10^{-4} \times 10^{(-1.22 \pm 0.10)\beta_{\mathrm{UV}}}
\end{equation}
This correlation exists because a bluer (more negative) $\beta_{\mathrm{UV}}$-slope indicates that there is less dust attenuation and/or a smaller contribution from the nebular continuum, i.e., less \ion{H}{1} gas, allowing more ionizing radiation to escape.

We measure $\beta_{\mathrm{UV}}$ for the CLASSY galaxies by fitting a power law to regions of the spectra between 1300\AA\ and 1800\AA\ that lack strong ISM features, specifically windows 1-7 from Table 2 of \citet{Calzetti_1994}. We used the \texttt{scipy.optimize.curve\_fit} package to perform these fits and estimate the uncertainty in $\beta_{\mathrm{UV}}$ using bootstrapping with 1000 Monte Carlo runs. The lower panel of Figure \ref{fig:SPS} shows an example of this fit for J1129+2034, where a dashed line shows the best-fit model and the MC uncertainty is shown in cyan. The windows used for these fits are shaded in pale orange.


\begin{figure}
    \centering
    \includegraphics[width=0.9\linewidth]{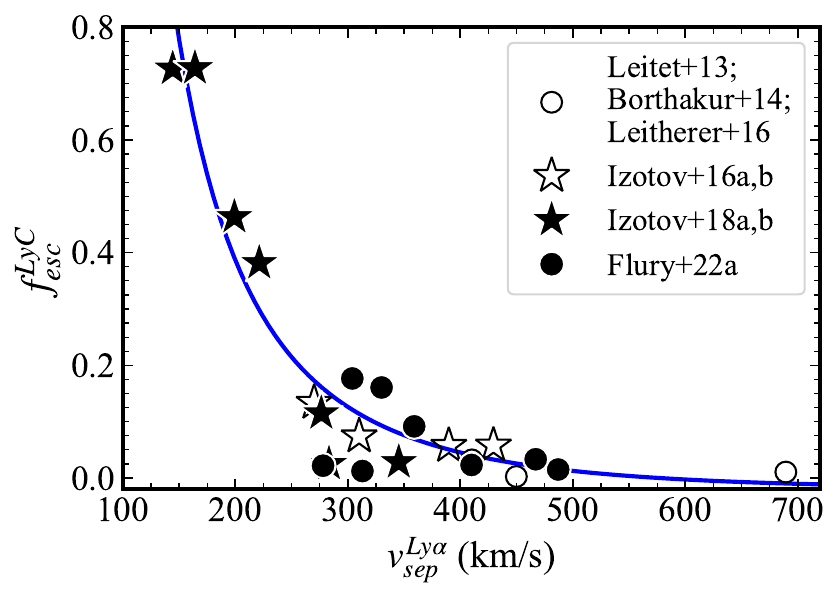}
     \caption{The trend between direct measurements of $f_{\mathrm{esc}}^{\: \mathrm{LyC}}$ and the peak separation of Ly$\alpha$, $\mathrm{v}_{\mathrm{sep}}^{\mathrm{Ly}\alpha}$. Individual measurements of $\mathrm{v}_{\mathrm{sep}}^{\mathrm{Ly}\alpha}$ from \cite{Leitet_2013}, \cite{Borthakur_2014}, and \cite{Leitherer_2016}, with escape fractions derived by \cite{Chisholm_2017}, are shown by empty circles. Measurements from \cite{Izotov_2016a} and \cite{Izotov_2016b} are shown with empty stars, while those from \cite{Izotov_2018a} and \cite{Izotov_2018b} are filled stars. Lastly, the measurements from \cite{Flury_2022a} for the eight LzLCS+ galaxies with double-peaked Ly$\alpha$ emission are shown by filled circles. The best-fit curve for these points is listed in Equation \ref{eq:vsep} and is in close agreement with the one reported by \cite{Izotov_2018b}.}
    \label{fig:vsep}
\end{figure}


\subsection{$f_{\mathrm{esc}}^{\: v_{\mathrm{sep}}}$: from Ly$\alpha$ peak separation} \label{sub:method3}

Ly$\alpha$ emission provides a powerful indirect diagnostic of the conditions that regulate the escape of ionizing photons \citep[e.g.,][]{Verhamme_2006, Dijkstra_2014, Hayes_2015}. Because Ly$\alpha$ photons are resonantly scattered by neutral hydrogen, their propagation through the ISM is highly sensitive to both the neutral gas column density and its velocity structure \citep[e.g.,][]{Neufeld_1990, Ahn_2003, Verhamme_2008}. This process can result in a double-peaked Ly$\alpha$ emission profile, where the velocity separation between these peaks ($\mathrm{v}_{\mathrm{sep}}^{\mathrm{Ly}\alpha}$) is linked to the degree of scattering among the escaping Ly$\alpha$ photons \citep{Verhamme_2015, Orlitova_2018, Izotov_2021}.

Smaller peak separations indicate lower \ion{H}{1} column densities or a more porous ISM, which are also more conducive to LyC escape. This makes $\mathrm{v}_{\mathrm{\mathrm{sep}}}^{\mathrm{Ly}\alpha}$ a promising predictor of the LyC escape fraction. Theoretical radiative transfer models have established the link between Ly$\alpha$ kinematics and gas properties \citep{Verhamme_2006,Verhamme_2015,Schaerer_2011}, and observational studies have since confirmed a strong empirical correlation between $f_{\mathrm{esc}}^{\: \mathrm{LyC}}$ and $\mathrm{v}_{\mathrm{\mathrm{sep}}}^{\mathrm{Ly}\alpha}$ among known LyC leakers \citep[e.g.,][]{Verhamme_2017,Izotov_2018b,Izotov_2021,Izotov_2022,Kakiichi_2021,Flury_2022b,Choustikov_2024}.

To refine the empirical relationship between the LyC escape fraction and the Ly$\alpha$ peak velocity separation (v$_{\rm sep}^{\rm Ly\alpha}$), we re-fit the trend using an expanded sample of known LyC leakers. 
We compile literature values of $f{\mathrm{esc}}^{\: \mathrm{LyC}}$ and $\mathrm{v}_{\mathrm{sep}}^{\mathrm{Ly}\alpha}$ from several studies \citep{Izotov_2016a, Izotov_2016b, Izotov_2018a, Izotov_2018b, Flury_2022a}, including eight galaxies from the LzLCS+ sample. 
The resulting best-fit relation is given by:
\begin{equation} \label{eq:vsep}
    f_{\rm esc}^{\rm v_{\rm sep}} = \frac{2.37 \times 10^4}{({\rm v_{sep}^{Ly\alpha}})^2} - \frac{41.0}{\rm v_{sep}^{Ly\alpha}},
\end{equation}
where ${\rm v_{sep}^{Ly\alpha}}$ is in km s$^{-1}$. 
The measurements and the best-fit curve are shown in Figure~\ref{fig:vsep}. This revised equation is in close agreement with the earlier fit from \citet{Izotov_2018b}, but reflects the increased statistical power of the updated dataset. However, we note that several studies have found the $\mathrm{v}_{\mathrm{sep}}^{\mathrm{Ly}\alpha}$ method to be less effective at high redshift \citep{Pahl_2023,Marques-Chaves_2024,Choustikov_2024,Kerutt_2024}.

The Ly$\alpha$ profiles for the CLASSY galaxies were presented and analyzed in \cite{Hu_2023}.


\subsection{$f_{\mathrm{esc}}^{\: \mathrm{O}_{32}}$: based on O$_{32}$} \label{sub:method6}
The O$_{32}$ (= [\ion{O}{3}] \W5007 / [\ion{O}{2}] \W3727) emission line flux ratio is a widely used diagnostic for probing the potential escape of ionizing photons from galaxies. Although O$_{32}$ has been observed to correlate with $f_{\mathrm{esc}}^{\: \mathrm{LyC}}$ in several samples of star-forming galaxies \citep[e.g.,][]{Jaskot_2013,Izotov_2016b,Izotov_2018b,Nakajima_2020,Flury_2022a}, there can be significant scatter in this trend and numerous exceptions have been found, including non-leaking galaxies with high O$_{32}$ values \citep[e.g.,][]{Izotov_2016a,Izotov_2018a,Naidu_2018,Wang_2019}. 

A key reason for this scatter is that O${32}$ is sensitive to a wide range of physical conditions, including the ionization parameter, the metallicity of the gas phase, the electron density, and the geometry and distribution of the ionized gas. Because it is a volume-averaged global quantity, O$_{32}$ can also be influenced by emission scattered in or out of the line of sight. In contrast, $f_{\mathrm{esc}}^{\: \mathrm{LyC}}$ depends strongly on the line-of-sight neutral gas column density, making it more sensitive to directional anisotropies in the ISM. As a result, a single O$_{32}$ value can correspond to a wide range of LyC escape fractions depending on the underlying gas distribution.

Some studies suggest that galaxies with relatively isotropic, density-bounded gas distributions exhibit tighter correlations between O$_{32}$ and $f_{\rm sc}^{\rm LyC}$, while galaxies with clumpy or porous ISM structures may show strong directional variation in LyC escape even at fixed O$_{32}$ \citep[e.g.,][]{Bassett_2019}. Both observational studies \citep[e.g.,][]{Jaskot_2013, Izotov_2018a, Gazagnes_2020, Flury_2022a, Naidu_2022} and simulation-based studies \citep[e.g.,][]{Barrow_2020, Katz_2020} converge to the conclusion that high O$_{32}$ values are not sufficient indicators of LyC leakage. However, elevated O$_{32}$ can still serve as a useful tracer of ionization conditions that are potentially conducive to LyC escape, particularly when combined with other diagnostics.

To predict the LyC escape fraction from the measured O$_{32}$ ratios of CLASSY galaxies, we use the empirical relation from \cite{Chisholm_2018}: 
\begin{equation} \label{eq:o32}
    f_{\mathrm{esc}}^{\: \mathrm{O}_{32}} = (0.0017 \pm 0.0004) \times \mathrm{O}_{32}^2 + (0.005 \pm 0.007).
\end{equation}


\begin{figure*}
    \centering
    \includegraphics[width=\linewidth]{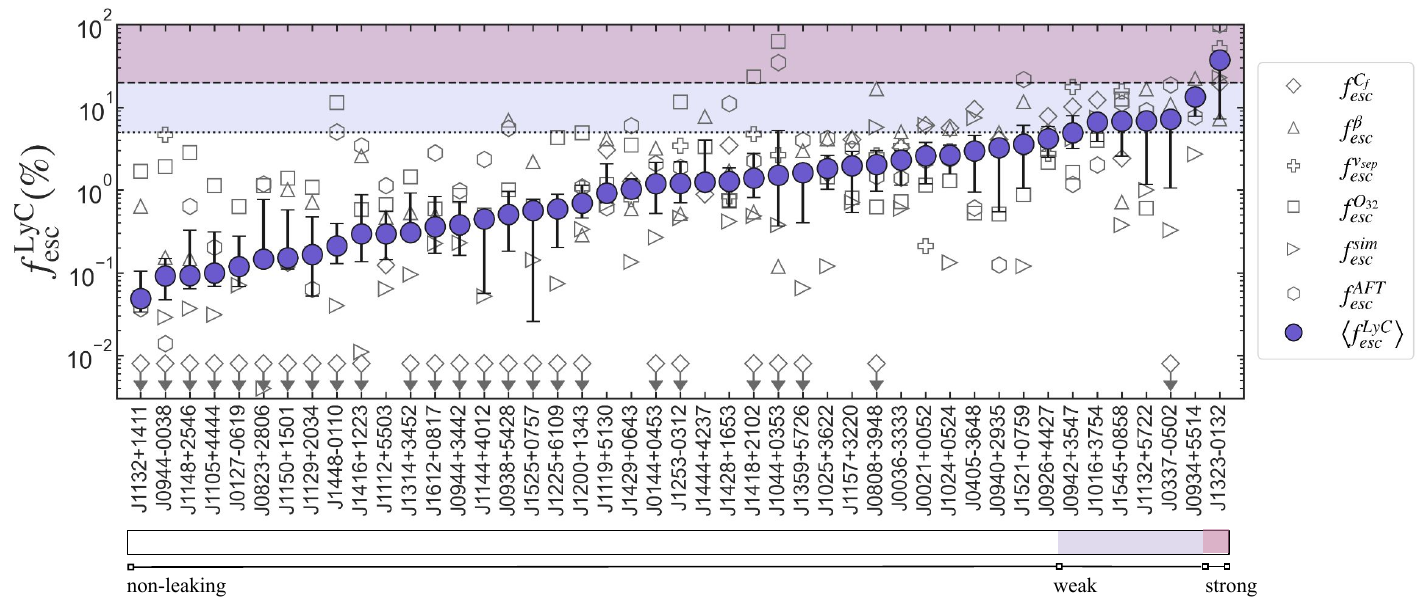}
     \caption{$f_{\mathrm{esc}}^{\: \mathrm{LyC}}$ from different indirect methods for CLASSY galaxies, as shown by different shapes of empty markers: diamonds ($f_{\mathrm{esc}}^{\: C_f}$), vertical triangles ($f_{\mathrm{esc}}^{\: \beta}$), pluses ($f_{\mathrm{esc}}^{\: v_{\mathrm{sep}}}$), squares ($f_{\mathrm{esc}}^{\: \mathrm{O}_{32}}$), horizontal triangles ($f_{\mathrm{esc}}^{\mathrm{sim}}$), and hexagons ($f_{\mathrm{esc}}^{\: \mathrm{AFT}}$).
     The medians of these estimates are shown by purple points, with uncertainties represented by the 16th and 84th percentiles from 300 MC variations of the individual $f_{\mathrm{esc}}^{\: \mathrm{LyC}}$ predictions based on their uncertainties. Overall, $\sim50\%$ of CLASSY galaxies have $\left<f_{\mathrm{esc}}^{\: \mathrm{LyC}}\right> > 1\%$, with multiple meeting our criterion for weak LyC leakers ($5\% \leq \left<f_{\mathrm{esc}}^{\: \mathrm{LyC}}\right> < 20\%$) and one that is a candidate for strong LyC leaking (J1323-0132; $\left<f_{\mathrm{esc}}^{\: \mathrm{LyC}}\right> \geq 20\%$).}
    \label{fig:esc_comp}
\end{figure*}


\subsection{$f_{\mathrm{esc}}^{\mathrm{sim}}$: based on \texttt{Ramses-RT} simulation} \label{sub:method5}

\citet{Gazagnes_2023} simulated 22,500 spectra of interstellar \ion{C}{2} \W1334 and \ion{Si}{2} \W1260 absorption across different times and along different lines-of-sight for a single galaxy with $\log$(M$_\star)$ = 9, using the zoom-in \texttt{Ramses-RT} simulation from \citet{Mauerhofer_2021}. They found that these mock profiles were able to accurately represent the observations for 35 (of the 45) CLASSY galaxies, spanning broad ranges in stellar mass (10$^6$–10$^9$ M$_\sun$) and metallicity (0.02–0.55 Z$_\sun$). 

To quantify the agreement between these simulated spectra and the CLASSY galaxies, we use the same procedure as \citet{Gazagnes_2023}, where the goodness-of-fit is defined by
\begin{equation}
    \chi^2 = \frac{\chi_{\mathrm{C II}}^2 \: + \: \chi_{\mathrm{Si II}}^2}{n_{\mathrm{C II}}^{\mathrm{obs}} \: + \: n_{\mathrm{Si II}}^{\mathrm{obs}}}
\end{equation}
with
\begin{equation*}
    \chi_{\mathrm{ion}}^2 = \Sigma\left(\frac{F_{\lambda}^{\mathrm{obs}} - F_{\lambda}^{\mathrm{sim}}}{\sigma_{\lambda}^{\mathrm{obs}}}\right)^2
\end{equation*}
In this equation, $F_{\lambda}^{\mathrm{obs}}$ and $F_{\lambda}^{\mathrm{sim}}$ are the fluxes in the \ion{C}{2} and \ion{Si}{2} spectral windows, $\sigma_{\mathrm{ion}}^\mathrm{obs}$ is the uncertainty of the observed flux and $n_{\mathrm{ion}}^\mathrm{obs}$ is the number of bins in the spectrum for each ion. When making this comparison between the mock and observed spectra, they are normalized using the median of the flux in a feature-free interval near each line, with the mock spectra degraded to the resolution of the observations \citep[Table 3 in][]{Berg_2022}.

We estimate $f_{\mathrm{esc}}^{\mathrm{sim}}$ and its uncertainties as the median, 16th, and 84th percentiles of the escape fractions for all models with a reduced $\chi^2 < 1.5$ (or $\chi^2 < 3$ if none met this more stringent criterion). These are the same thresholds used by \cite{Gazagnes_2024} and lead to estimates of $f_{\mathrm{esc}}^{\mathrm{sim}}$ for 33 (of the 45) CLASSY galaxies using the more stringent $\chi^2$ criterion and an additional two that met $\chi^2 < 3$.
We note that the predicted $f_{\mathrm{esc}}^{\: \mathrm{LyC}}$ from these simulated spectra are not necessarily independent of those predicted using the LIS covering fractions, as both utilize interstellar metal lines.


\subsection{$f_{\mathrm{esc}}^{\: \mathrm{AFT}}$: using \texttt{LyCsurv}} \label{sub:method4}

The \texttt{LyCsurv}\footnote{https://github.com/sflury/LyCsurv} code \citep{Flury_lycsurv} provides a multivariate framework to predict the LyC escape fraction indirectly. Using survival analysis techniques, it incorporates both detections and upper limits of the LzLCS+ sample to predict $f_{\mathrm{esc}}^{\: \mathrm{LyC}}$ from galaxy-scale physical properties \citep{Jaskot2024a, Jaskot2024b}. When applying \texttt{LyCsurv} to the CLASSY sample, we use nine input observables from Tables~\ref{tab:fit_props} and \ref{tab:results} (in the Appendix): stellar mass (M$_\star$), UV and nebular dust attenuation [$E(B-V)_\star$, $E(B-V)_{\mathrm{neb}}$], total SFR, SFR surface density [$\Sigma_{\mathrm{SFR}}$ = SFR/(2$\pi \times r_{\mathrm{opt}}$)], absolute UV magnitude (M$_{1500}$), UV slope ($\beta_{\mathrm{UV}}$), ionization parameter (O$_{32}$), and H$\beta$ equivalent width [$W_\lambda$(H$\beta$)]. Because some of these observables are also used in other indirect methods presented in this work, the \texttt{LyCsurv} predictions are not entirely independent. However, the multivariate nature of this method helps mitigate biases associated with any single diagnostic and is expected to yield more robust overall estimates. However, it is important to note that several CLASSY galaxies fall outside the original parameter space of the LzLCS+ sample, specifically including fainter and less actively star-forming galaxies. As a result, \texttt{LyCsurv} must extrapolate to predict $f_{\mathrm{esc}}^{\: \mathrm{LyC}}$ for this portion of the sample, which could introduce additional uncertainties.

We employ the survival analysis models available through \texttt{LyCsurv}: the Cox proportional hazards model \citep{Cox_1972}, a semi-parametric approach, and the accelerated failure time model (AFT), which assumes a parametric form for the survival function. \citet{Flury_lycsurv} find that the AFT model typically predicts higher $f_{\mathrm{esc}}^{\: \mathrm{LyC}}$ values than the Cox model, probably because it is more sensitive to variations in line-of-sight properties, such as anisotropies in the gas distribution. For each CLASSY galaxy, \texttt{LyCsurv} returns a Cox model classification flag, which indicates whether the predicted survival curve lies consistently below (–1) or above (+1) the median of the predicted distribution. We find that this flag is +1 for the majority of the CLASSY sample, meaning the Cox model failed to predict a specific value confidently and instead returned an upper limit estimate of $f_{\mathrm{esc}}^{\mathrm{Cox}} = 0.03$\%.  This value should not be interpreted as a physical limit, but rather as an artifact of the model's assumptions when constraints are weak. In contrast, the AFT model consistently returns a range of values across the CLASSY sample, including escape fractions below 0.03\%, and does not rely on a fixed cutoff. For the few galaxies where the Cox model does not return a default value, the Cox and AFT predictions are in close agreement. Due to its broader applicability and consistency, we adopt AFT-based predictions ($f_{\mathrm{esc}}^{\: \mathrm{AFT}}$) as preferred \texttt{LyCsurv} results throughout this work.


\subsection{Median predictions $\left<f_{\mathrm{esc}}^{\: \mathrm{LyC}}\right>$} \label{sub:avg}

The indirect methods used here utilize distinct observables that are associated with different aspects of ionizing photon escape, and so are not necessarily expected to agree intrinsically. As a result, these methods often exhibit substantial scatter, particularly at low escape fractions \citep[e.g., ][]{Roy_2024,Citro_2024}. Furthermore, since the CLASSY galaxies lack direct measurements of $f_{\mathrm{esc}}^{\: \mathrm{LyC}}$, we cannot conclusively determine the accuracy of any given method. When considering systematic differences between these methods, we, therefore, are considering the relative biases between these predictions rather than an intrinsic offset from the true escape fraction.

We use the median of the six indirect $f_{\mathrm{esc}}^{\: \mathrm{LyC}}$ predictions, hereafter denoted as $\left<f_{\mathrm{esc}}^{\: \mathrm{LyC}}\right>$, in an attempt to estimate a general prediction of the escape fraction that is less affected by the biases of the individual methods. This is meant to mitigate the influence of any one method that may systematically under- or over-predict the escape fraction due to its physical assumptions or sensitivities (e.g., to dust, geometry, or outflow orientation). However, we acknowledge that not all methods are independent. For instance, both the covering fraction method and the RAMSES-based simulation method rely in part on similar neutral gas diagnostics (e.g., LIS absorption lines), which introduces correlated uncertainties. As such, $\left<f_{\mathrm{esc}}^{\: \mathrm{LyC}}\right>$ is not a fully independent summary statistic of the escape fraction, but it does offer a way to compare galaxies internally within CLASSY, under a consistent framework of indirect diagnostics. In the next section, we assess the consistency of these predictions across methods by examining trends in dust and neutral gas content.

These median estimates, $\left<f_{\mathrm{esc}}^{\: \mathrm{LyC}}\right>$, along with those from individual methods, are listed in Table \ref{tab:results} of the Appendix. Overall, all 45 CLASSY galaxies have predicted escape fractions from the $\beta$-slope, multivariate AFT, and O$_{32}$ methods, whereas only 20 have non-zero $f_{\mathrm{esc}}^{\: \mathrm{LyC}}$ estimates from their LIS covering fractions, 26 have double-peaked Ly$\alpha$ emission, and 35 have estimates from the simulation method. In total, 11 galaxies have non-zero estimates of $f_{\mathrm{esc}}^{\: \mathrm{LyC}}$ from all six methods while 39 have non-zero estimates from at least four of the methods.

Figure \ref{fig:esc_comp} shows the $\left<f_{\mathrm{esc}}^{\: \mathrm{LyC}}\right>$ estimates (purple circles) for the CLASSY galaxies, along with the estimates from the various indirect methods, in order of increasing $\left<f_{\mathrm{esc}}^{\: \mathrm{LyC}}\right>$ on the $x$-axis. The shaded regions represent the ranges that we define as weak LyC leakers (blue) and strong LyC leakers (purple). We find that five of the CLASSY galaxies appear to be weak LyC leaker candidates and that one (J1323-0312) is likely a strong LyC leaker.


\section{Comparing Escape Fraction Predictions} \label{results}

In this section, we present the results of the indirect methods used to estimate $f_{\mathrm{esc}}^{\: \mathrm{LyC}}$ and consider how these techniques can be used in conjunction to more robustly constrain the ionizing escape fraction in galaxies where it cannot be directly measured. Given the general lack of direct measurements of LyC escape for the CLASSY sample, our comparisons in this section are limited to cross-method consistency and systematic differences, not to tests of accuracy. This makes it essential to identify which galaxy properties may bias or drive divergence between methods.


\subsection{Haro 11: a direct measurement} \label{sub:direct}

None of the CLASSY spectra extend blue enough with sufficient S/N to directly measure the LyC. As a result, Haro 11 (J0036-3333) is the only CLASSY galaxy with a confirmed LyC detection.
This detection was made using the Far Ultraviolet Spectroscopic Explorer (FUSE), which had a large aperture (30\arcsec$\times$30\arcsec) and thus measured the integrated LyC escape fraction across the entire galaxy. The FUSE data revealed a direct escape fraction of $f_{\mathrm{esc}}^{\: \mathrm{LyC}} = 3.3 \pm 0.7$\% \citep{Bergvall_2006, Leitet_2011, Leitet_2013}. In addition, \citet{Komarova_2024} performed a spatially resolved LyC analysis of Haro 11 with COS/G140L, measuring escape fractions for each of its three bright UV knots. CLASSY COS observations only target one of these knots (knot C) in Haro 11, for which \citet{Komarova_2024} measured a LyC escape fraction of 5.1 $\pm$ 4.3\%.

The predicted escape fraction of our six indirect estimators is $\langle f_{\mathrm{esc}}^{\: \mathrm{LyC}} \rangle = 2.32_{-1.18}^{+0.68}$\%, which is in excellent agreement with both the integrated FUSE measurement and the estimate for knot C specifically. Individual predictions from each method are listed in Table~\ref{tab:results} in the appendix, and all six are notably consistent for Haro 11, more so than for many other galaxies in our sample. Although uncertainties in the indirect estimates can be significant, this agreement provides strong validation for the ensemble approach used here.

\begin{figure*}
    \centering
    \includegraphics[width=0.9\linewidth]{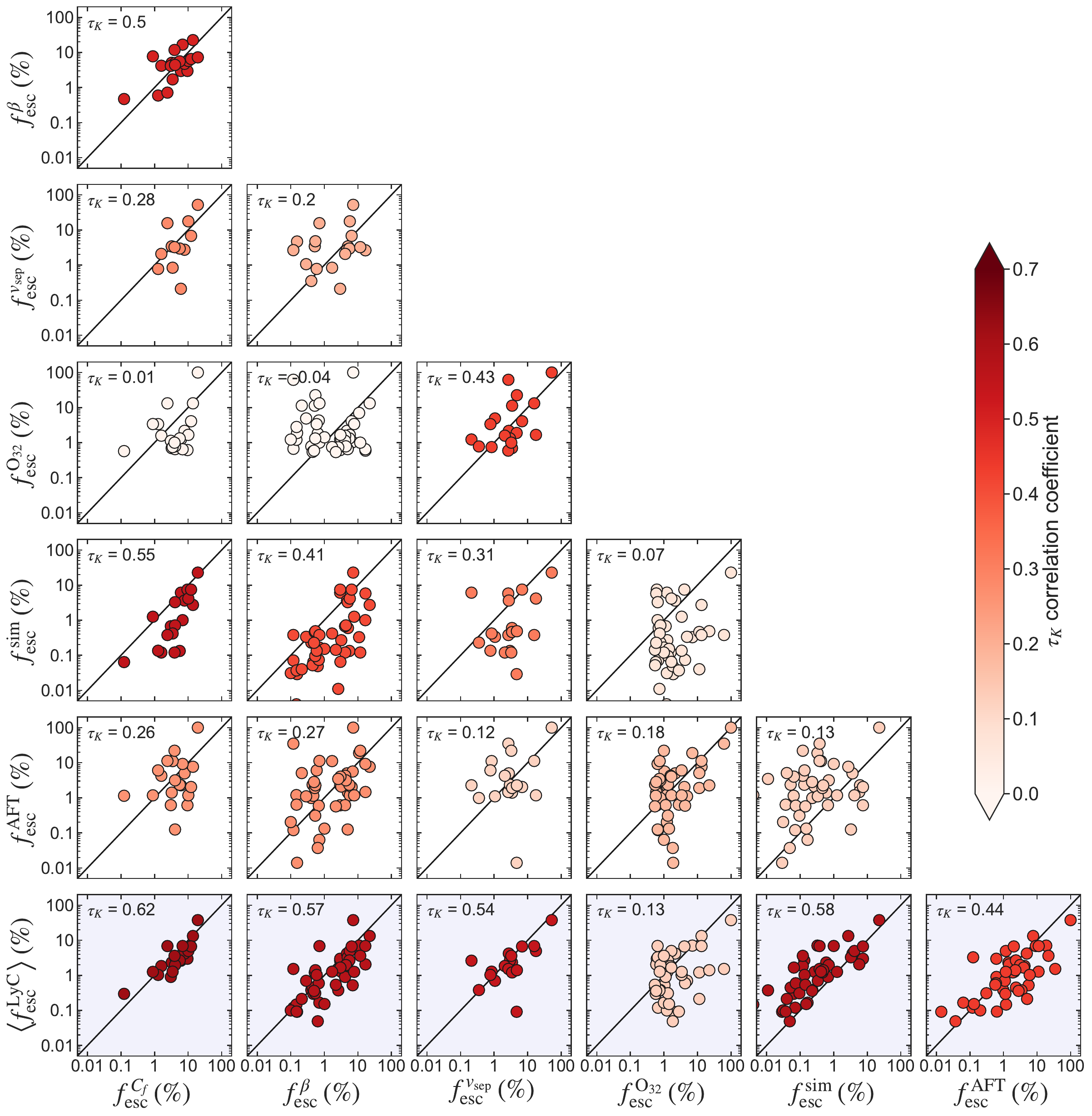}
     \caption{Pairwise comparison of six indirect estimates of $f_{\mathrm{esc}}^{: \mathrm{LyC}}$ for the CLASSY sample, as well as the median of these predictions $\left<f_{\mathrm{esc}}^{\: \mathrm{LyC}}\right>$. Each panel displays the Kendall’s tau ($\tau_K$) correlation coefficient between two methods, with the points color-coded by the $\tau_K$ value to visually highlight the strength of the correlation. Strongest correlations are found between methods that depend on shared observables (e.g., LIS absorption and dust), while weaker correlations appear for physically distinct methods such as O$_{32}$ or Ly$\alpha$ kinematics. This figure emphasizes both the areas of internal agreement and the systematic differences among the predictors.}
    \label{fig:coeffs}
\end{figure*}

\begin{figure}
    \centering
    \includegraphics[width=0.95\linewidth]{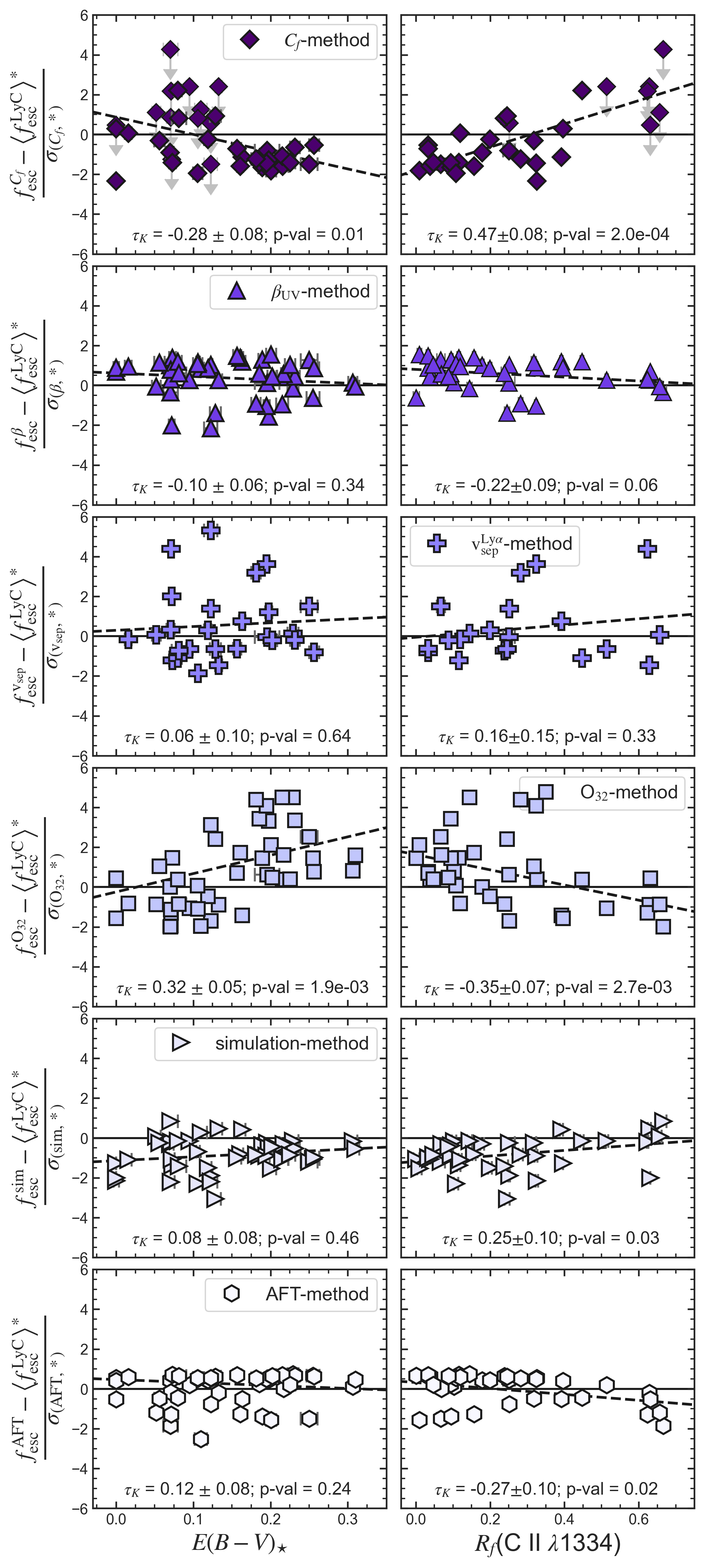}
     \caption{
     The difference between each indirect $f_{\mathrm{esc}}^{\: \mathrm{LyC}}$ measurement and the average of the other five indirect methods $\left<f_{\mathrm{esc}}^{\: \mathrm{LyC}}\right>^*$ as a function of $E(B-V)_\star$ for the left column and the \ion{C}{2} \W1334 residual flux for the right column. The $y$-axes correspond to how closely each estimate agrees with the other five, considering their uncertainties $\sigma$. This figure shows how different methods may be biased based on the amount of dust (left) and/or the amount of neutral gas (right) are present along the line-of-sight. Each panel includes a linear fit (dashed line) and lists the Kendall's tau correlation coefficient and corresponding $p$-value. The twelve galaxies with $C_f$(\ion{H}{1}) $< 0.9$ are shown as upper limits in the first row.}
    \label{fig:fesc_syst}
\end{figure}


\subsection{Exploring systematics between $f_{\mathrm{esc}}^{\: \mathrm{LyC}}$ predictions} \label{sub:comp_fesc}

Understanding the uncertainties and calibration of indirect methods is critical for their application at high redshift \citep[e.g.,][]{Mascia_2023,Mascia_2024}. Oftentimes, only one or two indirect methods can be applied for a given data set to predict the escape fraction, which makes it necessary to identify any systematic differences that may exist between the predictions from these various methods. The broad wavelength coverage and high resolution of the CLASSY data provide the unique opportunity to use multiple of these techniques across the same sample and identify systematic differences between their predictions. Although we lack direct measurements of the LyC escape fraction for CLASSY, comparing the internal consistency between indirect methods can still reveal important physical trends and potential biases, particularly since some methods are expected to be more sensitive to different properties of the ISM, such as ionization versus dust content.

To assess the internal consistency among our six indirect estimators of $f_{\mathrm{esc}}^{\: \mathrm{LyC}}$, Figure \ref{fig:coeffs} presents a pairwise comparison between each of the methods, along with the median $\left<f_{\mathrm{esc}}^{: \mathrm{LyC}}\right>$ for the six diagnostics (bottom row). Each panel lists the corresponding Kendall's tau ($\tau_K$) correlation coefficient for each comparison, which quantifies the degree of rank-order agreement where 1 (-1) corresponds to a monotonically increasing (descreasing) relationship and 0 indicates a complete lack of correlation. The points are also color-coded by these values of $\tau_K$.

Several clear patterns emerge from this comparison in Figure \ref{fig:coeffs}:
\begin{itemize}
    \item The covering fraction method is strongly correlated with both the simulation ($\tau_K = 0.55$) and $\beta_{UV}$ predictions ($\tau_K = 0.28$). This reflects the shared dependence on dust content between $f_{\mathrm{esc}}^{\: C_f}$ and $f_{\mathrm{esc}}^{\: \beta}$, and on the optical LIS absorption profiles between $f_{\mathrm{esc}}^{\: C_f}$ and $f_{\mathrm{esc}}^{\: \mathrm{sim}}$.
    \item The O$_{32}$ method shows notably weak correlations with most other predictions, likely due to its independence from line-of-sight tracers like LIS absorption and dust attenuation. Its strongest (though still modest) correlation is with the Ly$\alpha$ peak separation method ($\tau_K = 0.43$), despite their different physical origins. This may reflect a partial link between high ionization and Ly$\alpha$ kinematics in low-\ion{H}{1} systems.
    \item The AFT method shows only mild correlations with other diagnostics, 
    consistent with the fact that it is not directly determined by any single physical tracer but rather a diverse array of observables.
    \item The median prediction $\left<f_{\mathrm{esc}}^{: \mathrm{LyC}}\right>$ generally aligns most closely with the simulation, covering fraction, and $\beta_{UV}$ predictions (all with $\tau_K \gtrsim 0.54$), reflecting their relatively higher mutual agreement.
\end{itemize}
These comparisons highlight the complementarity and potential biases of different methods, underscoring the importance of using a multitracer approach to capture both the global and line-of-sight aspects of LyC escape. Strong correlations where methods share physical dependencies are reassuring, while weaker correlations identify regimes where methods diverge, either due to probing different escape mechanisms or varying sensitivity to ISM geometry.

For further comparison between these methods, we calculate the residual between its predicted value and the median of the other five methods, denoted $\left<f_{\mathrm{esc}}^{\: \mathrm{LyC}}\right>^*$. This comparison reveals how much each technique may diverge from the ensemble of the others. Figure~\ref{fig:fesc_syst} shows the residuals between these individual predictions of $f_{\mathrm{esc}}^{\: \mathrm{LyC}}$ and $\left<f_{\mathrm{esc}}^{\: \mathrm{LyC}}\right>^*$, normalized by their respective uncertainties, as a function of $E(B-V)_\star$ (left column) and the residual flux $R_f$(\ion{C}{2} \W1334) (right column). We used residual flux rather than the \ion{C}{2} covering fraction in this comparison because, assuming high-resolution observations, it can be measured directly from the observed absorption profile. This allows us to evaluate systematic differences due to variations in the amount of dust and neutral gas along the line-of-sight.

To highlight any potential systematics with $E(B-V)_\star$ and $R_f$(\ion{C}{2} \W1334), each panel of Figure~\ref{fig:fesc_syst} includes a linear regression and lists Kendall's tau ($\tau_k$) correlation coefficients and associated $p$-values, as determined using the \texttt{Kendall}\footnote{https://github.com/sflury/kendall} package \citep{Flury_kendall}. This method can incorporate censored data and non-detections, making it particularly well-suited to our dataset. We adopt the convention that a $p$-value less than 0.05 corresponds to $>2\sigma$ significance. 

Among the six diagnostics, the $\beta_{\mathrm{UV}}$, \texttt{LyCsurv}, and $v_{\mathrm{sep}}^{\mathrm{Ly}\alpha}$ methods show no statistically significant ($> 2\sigma$) systematic dependence on either $E(B-V)_\star$ or $R_f$(\ion{C}{2}), suggesting that these predictors are relatively robust for a range of dust and neutral gas conditions. In contrast, the $C_f$ method predicts higher $f_{\mathrm{esc}}^{\: \mathrm{LyC}}$ values than the other methods in galaxies with low $E(B-V)_\star$ and weak \ion{C}{2} absorption (i.e., high $R_f$), and predicts lower $f_{\mathrm{esc}}^{\: \mathrm{LyC}}$ values for galaxies with more dust and/or gas along the line-of-sight. The O$_{32}$ method, however, tends to predict systematically higher values than the median of other indicators in dusty or gas-rich systems, which may reflect a breakdown in the assumptions behind this method, possibly because O$_{32}$ can remain high even in cases where dust attenuation suppresses the emergent LyC flux. These findings reinforce the view that neither method captures the full complexity of LyC escape physics when used in isolation: the $C_f$ method is strongly dependent on assumptions about ISM geometry and dust homogeneity, while O$_{32}$ is sensitive to ionization conditions. Simulation-based predictions show a modest overall offset towards lower $f_{\mathrm{esc}}^{\: \mathrm{LyC}}$ values, though without a strong correlation with either $E(B-V)_\star$ or $R_f$(\ion{C}{2}). 

While the trends in Figure~\ref{fig:fesc_syst} do not represent a direct validation (due to the lack of direct $f_{\mathrm{esc}}^{\mathrm{LyC}}$ measurements for CLASSY galaxies), they serve as internal consistency checks between indicators. For example, for galaxies with more dust and/or neutral gas, predictions that neglect these effects (e.g., O$_{32}$) may systematically overestimate the amount of ionizing radiation that is able to escape. In contrast, diagnostics that directly incorporate line-of-sight opacity (e.g., $C_f$) may yield more conservative estimates.

As discussed in \citet{Jaskot2024a,Jaskot2024b}, this distinction is physically meaningful: indicators that are not sensitive to the line-of-sight column density (such as O${32}$ or $\beta{\mathrm{UV}}$) tend to recover a more global LyC escape fraction, while those incorporating LIS absorption features or \ion{H}{1} column density more directly trace the line-of-sight escape. Their work shows that combining both types of indicator is often necessary to reconcile observed differences and more accurately recover the escape fraction measured from direct detections. The systematic differences between the predictions of these methods therefore emphasize the importance of using multifaceted approaches when indirectly predicting $f_{\mathrm{esc}}^{\mathrm{LyC}}$, as no single method is universally reliable across all dust geometries, ISM conditions, or orientations.


\section{Discussion} \label{discuss}

In this section, we present the trends we observe between ISM and stellar population properties and our indirect predictions of ionizing photon escape and propose a physical interpretation for how recent burst episodes relate to different pathways for LyC escape in the CLASSY sample.


\subsection{Connecting to Reionization} \label{sub:reion}

To contextualize the predicted escape fractions for CLASSY galaxies to underlying models of cosmic reionization, we divide the sample into three regimes: non-leakers ($f_{\mathrm{esc}}^{\: \mathrm{LyC}} < 5\%$), weak leakers ($5\% \leq f_{\mathrm{esc}}^{\: \mathrm{LyC}} < 20\%$), and strong leakers ($f_{\mathrm{esc}}^{\: \mathrm{LyC}} \geq 20\%$). These thresholds are chosen to distinguish between galaxies that likely played minimal roles in reionization, those that could have contributed meaningfully if numerous enough, and those capable of rapidly reionizing the IGM.

Galaxies with $f_{\mathrm{esc}}^{\: \mathrm{LyC}} < 5\%$ likely have insufficient leakage to contribute meaningfully to reionization, unless they are both extremely numerous and intrinsically bright. Even then, if $f_{\mathrm{esc}}^{\: \mathrm{LyC}} \approx 0$, they would not contribute to EoR regardless of abundance. Several simulations \citep[e.g., SPHINX, THESAN;][]{Rosdahl_2015, Kannan_2022} suggest that average escape fractions in the weak-leaker regime (5–20\%) are required to maintain reionization over extended timescales without exceeding observational constraints.

Strong leakers, in contrast, are capable of ionizing large volumes of the IGM efficiently and may be responsible for rapid or bursty phases of reionization \citep{Naidu_2020,Sharma_2016}. However, recent simulations indicate that a large population of galaxies with both high $f_{\mathrm{esc}}^{\: \mathrm{LyC}}$ and high $\xi_{\mathrm{ion}}$, at abundances matching those observed by JWST at $z > 6$, would risk overproducing the ionizing photon budget and cause the universe to reionize too quickly \citep[e.g.,][]{Munoz_2024, Atek_2024}. Alternatively, some find that the $\xi_{\mathrm{ion}}$ values from \citet{Simmonds_2024}-- which are adopted in many recent simulations-- are biased to be too high \citep{Pahl_2025}. As such, models that rely on strong leakers that are as numerous as currently observed may overestimate the total ionizing budget unless these revised $\xi_{\mathrm{ion}}$ values are taken into account.

This suggests that observed leakers, which often have extreme physical conditions, may not represent the full diversity of LyC-emitting galaxies. Reionization models calibrated solely on strong leakers may overestimate the typical escape fraction or miss delayed-escape systems altogether. Including both early and late escape modes-- and their demographic weights-- is therefore essential for producing realistic ionizing emissivity histories. For instance, \citet{Mascia_2024} demonstrate that the assumed LyC escape behavior of the more abundant, low-luminosity galaxies has a significant impact on the resulting reionization history, with markedly different outcomes depending on whether these galaxies are predominantly weak or strong leakers. Our classification framework, therefore, has implications for the broader context of reionization efficiency, galaxy abundance, and ionizing photon production efficiency $\xi_{\mathrm{ion}}$.

\begin{figure*}
    \centering
    \includegraphics[width=0.8\linewidth]{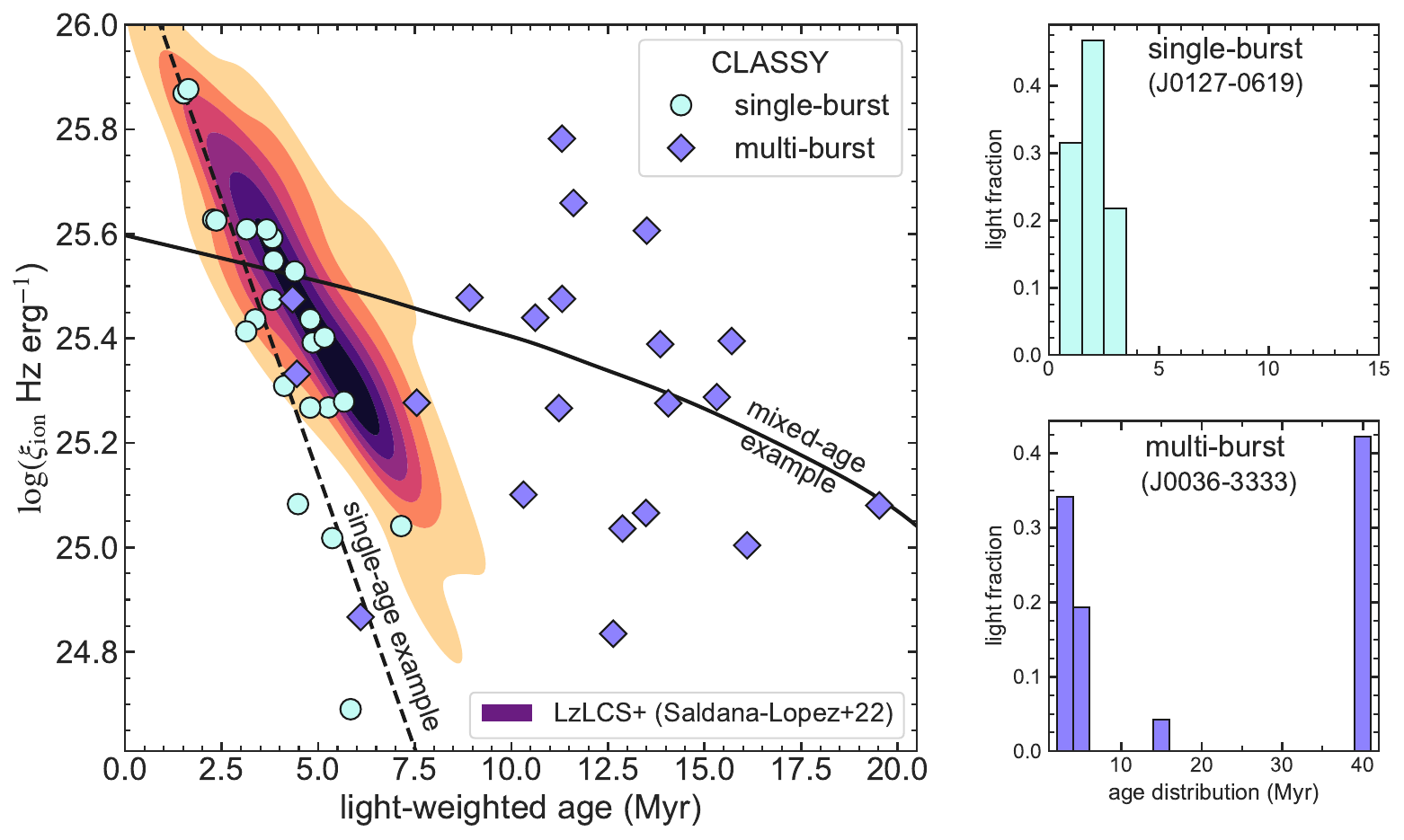}
     \caption{\textit{Left:} The predicted ionizing photon production, $\log(\xi_{\mathrm{ion}})$, for CLASSY galaxies as a function of their light-weighted stellar population ages. The dashed and solid lines show example trends from \cite{Chisholm_2019} that describe a single-age stellar population and a mixed-age stellar population, respectively, merely to demonstrate behavior that is typical for either SF history on this plot. We classify the CLASSY galaxies that lie near the single-age model as being "single-burst" systems (light blue circles) while those that deviate towards a flatter trend as "multi-burst" systems (purple diamonds). We note that some observations of CLASSY galaxies only include a portion of the galaxy itself, due to the size of the COS aperture, and therefore our classification of "single-burst" versus "multi-burst" is with regards to the stellar populations captured in the COS aperture and not necessarily the classification for the galaxy as a whole. The LzLCS+ galaxies are also overplotted as a density map \citep{Saldana-Lopez_2022}, showing that, unlike the CLASSY galaxies, the vast majority are in general agreement with a single-burst stellar population. \textit{Right:} Histograms with the population ages (based on the \texttt{SB99} fits) for two example CLASSY galaxies that are classified as "single-burst" (top panel) and "multi-burst" (bottom panel). Similar histograms were used to classify galaxies that could be in either sample based on the left panel.}
    \label{fig:subsamples}
\end{figure*}

\begin{figure*}
    \centering
    \includegraphics[width=\linewidth]{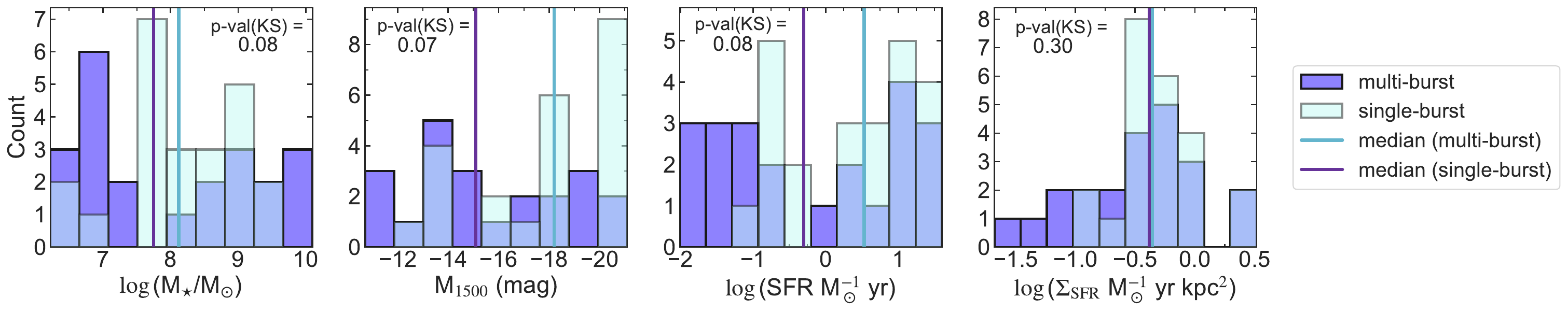}
     \caption{Distributions of large-scale properties of the "single-burst" (light blue) and "multi-burst" (purple) CLASSY populations. \textit{Left-to-right:} total stellar masses, 1500\AA\ AB magnitudes (M$_{1500}$), total star formation rates, and star formation surface density ($\Sigma_{\mathrm{SFR}}$). Solid lines of the corresponding color show medians of each distribution. The $p$-value according to a 2-sample KS test is listed in each panel, where a value $\leq 0.05$ corresponds to $\geq 2\sigma$ likelihood that the two subsets ("single-burst" versus "multi-burst") have different underlying distributions for a given property. With $p$-values of $\sim$0.08, the stellar masses, M$_{1500}$, and SFRs of the two subsets differ at nearly 2$\sigma$ significance. This suggests that a difference may intrinsically exist between these properties for the "single-burst" and "multi-burst" samples, but that it may be relatively weak or exhibit significant scatter. However, based on the results of these KS tests, it does indicate that the "single-burst" galaxies tend to be slightly more massive, brighter, and more actively star-forming than the "multi-burst" galaxies. There does not seem to be a difference between $\Sigma_{\mathrm{SFR}}$ in these samples based on the KS test, but it may be worth noting that the CLASSY galaxies with the lowest $\Sigma_{\mathrm{SFR}}$ are all in the "multi-burst" sample.}
    \label{fig:burst_hist}
\end{figure*}


\subsection{Recent burst history: "single-burst" vs. "multi-burst"} \label{sub:subsamples}

Using the \texttt{SB99} stellar continuum fits discussed in Section \ref{subsub:sps}, we consider the contribution of light from the observed spectra that is produced by stellar populations with varying ages and metallicities. The ages of the stellar populations responsible for significant fractions of the observed light can provide insights into the processes that are dominant in their local environments. For example, very massive stars (100-300 M$_\odot$) can reside in populations younger than $\lesssim$4 Myr. In contrast, these same stars will have ejected energy into their environments in the form of supernova explosions after $\gtrsim$7 Myr, creating cavities in the ISM.   

We estimated the luminosity-weighted ionizing photon production, $\log(\xi_{\mathrm{ion}})$, for the CLASSY galaxies using
\begin{equation} \label{eq:xi}
    \xi_{\mathrm{ion}} = \sum_{\mathrm{age}, \mathrm{Z}_\star} \mathrm{F(\mathrm{age}, \mathrm{Z}_\star)} \times \left[(1.6 \times 10^{26}) \times 10^{-0.21 \times \mathrm{age} - 0.28 \times \mathrm{Z}_{\star}}\right]
\end{equation}
where F(age, Z$_\star$) are the single-age predictions of $\xi_{\mathrm{ion}}$ weighted by their associated light fractions, and summed, based on Equation 11 from \cite{Chisholm_2019}, across the array of ages and metallicities from the \texttt{SB99} fits. The units for $\xi_{\mathrm{ion}}$, age, and metallicity are Hz erg$^{-1}$, Myr, and Z$_\odot$, respectively. 

The left panel of Figure \ref{fig:subsamples} shows the $\log(\xi_{\mathrm{ion}})$ predictions versus the light-weighted population ages for CLASSY, along with those corresponding to the LzLCS+ galaxies (purple shading; \citealt{Saldana-Lopez_2022}). The dashed and solid lines show examples for what is expected for a single-age stellar population and a mixed-age stellar population, respectively, based on empirical trends from \cite{Chisholm_2019}. These two models are overplotted to demonstrate the general behavior we expected from a single-age versus mixed-age stellar population, where single-age populations have a steep, decreasing trend between $\xi_{\mathrm{ion}}$ and the population's age. In contrast, a mixed-age population produces significant amounts of ionizing radiation over more extended periods of time, but produces fewer ionizing photons at very young ages ($\lesssim3$ Myr).

Comparing galaxy samples in Figure~\ref{fig:subsamples}, the LzLCS+ galaxies are best described by a single-age stellar population, whereas the CLASSY galaxies exhibit more diversity in behavior, with some lying near the single-age trend and others distinctly offset to the right, suggesting a mixed-age stellar population. Since there is overlap between these models at lower population ages, for the galaxies with light-weighted ages $<$ 10 Myr, we also considered the distribution of ages with non-zero light fractions from the \texttt{SB99} fits. The right panel of Figure \ref{fig:subsamples} shows an example of this for a galaxy we classify as "single-burst" (top panel) and one we classify as "multi-burst" (bottom panel). To make this distinction, we consider the difference between the youngest and oldest stars in a galaxy's stellar population. Galaxies with large differences in population ages ($\geq 20$ Myr) were classified as "multi-burst." The final classifications for CLASSY are distinguished in the left panel of Figure \ref{fig:subsamples} by light blue circles for the "single-burst" subset and purple diamonds for the "multi-burst" subset. We use these classifications ("single-burst" or "multi-burst") in the following analysis to investigate how the recent burstiness may relate to other galactic properties. Overall, we find that 23 CLASSY galaxies fall into this "single-burst" category, while the remaining 22 are classified as "multi-burst."

In Figure \ref{fig:burst_hist}, we compare some overall properties of the CLASSY galaxies in the "single-burst" and "multi-burst" subsets, specifically (from left to right) the stellar mass, the UV magnitude (M$_{1500}$), the SFR, and the SFR surface density ($\Sigma_{\mathrm{SFR}}$). The distribution of these global properties is shown in light blue for the "single-burst" sample and in purple for the "multi-burst" sample. Vertical lines of the same color show the medians of each distribution. To evaluate the likelihood that these distributions differ between the two subsets, we used a 2-sample Kolmogorov-Smirnov (KS; \texttt{scipy.stats.2samp\_ks}) test. The $p$-value from this test is listed in each panel, where a value $\leq 0.05$ corresponds to a $\geq 2\sigma$ confidence that they have different underlying distributions for a given property. The 2-sample KS-tests result in $p$-values of 0.07-0.08 for the left three panels, which indicates that there is likely a statistically significant difference between the stellar masses, M$_{1500}$, and SFRs of these populations, but that it is not quite 2$\sigma$ and so significant scatter may be present.

Overall, we find that the "single-burst" galaxies in the CLASSY sample tend toward slightly higher stellar masses, though the "multi-burst" exhibit a somewhat bimodal distribution, containing multiple very low-mass galaxies and high-mass galaxies but few around $\log$(M$_\star$) $\sim 8$, where the "single-burst" galaxies tend to be more common. The most significant difference between the properties in this figure seems to be with the UV magnitude, where the "single-burst" systems are generally more UV-luminous than the "multi-burst" systems. This suggests more intense and concentrated star formation in these single-aged galaxies. The "single-burst" galaxies also tend to have higher SFRs (and potentially higher $\Sigma_{\mathrm{SFR}}$) than the "multi-burst" galaxies. The KS test does not indicate that there is a statistical difference between $\Sigma_{\mathrm{SFR}}$ in these samples, but it is worth noting that the CLASSY galaxies with the lowest $\Sigma_{\mathrm{SFR}}$ are all in the "multi-burst" sample.


\begin{figure*}
    \centering
    \includegraphics[width=\linewidth]{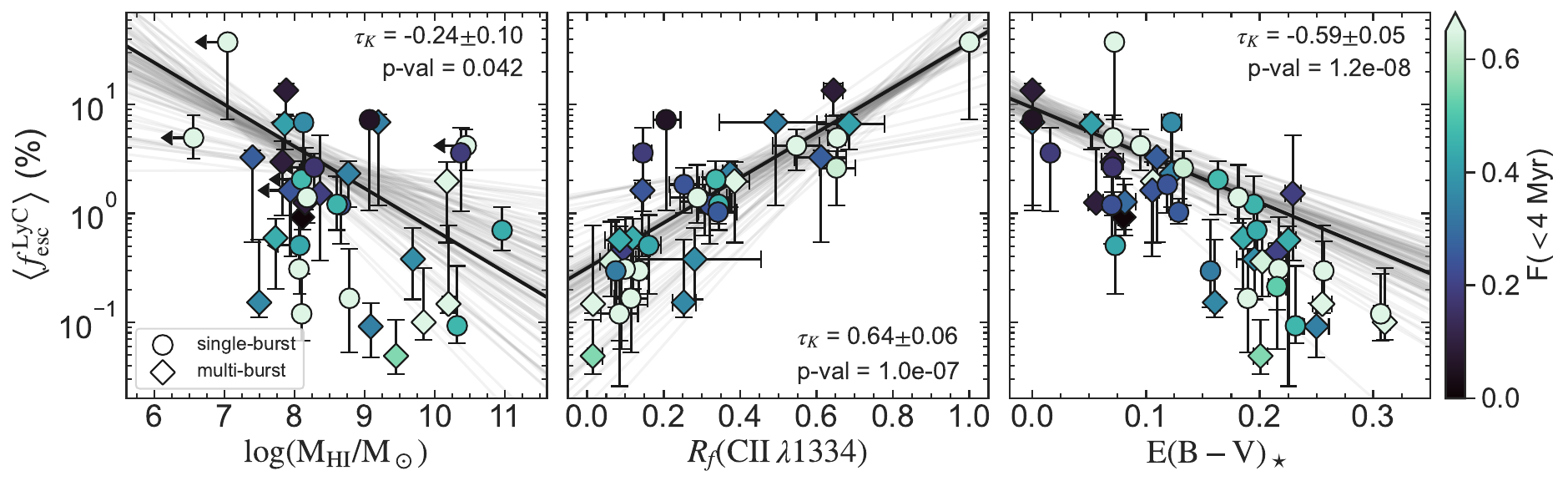}
     \caption{The median predicted ionizing photon escape fraction, $\left<f_{\mathrm{esc}}^{\: \mathrm{LyC}}\right>$, is shown as a function of three key galaxy properties: total \ion{H}{1} gas mass ($\log(M_{\mathrm{HI}}/M_\odot)$; left), residual flux of the \ion{C}{2} $\lambda1334$ absorption line ($R_f$(\ion{C}{2}); middle), and stellar continuum dust attenuation ($E(B-V)_\star$; right) for the "single-burst" and "multi-burst" subsets (circles and diamonds, respectively). Note: For the comparison with $R_f$(\ion{C}{2}), we exclude the $f_{\mathrm{esc}}^{\: C_f}$ predictions when calculating $\left<f_{\mathrm{esc}}^{\: \mathrm{LyC}}\right>$, since the covering-fraction method directly utilizes $R_f$(\ion{C}{2}). Points are color-coded by the fraction of UV light contributed by stellar populations younger than 4 Myr, as determined from the best-fit \texttt{SB99} models. The black line corresponds to the best fit for each panel while the grey lines represent variations in this fit based on the $x$ and $y$ uncertainties. Overall, this demonstrates that our predicted escape fractions tend to be higher in galaxies with less total \ion{H}{1} gas, less coverage of \ion{C}{2}-absorbing gas along the line-of-sight (i.e., a higher residual flux), and less dust.}
    \label{fig:esc_CII}
\end{figure*}


\subsection{Trends with $\left<f_{\mathrm{esc}}^{\: \mathrm{LyC}}\right>$} \label{sub:global}

The high S/N and high resolution of the CLASSY spectra provide the opportunity to investigate the conditions that enable LyC escape in ways that are often not possible with direct observations of LyC emitters. This allows us to explore how different host galaxy properties relate to ionizing photon escape, particularly with regard to ISM observables (\S~\ref{subsub:trends}, \S~\ref{subsub:dust_vs_hi}) and their recent star formation histories (\S~\ref{subsub:age_trends}). 


\subsubsection{With ISM observables} \label{subsub:trends}

This subsection explores how the prediction of the median escape fraction $\left<f_{\mathrm{esc}}^{\: \mathrm{LyC}}\right>$ varies with three key ISM diagnostics: the total \ion{H}{1} mass, the residual flux of interstellar \ion{C}{2} \W1334 absorption, and the UV dust attenuation $E(B-V)_\star$. The trends between these properties and $\left<f_{\mathrm{esc}}^{\: \mathrm{LyC}}\right>$ are shown in Figure \ref{fig:esc_CII}. The points are color-coded by the fraction of stellar continuum light originating from very young stars ($<$4 Myr), tracing the influence of O-star populations. In each panel, the black line corresponds to the best-fit model, and we report $\tau_k$ and its corresponding $p$-value.

We find that $\left<f_{\mathrm{esc}}^{\: \mathrm{LyC}}\right>$ exhibits a weak and marginally significant anti-correlation with total \ion{H}{1} mass ($\tau_K = -0.24$, $p$-value = 0.04), suggesting that larger reservoirs of neutral gas may modestly suppress LyC escape, although with considerable scatter. A stronger correlation is seen with the residual \ion{C}{2} flux ($\tau_K = 0.64$, $p$-value $\ll 0.05$), indicating that galaxies with stronger \ion{C}{2} \W1334 absorption (i.e., lower residual fluxes) tend to have lower LyC escape fractions. The best-fit trend between these measurements is reported below (Equation \ref{eq:fesc_fit}).
\begin{equation}\label{eq:fesc_fit}
    f_{\mathrm{esc}}^{\: \mathrm{LyC}} (\%) = (0.31 \pm 0.10) \: \times 10^{(2.07 \pm 0.14) \times R_f(\mathrm{C \: II})}
\end{equation}

The use of the \ion{C}{2} residual flux as a proxy for LyC escape has been explored in prior works with varying reliability. For example, \citet{Vasei_2016} and \citet{Chisholm_2018} found that the depth of \ion{C}{2} absorption alone may not reliably trace $f_{\mathrm{esc}}^{\: \mathrm{LyC}}$, due to the line's sensitivity to saturation, dust depletion, and ionized gas contamination. However, the CLASSY sample benefits from higher spectral resolution COS data (G130M and G160M), which allows more robust Voigt profile fitting and mitigates some of the limitations that affect these low-resolution studies. In particular, our measurements for $C_f$(\ion{C}{2}) come from simultaneous fits to multiple LIS ions and not a single \ion{C}{2} line. This decreases the effect that line saturation and contamination from higher energy absorption can have on estimates of $C_f$(\ion{C}{2}). Furthermore, our results are consistent with more recent corrections proposed by \citet{Jennings_2025} and \citet{Flury_2024}, who accounted for COS line-spread function effects and spectral stacking. Taken together, this suggests that residual flux in CLASSY provides a useful probe of neutral gas geometry in galaxies where direct \ion{H}{1} measurements are unavailable.

The strongest trend of the three shown in Figure \ref{fig:esc_CII} is with $E(B-V)_\star$ ($\tau_K = -0.59$, $p$-value $\ll 0.05$), indicating that galaxies that seem to have higher predicted LyC escape fractions have significantly less dust than those with lower predicted escape fractions. This suggests that dust plays an important role in establishing the amount of ionizing radiation that can escape, which is consistent with the findings from \citet{Saldana-Lopez_2022} and \citet{Chisholm_2022}. Equation \ref{eq:fesc_fit_ebv} reports the best-fit trend that we find between $E(B-V)_\star$ and our estimates of $\left<f_{\mathrm{esc}}^{\: \mathrm{LyC}}\right>$.
\begin{equation}\label{eq:fesc_fit_ebv}
    f_{\mathrm{esc}}^{\: \mathrm{LyC}} (\%) = (9.3 \pm 2.6) \: \times 10^{(-4.3 \pm 1.7) \times E(B-V)_\star}
\end{equation}

The color-coding in the left-most panel shows a general lack of a correlation between F($<$ 4 Myr) and the \ion{H}{1} mass when considering the full CLASSY sample ($\tau_K$ = 0.05; $p$-value = 0.67). For the "multi-burst" subset alone, however, there is a moderate positive correlation ($\tau_K$ = 0.43; $p$-value = 0.01), indicating that the more gas-rich "multi-burst" galaxies are more likely to have significant contributions from O-stars. This might reflect that these galaxies retain more gas to fuel repeated bursts, rather than undergoing gas blowout after a single episode. This reinforces the idea that "multi-burst" systems undergo ongoing or recurrent star formation rather than a rapid, burst-and-escape feedback cycle.

With regards to the middle panel of Figure \ref{fig:esc_CII}, there is a lack of a correlation between $R_f$(\ion{C}{2}) and F($<$ 4 Myr) for the full CLASSY sample ($\tau_K$ = 0.06; $p$-value = 0.6), as well as for the individual subsets: $\tau_K$ = (0.04, -0.06) with $p$-value = (0.8, 0.7) for "single-burst" and "multi-burst", respectively. This indicates that the \ion{C}{2} 1334$\AA$ residual flux is not directly linked to the fraction of light produced by O-stars, and hence may not be sensitive to short-term fluctuations in star formation activity.

Meanwhile, the color-coding in the right-most panel shows that the CLASSY galaxies with higher fractions of very young stars tend to have more dust present than those with a more significant contribution of light from older stellar populations. This relationship between $E(B-V)_\star$ and F($<$4 Myr) is statistically significant for the full CLASSY sample ($\tau_K$ = 0.37; $p$-value $<<$ 0.05), as well as for the "single-burst" ($\tau_K$ = 0.47; $p$-value = 0.002) and "multi-burst" samples ($\tau_K$ = 0.40; $p$-value = 0.009) individually. This makes sense with a physical picture where the youngest stellar populations are often still embedded in their natal dust clouds.

Together, these trends underscore that the escape of ionizing radiation is shaped by a combination of ISM structure, dust content, and stellar population age-- factors that must be considered jointly when interpreting LyC escape in both local analogs and high-redshift galaxies.


\begin{figure}
    \centering
    \includegraphics[width=\linewidth]{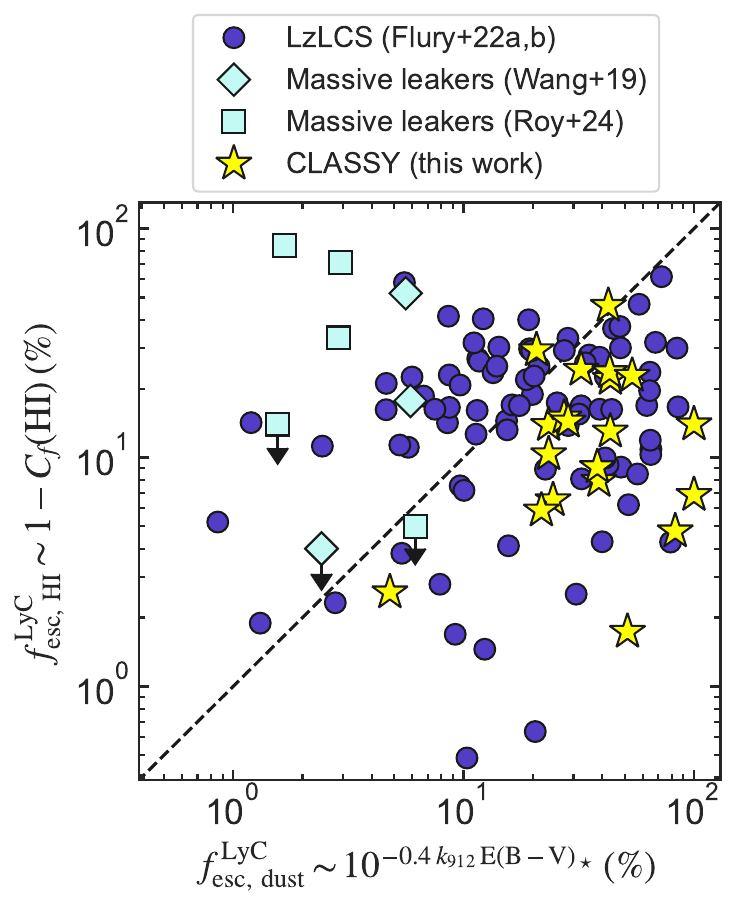}
     \caption{Estimates of the relative contribution of dust vs. \ion{H}{1} gas to the total LyC escape fraction, as predicted from the two components of Equation \ref{eq:cf_fesc}: $f_{\mathrm{esc,HI}}^{\: \mathrm{LyC}} \sim$ 1 - $C_f$(\ion{H}{1}) and $f_{\mathrm{esc,dust}}^{\: \mathrm{LyC}} \sim 10^{-0.4 \: k_{912} \: E(B-V)_\star}$. The dashed line indicates a one-to-one relationship. While the LzLCS+ galaxies (purple circles) have large scatter around this one-to-one line, the massive LyC leakers (light blue diamonds: Wang+19 and light blue squares: Roy+24) tend to lie above this line, while the CLASSY galaxies (yellow stars) are nearly all below this line. This suggests that the dust in these massive leaker galaxies significantly reduces their escaping LyC, while the neutral gas coverage plays a larger role in reducing the escape fraction for the CLASSY galaxies. This could indicate differences in the geometry of the neutral gas and dust between these galaxies, with a more porous distribution among those where $f_{\mathrm{esc,HI}}^{\: \mathrm{LyC}} > f_{\mathrm{esc,dust}}^{\: \mathrm{LyC}}$.}
    \label{fig:dust_vs_HI}
\end{figure}


\subsubsection{Weighing the effects of dust versus gas} \label{subsub:dust_vs_hi}

Based on the growing collection of known LyC leakers in the local universe, the LyC escape fraction can be approximated as a product of two main components: attenuation by neutral hydrogen and by dust, i.e., $f_{\mathrm{esc}}^{\: \mathrm{LyC}} \sim f_{\mathrm{esc,HI}}^{\: \mathrm{LyC}} \times f_{\mathrm{esc,dust}}^{\: \mathrm{LyC}}$ \citep[e.g.,][]{Reddy_2016,Gazagnes_2020,Ma_2020,Roy_2024}. Figure \ref{fig:dust_vs_HI} shows $f_{\mathrm{esc,HI}}^{\: \mathrm{LyC}}$ versus $f_{\mathrm{esc,dust}}^{\: \mathrm{LyC}}$ for the CLASSY galaxies (yellow stars), the LzLCS+ (purple circles), and samples of massive and dusty LyC leakers from \cite{Roy_2024} and \cite{Wang_2019} (light blue squares and diamonds, respectively). \cite{Roy_2024} presents a similar plot (their Figure 7), though without the addition of CLASSY. We approximate the escape fraction contributions as $f_{\mathrm{esc,HI}}^{\: \mathrm{LyC}} \sim$ 1 - $C_f$(\ion{H}{1}) and $f_{\mathrm{esc,dust}}^{\: \mathrm{LyC}} \sim 10^{-0.4 \: k_{912} \: E(B-V)_\star}$, which are the two components in Equation \ref{eq:cf_fesc} that we used to predict $f_{\mathrm{esc}}^{C_f}$. However, since we lack estimates of the \ion{H}{1} covering fractions for the massive leaker sample, we instead used the published values of $f_{\mathrm{esc,HI}}^{\: \mathrm{LyC}}$ for these galaxies, which were determined from the ratio of the dust-corrected versus the intrinsic estimates of F$_{\mathrm{LyC}}$/F$_{\lambda1100}$.

The dashed line in Figure \ref{fig:dust_vs_HI} corresponds to a one-to-one relationship. Although the LzLCS+ galaxies seem to be scattered relatively symmetrically on either side of this one-to-one line, the massive LyC leakers tend to lie above this line, while the CLASSY galaxies are nearly all below this line. This suggests that the dust in these massive leaker galaxies significantly reduces their escaping LyC. In contrast, the neutral gas coverage plays a larger role in reducing the escape fraction for the CLASSY galaxies, which points towards differences in the geometry of the neutral gas and dust between these samples, with a more homogeneous distribution of gas and dust among the CLASSY galaxies compared to those in the massive leaker samples from \cite{Wang_2019} and \cite{Roy_2024}.

\begin{figure}
    \centering
    \includegraphics[width=\linewidth]{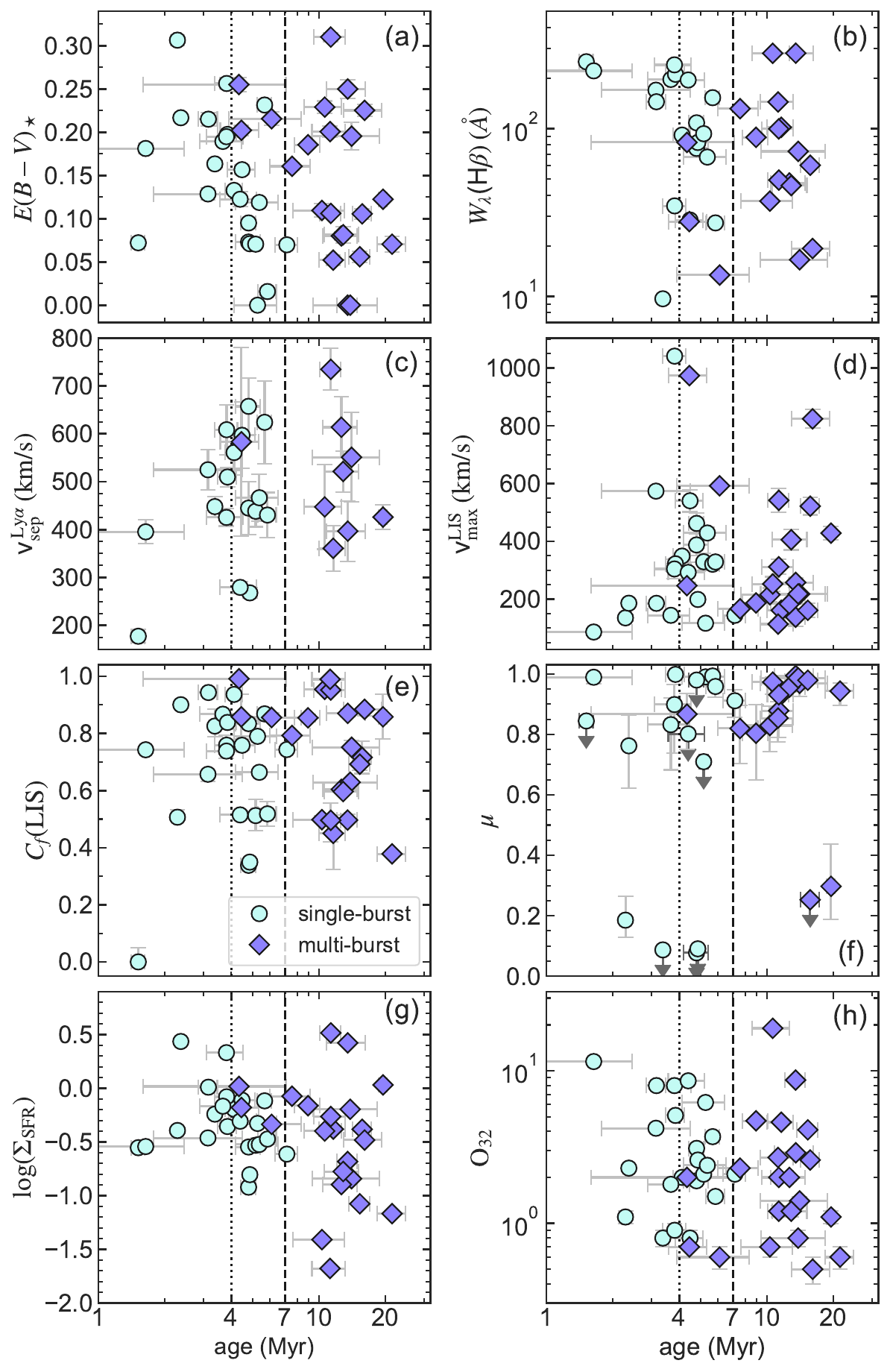}
     \caption{CLASSY galaxy properties as a function of the light-weighted stellar population age, distinguishing between the "single-burst" sample (light blue circles) and the "multi-burst" sample (purple diamonds). \textit{Left column:} $E(B-V)_\star$, Ly$\alpha$ peak separation $\mathrm{v}_{\mathrm{\mathrm{sep}}}^{\mathrm{Ly}\alpha}$, $C_f$(LIS), and star formation rate surface density $\log \Sigma_{\mathrm{SFR}}$. \textit{Right column:} H$\beta$ equivalent widths, the maximum outflow velocity based on LIS absorption lines $\mathrm{v}_{\mathrm{max}}^{\mathrm{LIS}}$, the neutral gas fraction $\mu$, and the O$_{32}$ ratio for CLASSY galaxies. Vertical lines denote age transitions at 4 Myr (dotted) and 7 Myr (dashed), approximating the timescales for O-star and supernova-driven feedback, respectively.}
    \label{fig:sed_pops}
\end{figure}

\begin{deluxetable*}{l|cc|cc|cc|cc|cc|cc|cc|cc|cc}
	 \tablewidth{0pt}
	 \setlength{\tabcolsep}{3pt}
	 \tabletypesize{\footnotesize}
	 \tablecaption{Kendall's $\tau_K$ correlation coefficents and associated $p$-values for the trends in Figures \ref{fig:sed_pops} and \ref{fig:sed_fesc}.}
	 \tablehead{\CH{Subset} \vline & \multicolumn{2}{c|}{$E(B-V)_\star$} & \multicolumn{2}{c|}{v$_{\mathrm{sep}}^{\mathrm{Ly}\alpha}$} & \multicolumn{2}{c|}{$C_f$(LIS)} & \multicolumn{2}{c|}{$\Sigma_{\mathrm{SFR}}$} & \multicolumn{2}{c|}{W$_\lambda$(H$\beta$)} & \multicolumn{2}{c|}{v$_{\mathrm{max}}^{\mathrm{LIS}}$} & \multicolumn{2}{c|}{$\mu$} & \multicolumn{2}{c|}{O$_{32}$} & \multicolumn{2}{c}{$\left<f_{\mathrm{esc}}^{\: \mathrm{LyC}}\right>$} \\ 
    \CH{} \vline & \CH{$\tau_K$} & \CH{$p$-value} \vline  & \CH{$\tau_K$} & \CH{$p$-value} \vline & \CH{$\tau_K$} & \CH{$p$-value} \vline & \CH{$\tau_K$} & \CH{$p$-value} \vline & \CH{$\tau_K$} & \CH{$p$-value} \vline & \CH{$\tau_K$} & \CH{$p$-value} \vline & \CH{$\tau_K$} & \CH{$p$-value} \vline & \CH{$\tau_K$} & \CH{$p$-value}\vline & \CH{$\tau_K$} & \CH{$p$-value}}
    \startdata
    full sample & -0.21 & 0.04 & 0.10 & 0.49 & -0.03 & 0.74 & -0.18 & 0.08 & -0.25 & 0.03 & 0.01 & 0.96 & 0.28 & 0.02 & -0.18 & 0.09 & 0.10 & 0.35 \\ 
    single-burst & -0.44 & 3.4e-03 & 0.15 & 0.41 & -0.04 & 0.81 & -0.18 & 0.23 & -0.37 & 0.03 & 0.13 & 0.40 & 0.31 & 0.08 & -0.13 & 0.37 & 0.19 & 0.21 \\ 
    multi-burst & -0.31 & 0.04 & -0.22 & 0.40 & -0.21 & 0.17 & -0.20 & 0.19 & -0.08 & 0.62 & 4.3e-03 & 0.98 & 0.39 & 0.02 & -0.10  & 0.52 & 0.30 & 0.05 \\ 
    \enddata
    \label{tab:corr_coeffs}
    \tablecomments{Positive values (with a maximum of $+1$) of $\tau_K$ correspond to an increasing trend with age, negative values (with a minimum of $-1$) correspond to a decreasing relationship, and values near zero indicate a lack of correlation. The $p$-values denote the confidence in each value of $\tau_K$, where values $<0.05$ are equivalent to a $>2\sigma$ significance.}
\end{deluxetable*}

\begin{figure}
    \centering
    \includegraphics[width=0.9\linewidth]{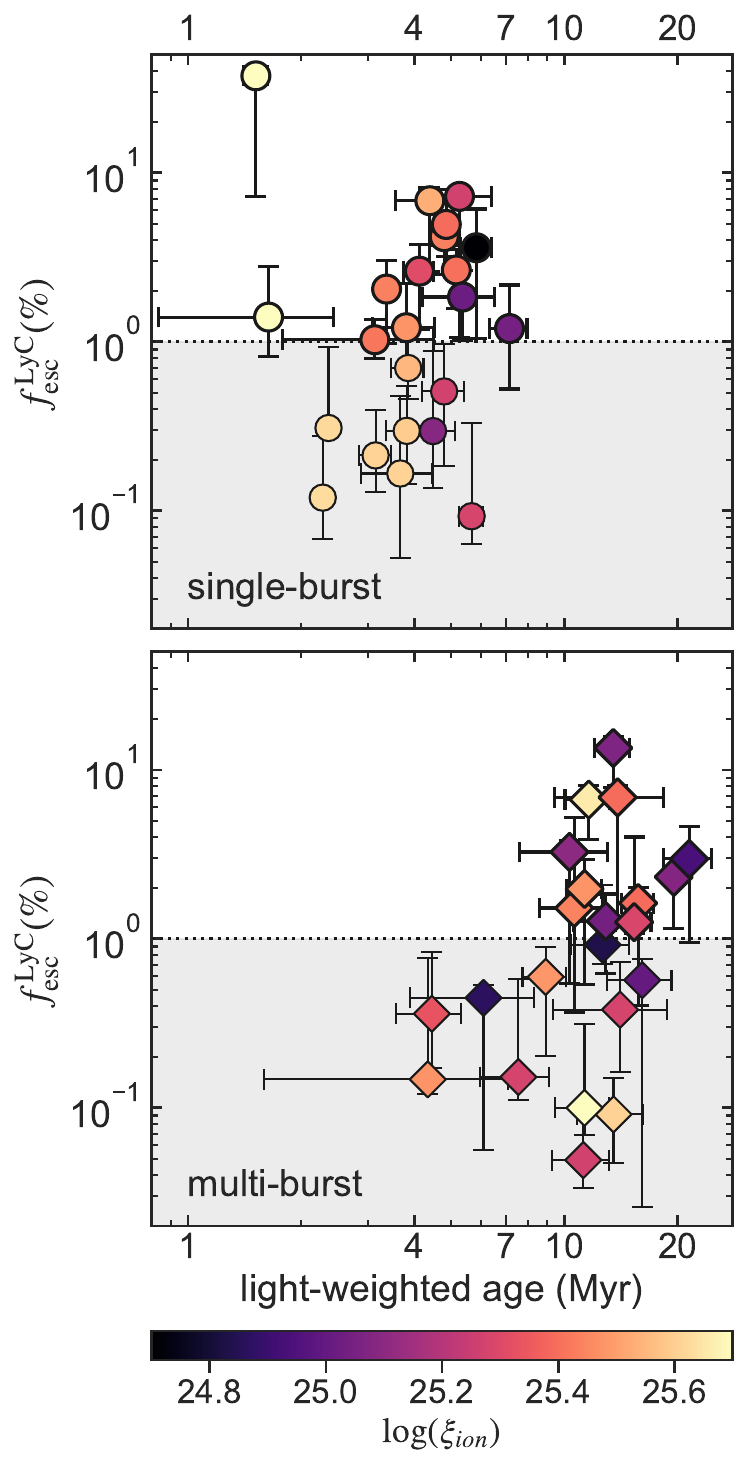}
     \caption{The median predicted LyC escape fraction $\langle f_{\mathrm{esc}}^{\: \mathrm{LyC}} \rangle$ for the "single-burst" (top panel; circles) and the "multi-burst" (bottom panel; diamonds) CLASSY galaxies, as a function of the light-weighted stellar population age. The color-coding in both panels corresponds to $\log(\xi_{\mathrm{ion}})$. 
     Overall, this figure demonstrates that the "single-burst" CLASSY galaxies can have larger LyC escape fractions ($> 1\%$) at young ages ($< 10$ Myr), while the "multi-burst" sample only have higher escape fractions after $\sim 10$ Myr. The clear gradient with $\log(\xi_{\mathrm{ion}})$ for the "single-burst" galaxies indicates that young massive stars are driving feedback and the ionizing radiation in these galaxies, while the lack of a trend for the "multi-burst" galaxies suggests that other mechanisms are responsible, likely the onset of significant SN contributions based on the increase in $\langle f_{\mathrm{esc}}^{\: \mathrm{LyC}} \rangle$ after $\sim 10$ Myr.}
    \label{fig:sed_fesc}
\end{figure}


\subsubsection{With stellar population ages} \label{subsub:age_trends}

In Figure~\ref{fig:sed_pops}, we show how a range of ISM and feedback properties evolve with the light-weighted age of their stellar populations, both for the "single-burst" (light blue circles) and "multi-burst" (purple diamonds) samples. The vertical dotted and dashed lines mark ages of 4 Myr and 7 Myr, roughly corresponding to transitions between stellar populations dominated by O-stars and those in which supernova feedback becomes significant. For each panel, the $\tau_K$ correlation coefficients and their associated $p$-values are listed in Table \ref{tab:corr_coeffs} for the full CLASSY sample, the "single-burst" subset, and the "multi-burst" subset.

We plot $E(B-V)_\star$ in panel (a), finding that it decreases with increasing age in the "single-burst" population, suggesting that dust is being cleared or destroyed over time in these galaxies. There may be a similar, but potentially less significant, trend for the "multi-burst" galaxies, though notably this sample contains galaxies with more dust at older populations than is observed among the "single-burst" galaxies (for example, galaxies with light-weighted ages $>$ 7 Myr and $E(B-V)_\star$ $\gtrsim$ 0.2).

For both the "single-burst" and "multi-burst" subsets, we find several ISM and feedback-sensitive tracers that show a clear age dependence. Galaxies with the youngest light-weighted ages tend to have larger $W_\lambda$(H$\beta$) equivalent widths (panel b), smaller Ly$\alpha$ peak separations ($\mathrm{v}_{\mathrm{\mathrm{sep}}}^{\mathrm{Ly}\alpha}$) (panel c), lower maximum outflow velocities ($\mathrm{v}_{\mathrm{max}}^{\mathrm{LIS}}$; panel d), and higher O$_{32}$ ratios (panel h). Measurements of $\mathrm{v}_{\mathrm{max}}^{\mathrm{LIS}}$-- the highest velocity of outflowing gas-- correspond to the maximum velocity offset of the LIS absorption, which we measured from the blueward side of the absorption profile (specifically where it crosses the threshold of 0.99 in terms of the normalized flux). However, for the other properties in Figure~\ref{fig:sed_pops}-- $E(B-V)_\star$ (panel a), $C_f$(LIS) (panel e), $\mu$ (panel f), and $\log(\Sigma_{\mathrm{SFR}})$ (panel g)-- the trends with light-weighted age can differ between the two subsets.  

Although there does not appear to be much of a correlation between $C_f$(LIS) and the light-weighted age for the "single-burst" sample (panel e), the "multi-burst" galaxies show some indication of a decrease in the covering fraction with older populations ($p$-value = 0.17). The "multi-burst" galaxies with ages $\lesssim$10 Myr have higher covering fractions ($C_f$(LIS) $>$ 0.8) than many of those in the "single-burst" galaxies, despite their older populations. This suggests that the onset of supernovae is essential to driving down the LIS covering fraction in the "multi-burst" galaxies. Overall, this panel shows that the covering fractions for the "single-burst" galaxies can vary and reach low values even for those with very young stellar populations. At the same time, they remain high until after $\sim$10 Myr for the "multi-burst" galaxies.

In panel (c), we plot the maximum outflow velocity, $\mathrm{v}_{\mathrm{max}}^{\mathrm{LIS}}$, versus the light-weight population age, finding that a significant fraction of the "multi-burst" galaxies have lower outflow velocities ($\lesssim$ 300 km/s) than we might expect from the trend established by the "single-burst" galaxies (where the outflow velocity increases with age). This may arise because of the more extended star formation in the "multi-burst" systems, which could dilute the dynamical impact of individual bursts. As a result, mechanical feedback may be less coherent and produce slower outflows than in the "single-burst" systems, where a more synchronized energy output can drive stronger gas acceleration.

In panel (d), the neutral gas fraction, defined as $\mu = \mathrm{M}_{\mathrm{HI}} / (\mathrm{M}_{\mathrm{HI}} + \mathrm{M}_\star)$, is generally high across both galaxy subsets, indicating that most CLASSY galaxies remain globally gas-rich. However, important distinctions emerge when considering the low-$\mu$ and non-detected cases, which are mainly present in the "single-burst" group. This could indicate early stages of efficient star formation or gas removal in these more extreme systems. However, the removal of \ion{H}{1} gas may be slower in the "multi-burst" sample.

Panel (g) examines the SFR surface density, $\log(\Sigma_{\mathrm{SFR}})$, in the "single-burst" and "multi-burst" samples. For the "single-burst" sample, it is relatively constant around $\log(\Sigma_{\mathrm{SFR}}) \sim 0$, while more variation is present among the "multi-burst" galaxies, particularly extending to lower values. This suggests that star formation in some of the "multi-burst" systems is more spatially extended or less concentrated, leading to weaker local feedback pressure and less efficient gas clearing.

Finally, the two subsets show different behaviors in their predicted LyC escape fractions over time, as shown in Figure \ref{fig:sed_fesc} for the "single-burst" (top panel; circles) and "multi-burst" (bottom panel; diamonds) CLASSY galaxies. The points are color-coded by the ionizing photon efficiency, $\log(\xi_{\mathrm{ion}})$. The grey shading highlights the region of the panels where $\langle f_{\mathrm{esc}}^{\: \mathrm{LyC}} \rangle$ $< 1$\%.

Overall, Figure~\ref{fig:sed_fesc} demonstrates that the "single-burst" CLASSY galaxies can have larger predicted LyC escape fractions ($>1\%$) at young ages ($<10$ Myr), similar to the findings for the LzLCS+ presented by \cite{Flury_2025}, while the "multi-burst" sample only have higher escape fractions after $\sim10$ Myr. This indicates that older stellar populations play a key role in facilitating LyC escape in the "multi-burst" systems, potentially related to the onset of significant contributions of energy from supernovae or evolved stars. This is supported by the lack of a clear gradient between $\log(\xi_{\mathrm{ion}})$ and age among this subset, which agrees with the findings from \citet{Chisholm_2019}, for which galaxies with more complex or extended star formation histories show a more gradual decline and increased scatter in $\xi_{\mathrm{ion}}$ with age.

The "single-burst" galaxies present a different case, with large LyC escape fractions at much younger ages and a clear gradient between $\log(\xi_{\mathrm{ion}})$ and age. This agrees with the steep exponential decline in $\log(\xi_{\mathrm{ion}})$ with age for single-burst populations, as reported by \citet{Chisholm_2019}, which results from the short lifetimes of the most massive stars. Overall, this suggests that young massive stars are the main drivers for feedback and LyC radiation among the "single-burst" galaxies. 


\startlongtable
\begin{deluxetable}{|c|cc|}
\tablewidth{\textwidth}
\setlength{\tabcolsep}{8pt}
\tabletypesize{\normalsize}
\tablecaption{Possible LyC escape pathways. Based on measured properties of the stellar populations, we posit two models for LyC escape.} \label{tab:pathways}
\tabletypesize{\footnotesize}
\tablehead{
\CH{} & \CH{Early Escape Model} & \CH{Delayed Escape Model}
}
\startdata
SF history & a recent, intense burst & a series of bursts \\
\hline
Main feedback & O-star winds, & supernovae, \\
 mechanisms & radiation pressure & evolved stars \\
\hline
ISM geometry & more porous & more homogeneous \\
\hline
\enddata 
\end{deluxetable}

\begin{figure}
    \centering
    \includegraphics[width=\linewidth]{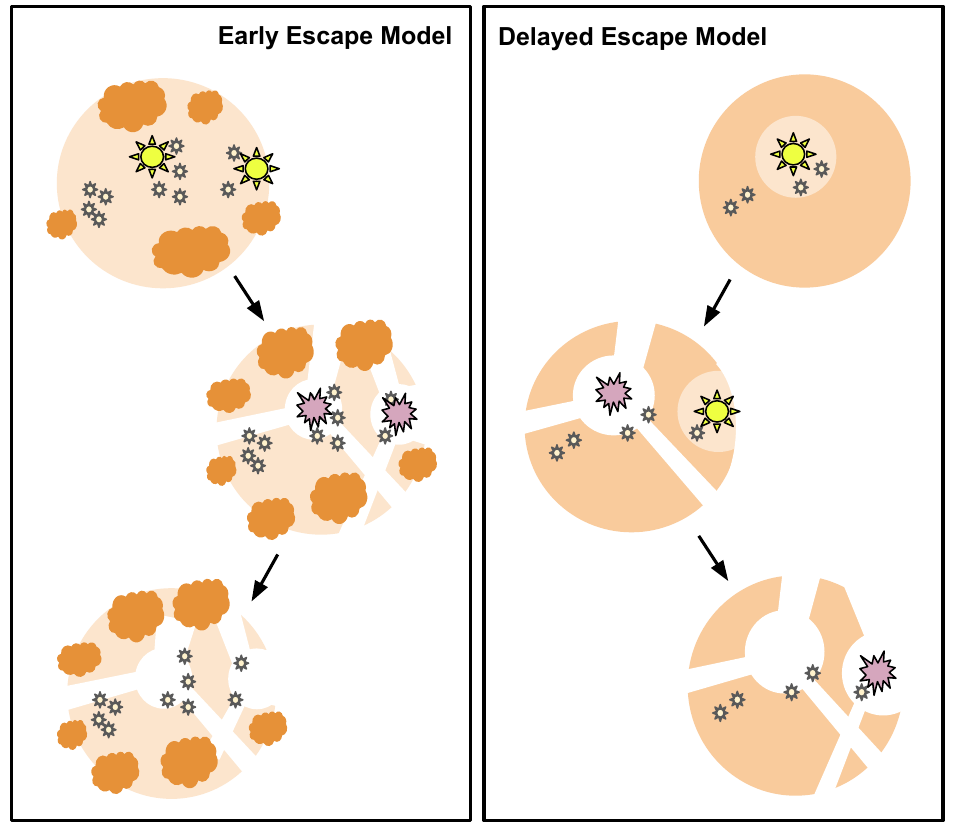}\\
    \includegraphics[width=\linewidth]{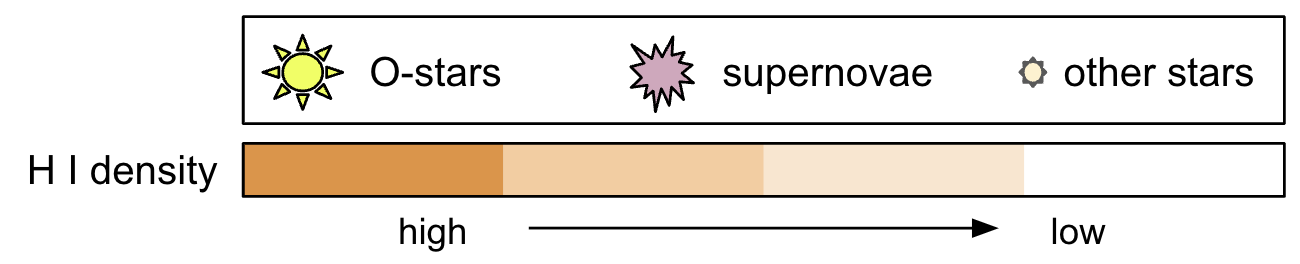}
     \caption{Diagrams of the two different pathways for LyC escape that seem to be present among the CLASSY galaxies. The `early escape' model (shown in the left panel) is more prevalent among the "single-burst" CLASSY galaxies while the `delayed escape' model (right panel) better describes the "multi-burst" CLASSY galaxies. In both panels, \ion{H}{1} gas is represented in orange, with darker shades corresponding to higher densities. O-stars, and their eventual supernovae explosions, are represented by large yellow stars and pale red blasts, respectively, in these diagrams. In the `early escape' model, the compact and intense bursts of star formation lead to a more porous ISM at early ages. The more extended bursts of star-formation in the `delayed escape' model are incapable of producing this same porosity, thereby requiring the onset of supernovae, or the continued longer-term production of ionizing radiation by binary stars, to allow higher fractions of LyC to eventually escape.}
    \label{fig:cartoon}
\end{figure}


\subsection{Pathways for LyC escape: early versus delayed feedback scenarios} \label{sub:disc_SP}

The CLASSY sample exhibits a diverse range of indirect predictions for $f_{\mathrm{esc}}^{\: \mathrm{LyC}}$, driven in part by differences in ISM conditions and, seemly, the recent burst history ($<100$ Myr). Building on the empirical trends presented in the previous subsections, we now synthesize these findings into a physical framework for LyC escape among the CLASSY galaxies. In particular, we find evidence for two distinct escape pathways: one in galaxies dominated by a single, recent burst of star formation, and another in galaxies with multiple, time-separated bursts over the past $\sim$40 Myr. These two scenarios-- an "early" escape phase powered by very young massive stars, and a "delayed" escape phase driven by evolved feedback and ISM clearing-- appear to correlate with differences in outflow velocities, neutral gas and dust content, and star formation rates. In this section, we discuss the broader picture suggested by these trends.

The difference in trends between the predicted escape fractions, $\log(\xi_{\mathrm{ion}})$, and the stellar population ages of the "single-burst" and the "multi-burst" samples suggests that different feedback mechanisms are dominating the escape of ionizing radiation in these systems. Since the "single-burst" CLASSY galaxies can have high escape fractions at young ages, it suggests a more porous ISM at earlier times in these galaxies. In contrast, the ISM of the "multi-burst" sample remains more uniform until older-aged stellar populations are present. The trends with $C_f$(LIS) and $\mu$ in Figure \ref{fig:sed_pops} support this conclusion, showing that the "single-burst" galaxies can have lower covering fractions and gas fractions at younger ages than the "multi-burst" galaxies. 

Overall, we find that the "single-burst" CLASSY galaxies seem to favor an early model of LyC escape, where $f_{\mathrm{esc}}^{\: \mathrm{LyC}}$ is driven by intense but short-lived stellar feedback by young, massive stars, while the "multi-burst" CLASSY galaxies show evidence of more sustained feedback, resulting in delayed LyC escape (only significant after $\sim$10 Myr). Table \ref{tab:pathways} summarizes some of the main differences we find between the galaxies dominated by each of these pathways, which we hereafter refer to as the early escape model and delayed escape model.

We posit that the majority of LzLCS+ galaxies support the early model of LyC escape, as many have relatively high escape fractions ($>1$\%) at very young stellar population ages and display a gradient with $\log(\xi_{\mathrm{ion}})$ \citep{Flury_2024,Carr_2025}, similar to those in our "single-burst" sample. This suggests that LyC escape likely occurs rapidly during or shortly after an intense burst of star formation in these galaxies, where a porous ISM allows for significant pre-supernova feedback mechanisms (e.g., radiation pressure and stellar winds) to facilitate escape. This agrees with other findings regarding the dominant feedback mechanisms for known LyC emitters in the local universe \citep{Amorin_2024,Flury_2024,Carr_2025}.

The contributions from evolved stars (specifically binary and stripped stars) are absent from the \texttt{SB99} modeling but could be responsible for producing significant amounts of ionizing radiation in older stellar populations. Binary evolution can extend the period over which massive stars contribute to feedback, thus extending the LyC production of O-stars to older ages \citep[e.g.,][]{Eldridge_2017,Ma_2016,Ma_2020}. Theoretical models of stripped stars have found that their ionizing photon production can dominate the LyC output in populations older than $\sim$10 Myr \citep[e.g.,][]{Gotberg_2019}. This is around the age that the `delayed escape' galaxies can start having higher predicted LyC escape fractions, which may indicate that binary and stripped stars are playing an essential role in driving higher $f_{\mathrm{esc}}^{\: \mathrm{LyC}}$ in these galaxies.

We present an overall picture of the escape of ionizing radiation as a function of time in Figure \ref{fig:cartoon}. The early escape model (left panel) shows a porous ISM at early ages because of the radiation pressure and energetic stellar winds produced after an intense burst of star formation. Some of the ionizing radiation produced by O-stars (shown by large yellow stars in this diagram) can escape through the ISM via these early channels. As the population ages and the O-stars evolve, some of the surrounding ionized gas will be able to cool and recombine into \ion{H}{1}, which, along with the decreased production of LyC, could cause the escape fraction to decrease with time \citep[as found by][]{Flury_2024,Carr_2025}. However, the delayed escape model (right panel) has a more homogeneous ISM until after $\sim$10 Myr, due to more extended or less concentrated bursts of star formation. Because they lack the early channels that are present in the other model, these galaxies require the onset of supernova (and/or the contributions from binary and stripped stars) for more ionizing radiation to escape, leading to an increasing escape fraction as the stellar populations age.

Overall, we find evidence for two different star formation histories that can lead to LyC escape in nearby star-forming galaxies: (1) an early-time escape (common among galaxies with a single, recent, burst of star formation) that likely occurs through inhomogeneous ISM and pre-SN feedback, and (2) a delayed escape (common among galaxies that have had multiple bursts of star formation), which is more consistent with SN-driven outflows and time-dependent clearing. These findings support a dual-mode feedback model, where both early leakage in compact, young starbursts and delayed escape in more evolved, feedback-driven systems may contribute to cosmic reionization. It will be necessary to recognize this diversity when interpreting ionizing photon leakage in galaxies, both in the local universe and at high redshifts. 

Additionally, distinguishing between these two escape pathways is critical for interpreting ionizing emissivity at both low and high redshifts, as the dominant mechanism influences the timescale and stochasticity of LyC leakage. If early, short-lived pathways dominate, then LyC escape may be highly transient and linked to the youngest starburst phases; if delayed escape is more common, then feedback-regulated models with longer timescales become more applicable.

\section{Summary} \label{summary}
 
In this work, we have presented a detailed comparison of six indirect methods for estimating the escape fraction of ionizing photons, $f_{\rm esc}^{\rm LyC}$, leveraging  a variety observables, such as UV ISM absorption lines, the $\beta_{UV}$-slope, and the Ly$\alpha$ peak separation. We apply these methods to the sample of 45 nearby star-forming galaxies from the CLASSY survey, whose broad UV coverage, high-S/N, and high-spectral resolution, enables detailed comparisons between these diagnostics across a range of galactic properties. This analysis revealed several significant results:
\begin{itemize}
    \item {\it Few CLASSY galaxies are moderate-to-strong LyC leaker candidates.} We performed the first analysis of the escape fractions of the CLASSY sample, finding that about half have median $\left <f_{\rm esc}^{\rm LyC}\right>$ values exceeding 1\%, with five galaxies identified as candidate weak leakers (5\%–20\%) and one (J1323–0132) as a likely strong leaker ($\geq 20$\%).
    \item {\it There is significant variation between different $f_{\rm esc}^{\rm LyC}$ methods.} For the 45 CLASSY galaxies, we measured significant scatter between the escape fraction predictions of different methods, especially at low $f_{\rm esc}^{\rm LyC}$, since each technique is sensitive to different physical parameters and systematics. Despite these differences, there is general agreement between the methods on which galaxies are predicted to be the most efficient and least efficient LyC leakers.
    \item {\it The method of LyC escape depends on galaxy properties.} By linking our indirect $f_{\rm esc}^{\rm LyC}$ predictions to properties of the ISM and the stellar populations, we find compelling support for distinct physical pathways for LyC escape among the CLASSY galaxies: an "early" escape model and a "delayed" escape model.
    \begin{itemize}
        \item[1.] {\it Early LyC escape.} The early escape mode occurs when feedback from very young, UV-bright stellar populations clears channels in a porous ISM, allowing prompt photon leakage. This mode seems to dominate for the galaxies in our "single-burst" sample, which are experiencing a powerful burst that makes them appear very UV-bright and compact.
        \item[2.] {\it Delayed LyC escape.} The delayed escape mode is associated with older bursts where supernova-driven feedback clears paths for ionizing photons over time. This mode describes the galaxies in our "multi-burst" sample, which have experienced multiple subsequent bursts within the past few 100 Myr.
    \end{itemize}
    This dual-channel scenario is consistent with the diversity of observed galaxy properties and provides a unified interpretation of LyC escape in both local and high-redshift analogs. As future JWST and HST observations expand the census of LyC-emitting galaxies, these empirical frameworks will be critical for interpreting the mechanisms driving reionization and refining predictions of the cosmic ionizing budget.
    \item {\it Individual $f_{\rm esc}^{\rm LyC}$ methods produce biased results.} The diversity of escape mechanisms observed locally for the CLASSY survey and LzLCS underscores the need for caution when applying a single diagnostic or universal threshold when interpreting contributions of high-redshift galaxies to cosmic reionization. Our results reinforce the importance of using a multifaceted approach to constraining $f_{\rm esc}^{\rm LyC}$, especially when extending these techniques to the lower-S/N and spectral resolution JWST observations of early galaxies at $z>6$.
\end{itemize}  

In the future, ionizing photon escape diagnostics will be improved as larger numbers of direct LyC detections are observed, and the parameter space probed by both empirical samples and simulations is expanded. Such steps are critical for anchoring interpretations of ionizing photon escape at high redshift, where direct measurements remain elusive.


\begin{acknowledgements}
We would like to thank the referee for their thoughtful review and suggestions, which have greatly improved the quality of this work. KSP and DAB are grateful for the support for this program, HST-GO-15840, which was provided by NASA through a grant from the Space Telescope Science Institute, operated by the Associations of Universities for Research in Astronomy, Incorporated, under NASA contract NAS5-26555. The CLASSY collaboration thanks the COS team for all their assistance and advice in the reduction of the COS data. 
\end{acknowledgements}

\software{\texttt{LyCsurv} \citep{Flury_2024}, \texttt{kendall} \citep{Flury_2022b}, \texttt{Starburst99} \citep{Leitherer_1999}, \texttt{Cloudy} \citep{Ferland_2013,Ferland_2017}, \texttt{scipy.optimize.curve\_fit}, \texttt{lmfit} \citep{Newville_2014}, \texttt{mpfit} \citep{Markwardt_2009}, \texttt{scipy.stats.2samp\_ks} \citep{Virtanen_2020}}
\facilities{HST (COS)}

\bibliographystyle{aasjournal}
\bibliography{thbib.bib}

\begin{thebibliography}{}
\expandafter\ifx\csname natexlab\endcsname\relax\def\natexlab#1{#1}\fi
\providecommand{\url}[1]{\href{#1}{#1}}
\providecommand{\dodoi}[1]{doi:~\href{http://doi.org/#1}{\nolinkurl{#1}}}
\providecommand{\doeprint}[1]{\href{http://ascl.net/#1}{\nolinkurl{http://ascl.net/#1}}}
\providecommand{\doarXiv}[1]{\href{https://arxiv.org/abs/#1}{\nolinkurl{https://arxiv.org/abs/#1}}}

\bibitem[{{Ahn} {et~al.}(2003){Ahn}, {Lee}, \& {Lee}}]{Ahn_2003}
{Ahn}, S.-H., {Lee}, H.-W., \& {Lee}, H.~M. 2003, \mnras, 340, 863, \dodoi{10.1046/j.1365-8711.2003.06353.x}

\bibitem[{{Amor{\'\i}n} {et~al.}(2024){Amor{\'\i}n}, {Rodr{\'\i}guez-Henr{\'\i}quez}, {Fern{\'a}ndez}, {et~al.}}]{Amorin_2024}
{Amor{\'\i}n}, R.~O., {Rodr{\'\i}guez-Henr{\'\i}quez}, M., {Fern{\'a}ndez}, V., {et~al.} 2024, \aap, 682, L25, \dodoi{10.1051/0004-6361/202449175}

\bibitem[{{Atek} {et~al.}(2024){Atek}, {Labb{\'e}}, {Furtak}, {et~al.}}]{Atek_2024}
{Atek}, H., {Labb{\'e}}, I., {Furtak}, L.~J., {et~al.} 2024, \nat, 626, 975, \dodoi{10.1038/s41586-024-07043-6}

\bibitem[{{Barrow} {et~al.}(2020){Barrow}, {Robertson}, {Ellis}, {et~al.}}]{Barrow_2020}
{Barrow}, K. S.~S., {Robertson}, B.~E., {Ellis}, R.~S., {et~al.} 2020, \apjl, 902, L39, \dodoi{10.3847/2041-8213/abbd8e}

\bibitem[{{Bassett} {et~al.}(2019){Bassett}, {Ryan-Weber}, {Cooke}, {et~al.}}]{Bassett_2019}
{Bassett}, R., {Ryan-Weber}, E.~V., {Cooke}, J., {et~al.} 2019, \mnras, 483, 5223, \dodoi{10.1093/mnras/sty3320}

\bibitem[{{Becker} {et~al.}(2021){Becker}, {D'Aloisio}, {Christenson}, {et~al.}}]{Becker_2021}
{Becker}, G.~D., {D'Aloisio}, A., {Christenson}, H.~M., {et~al.} 2021, \mnras, 508, 1853, \dodoi{10.1093/mnras/stab2696}

\bibitem[{{Berg} {et~al.}(2022){Berg}, {James}, {King}, {et~al.}}]{Berg_2022}
{Berg}, D.~A., {James}, B.~L., {King}, T., {et~al.} 2022, \apjs, 261, 31, \dodoi{10.3847/1538-4365/ac6c03}

\bibitem[{{Bergvall} {et~al.}(2006){Bergvall}, {Zackrisson}, {Andersson}, {et~al.}}]{Bergvall_2006}
{Bergvall}, N., {Zackrisson}, E., {Andersson}, B.~G., {et~al.} 2006, \aap, 448, 513, \dodoi{10.1051/0004-6361:20053788}

\bibitem[{{Borthakur} {et~al.}(2014){Borthakur}, {Heckman}, {Leitherer}, {et~al.}}]{Borthakur_2014}
{Borthakur}, S., {Heckman}, T.~M., {Leitherer}, C., {et~al.} 2014, Science, 346, 216, \dodoi{10.1126/science.1254214}

\bibitem[{{Bosman} {et~al.}(2022){Bosman}, {Davies}, {Becker}, {et~al.}}]{Bosman_2022}
{Bosman}, S. E.~I., {Davies}, F.~B., {Becker}, G.~D., {et~al.} 2022, \mnras, 514, 55, \dodoi{10.1093/mnras/stac1046}

\bibitem[{{Bosman} {et~al.}(2018){Bosman}, {Fan}, {Jiang}, {et~al.}}]{Bosman_2018}
{Bosman}, S. E.~I., {Fan}, X., {Jiang}, L., {et~al.} 2018, \mnras, 479, 1055, \dodoi{10.1093/mnras/sty1344}

\bibitem[{{Calzetti} {et~al.}(1994){Calzetti}, {Kinney}, \& {Storchi-Bergmann}}]{Calzetti_1994}
{Calzetti}, D., {Kinney}, A.~L., \& {Storchi-Bergmann}, T. 1994, \apj, 429, 582, \dodoi{10.1086/174346}

\bibitem[{{Carr} {et~al.}(2025){Carr}, {Cen}, {Scarlata}, {et~al.}}]{Carr_2025}
{Carr}, C.~A., {Cen}, R., {Scarlata}, C., {et~al.} 2025, \apj, 982, 137, \dodoi{10.3847/1538-4357/adb72f}

\bibitem[{{Chisholm} {et~al.}(2018){Chisholm}, {Gazagnes}, {Schaerer}, {et~al.}}]{Chisholm_2018}
{Chisholm}, J., {Gazagnes}, S., {Schaerer}, D., {et~al.} 2018, \aap, 616, A30, \dodoi{10.1051/0004-6361/201832758}

\bibitem[{{Chisholm} {et~al.}(2019){Chisholm}, {Rigby}, {Bayliss}, {Berg}, {Dahle}, {Gladders}, \& {Sharon}}]{Chisholm_2019}
{Chisholm}, J., {Rigby}, J.~R., {Bayliss}, M., {et~al.} 2019, \apj, 882, 182, \dodoi{10.3847/1538-4357/ab3104}

\bibitem[{{Chisholm} {et~al.}(2022){Chisholm}, {Saldana-Lopez}, {Flury}, {et~al.}}]{Chisholm_2022}
{Chisholm}, J., {Saldana-Lopez}, A., {Flury}, S., {et~al.} 2022, \mnras, 517, 5104, \dodoi{10.1093/mnras/stac2874}

\bibitem[{{Chisholm} {et~al.}(2017){Chisholm}, {Tremonti}, {Leitherer}, {et~al.}}]{Chisholm_2017}
{Chisholm}, J., {Tremonti}, C.~A., {Leitherer}, C., {et~al.} 2017, \mnras, 469, 4831, \dodoi{10.1093/mnras/stx1164}

\bibitem[{{Choustikov} {et~al.}(2024){Choustikov}, {Katz}, {Saxena}, {et~al.}}]{Choustikov_2024}
{Choustikov}, N., {Katz}, H., {Saxena}, A., {et~al.} 2024, \mnras, 532, 2463, \dodoi{10.1093/mnras/stae1586}

\bibitem[{{Citro} {et~al.}(2024){Citro}, {Scarlata}, {Mantha}, {et~al.}}]{Citro_2024}
{Citro}, A., {Scarlata}, C.~M., {Mantha}, K.~B., {et~al.} 2024, arXiv e-prints, arXiv:2406.07618, \dodoi{10.48550/arXiv.2406.07618}

\bibitem[{Cox(2018)}]{Cox_1972}
Cox, D.~R. 2018, Journal of the Royal Statistical Society: Series B (Methodological), 34, 187, \dodoi{10.1111/j.2517-6161.1972.tb00899.x}

\bibitem[{{Davis} {et~al.}(2021){Davis}, {Gebhardt}, {Mentuch Cooper}, {et~al.}}]{Davis_2021}
{Davis}, D., {Gebhardt}, K., {Mentuch Cooper}, E., {et~al.} 2021, \apj, 920, 122, \dodoi{10.3847/1538-4357/ac1598}

\bibitem[{{Dayal} {et~al.}(2020){Dayal}, {Volonteri}, {Choudhury}, {et~al.}}]{Dayal_2020}
{Dayal}, P., {Volonteri}, M., {Choudhury}, T.~R., {et~al.} 2020, \mnras, 495, 3065, \dodoi{10.1093/mnras/staa1138}

\bibitem[{{de Barros} {et~al.}(2016){de Barros}, {Vanzella}, {Amor{\'\i}n}, {Castellano}, {Siana}, {Grazian}, {Suh}, {Balestra}, {Vignali}, {Verhamme}, {Zamorani}, {Mignoli}, {Hasinger}, {Comastri}, {Pentericci}, {P{\'e}rez-Montero}, {Fontana}, {Giavalisco}, \& {Gilli}}]{deBarros2016}
{de Barros}, S., {Vanzella}, E., {Amor{\'\i}n}, R., {et~al.} 2016, \aap, 585, A51, \dodoi{10.1051/0004-6361/201527046}

\bibitem[{{de Mello} {et~al.}(2000){de Mello}, {Leitherer}, \& {Heckman}}]{deMello_2000}
{de Mello}, D.~F., {Leitherer}, C., \& {Heckman}, T.~M. 2000, \apj, 530, 251, \dodoi{10.1086/308358}

\bibitem[{{Dijkstra}(2014)}]{Dijkstra_2014}
{Dijkstra}, M. 2014, \pasa, 31, e040, \dodoi{10.1017/pasa.2014.33}

\bibitem[{{Eldridge} {et~al.}(2017){Eldridge}, {Stanway}, {Xiao}, {et~al.}}]{Eldridge_2017}
{Eldridge}, J.~J., {Stanway}, E.~R., {Xiao}, L., {et~al.} 2017, \pasa, 34, e058, \dodoi{10.1017/pasa.2017.51}

\bibitem[{{Ferland} {et~al.}(2013){Ferland}, {Porter}, {van Hoof}, {Williams}, {Abel}, {Lykins}, {Shaw}, {Henney}, \& {Stancil}}]{Ferland_2013}
{Ferland}, G.~J., {Porter}, R.~L., {van Hoof}, P.~A.~M., {et~al.} 2013, \rmxaa, 49, 137.
\newblock \doarXiv{1302.4485}

\bibitem[{{Ferland} {et~al.}(2017){Ferland}, {Chatzikos}, {Guzm{\'a}n}, {Lykins}, {van Hoof}, {Williams}, {Abel}, {Badnell}, {Keenan}, {Porter}, \& {Stancil}}]{Ferland_2017}
{Ferland}, G.~J., {Chatzikos}, M., {Guzm{\'a}n}, F., {et~al.} 2017, \rmxaa, 53, 385.
\newblock \doarXiv{1705.10877}

\bibitem[{{Finkelstein} {et~al.}(2019){Finkelstein}, {D'Aloisio}, {Paardekooper}, {et~al.}}]{Finkelstein_2019}
{Finkelstein}, S.~L., {D'Aloisio}, A., {Paardekooper}, J.-P., {et~al.} 2019, \apj, 879, 36, \dodoi{10.3847/1538-4357/ab1ea8}

\bibitem[{{Fletcher} {et~al.}(2019){Fletcher}, {Tang}, {Robertson}, {et~al.}}]{Fletcher_2019}
{Fletcher}, T.~J., {Tang}, M., {Robertson}, B.~E., {et~al.} 2019, \apj, 878, 87, \dodoi{10.3847/1538-4357/ab2045}

\bibitem[{{Flury}(2023)}]{Flury_kendall}
{Flury}, S.~R. 2023, {kendall}, v1.0.0, \dodoi{10.5281/zenodo.14732912}

\bibitem[{{Flury} {et~al.}(2024{\natexlab{a}}){Flury}, {Jaskot}, \& {Silveyra}}]{Flury_lycsurv}
{Flury}, S.~R., {Jaskot}, A., \& {Silveyra}, A. 2024{\natexlab{a}}, {LyCsurv}, v0.1.0, \dodoi{10.5281/zenodo.11392540}

\bibitem[{{Flury} {et~al.}(2022{\natexlab{a}}){Flury}, {Jaskot}, {Ferguson}, {et~al.}}]{Flury_2022a}
{Flury}, S.~R., {Jaskot}, A.~E., {Ferguson}, H.~C., {et~al.} 2022{\natexlab{a}}, \apjs, 260, 1, \dodoi{10.3847/1538-4365/ac5331}

\bibitem[{{Flury} {et~al.}(2022{\natexlab{b}}){Flury}, {Jaskot}, {Ferguson}, {et~al.}}]{Flury_2022b}
---. 2022{\natexlab{b}}, \apj, 930, 126, \dodoi{10.3847/1538-4357/ac61e4}

\bibitem[{{Flury} {et~al.}(2024{\natexlab{b}}){Flury}, {Jaskot}, {Saldana-Lopez}, {et~al.}}]{Flury_2024}
{Flury}, S.~R., {Jaskot}, A.~E., {Saldana-Lopez}, A., {et~al.} 2024{\natexlab{b}}, arXiv e-prints, arXiv:2409.12118.
\newblock \doarXiv{2409.12118}

\bibitem[{{Flury} {et~al.}(2025){Flury}, {Jaskot}, {Saldana-Lopez}, {Oey}, {Chisholm}, {Amor{\'\i}n}, {Bait}, {Borthakur}, {Carr}, {Ferguson}, {Giavalisco}, {Hayes}, {Heckman}, {Henry}, {Ji}, {Komarova}, {Leclercq}, {Le Reste}, {McCandliss}, {Marques-Chaves}, {{\"O}stlin}, {Pentericci}, {Ravindranath}, {Rutkowski}, {Scarlata}, {Schaerer}, {Thuan}, {Trebitsch}, {Vanzella}, {Verhamme}, {Wang}, {Worseck}, \& {Xu}}]{Flury_2025}
---. 2025, \apj, 985, 128, \dodoi{10.3847/1538-4357/adc305}

\bibitem[{{Gazagnes} {et~al.}(2018){Gazagnes}, {Chisholm}, {Schaerer}, {et~al.}}]{Gazagnes_2018}
{Gazagnes}, S., {Chisholm}, J., {Schaerer}, D., {et~al.} 2018, \aap, 616, A29, \dodoi{10.1051/0004-6361/201832759}

\bibitem[{{Gazagnes} {et~al.}(2020){Gazagnes}, {Chisholm}, {Schaerer}, {et~al.}}]{Gazagnes_2020}
---. 2020, \aap, 639, A85, \dodoi{10.1051/0004-6361/202038096}

\bibitem[{{Gazagnes} {et~al.}(2024){Gazagnes}, {Cullen}, {Mauerhofer}, {et~al.}}]{Gazagnes_2024}
{Gazagnes}, S., {Cullen}, F., {Mauerhofer}, V., {et~al.} 2024, \apj, 969, 50, \dodoi{10.3847/1538-4357/ad47a4}

\bibitem[{{Gazagnes} {et~al.}(2023){Gazagnes}, {Mauerhofer}, {Berg}, {et~al.}}]{Gazagnes_2023}
{Gazagnes}, S., {Mauerhofer}, V., {Berg}, D.~A., {et~al.} 2023, \apj, 952, 164, \dodoi{10.3847/1538-4357/acda2c}

\bibitem[{{G{\"o}tberg} {et~al.}(2019){G{\"o}tberg}, {de Mink}, {Groh}, {et~al.}}]{Gotberg_2019}
{G{\"o}tberg}, Y., {de Mink}, S.~E., {Groh}, J.~H., {et~al.} 2019, \aap, 629, A134, \dodoi{10.1051/0004-6361/201834525}

\bibitem[{{Hayes}(2015)}]{Hayes_2015}
{Hayes}, M. 2015, \pasa, 32, e027, \dodoi{10.1017/pasa.2015.25}

\bibitem[{{Hu} {et~al.}(2023){Hu}, {Martin}, {Gronke}, {et~al.}}]{Hu_2023}
{Hu}, W., {Martin}, C.~L., {Gronke}, M., {et~al.} 2023, \apj, 956, 39, \dodoi{10.3847/1538-4357/aceefd}

\bibitem[{{Inoue} {et~al.}(2014){Inoue}, {Shimizu}, {Iwata}, {et~al.}}]{Inoue_2014}
{Inoue}, A.~K., {Shimizu}, I., {Iwata}, I., {et~al.} 2014, \mnras, 442, 1805, \dodoi{10.1093/mnras/stu936}

\bibitem[{{Izotov} {et~al.}(2022){Izotov}, {Chisholm}, {Worseck}, {et~al.}}]{Izotov_2022}
{Izotov}, Y.~I., {Chisholm}, J., {Worseck}, G., {et~al.} 2022, \mnras, 515, 2864, \dodoi{10.1093/mnras/stac1899}

\bibitem[{{Izotov} {et~al.}(2016{\natexlab{a}}){Izotov}, {Orlitov{\'a}}, {Schaerer}, {et~al.}}]{Izotov_2016b}
{Izotov}, Y.~I., {Orlitov{\'a}}, I., {Schaerer}, D., {et~al.} 2016{\natexlab{a}}, \nat, 529, 178, \dodoi{10.1038/nature16456}

\bibitem[{{Izotov} {et~al.}(2016{\natexlab{b}}){Izotov}, {Schaerer}, {Thuan}, {et~al.}}]{Izotov_2016a}
{Izotov}, Y.~I., {Schaerer}, D., {Thuan}, T.~X., {et~al.} 2016{\natexlab{b}}, \mnras, 461, 3683, \dodoi{10.1093/mnras/stw1205}

\bibitem[{{Izotov} {et~al.}(2018{\natexlab{a}}){Izotov}, {Schaerer}, {Worseck}, {et~al.}}]{Izotov_2018a}
{Izotov}, Y.~I., {Schaerer}, D., {Worseck}, G., {et~al.} 2018{\natexlab{a}}, \mnras, 474, 4514, \dodoi{10.1093/mnras/stx3115}

\bibitem[{{Izotov} {et~al.}(2018{\natexlab{b}}){Izotov}, {Worseck}, {Schaerer}, {et~al.}}]{Izotov_2018b}
{Izotov}, Y.~I., {Worseck}, G., {Schaerer}, D., {et~al.} 2018{\natexlab{b}}, \mnras, 478, 4851, \dodoi{10.1093/mnras/sty1378}

\bibitem[{{Izotov} {et~al.}(2021){Izotov}, {Worseck}, {Schaerer}, {et~al.}}]{Izotov_2021}
---. 2021, \mnras, 503, 1734, \dodoi{10.1093/mnras/stab612}

\bibitem[{{James} {et~al.}(2022){James}, {Berg}, {King}, {et~al.}}]{James_2022}
{James}, B.~L., {Berg}, D.~A., {King}, T., {et~al.} 2022, \apjs, 262, 37, \dodoi{10.3847/1538-4365/ac8008}

\bibitem[{{Jaskot} \& {Oey}(2013)}]{Jaskot_2013}
{Jaskot}, A.~E., \& {Oey}, M.~S. 2013, \apj, 766, 91, \dodoi{10.1088/0004-637X/766/2/91}

\bibitem[{{Jaskot} {et~al.}(2024{\natexlab{a}}){Jaskot}, {Silveyra}, {Plantinga}, {et~al.}}]{Jaskot2024a}
{Jaskot}, A.~E., {Silveyra}, A.~C., {Plantinga}, A., {et~al.} 2024{\natexlab{a}}, \apj, 972, 92, \dodoi{10.3847/1538-4357/ad58b9}

\bibitem[{{Jaskot} {et~al.}(2024{\natexlab{b}}){Jaskot}, {Silveyra}, {Plantinga}, {et~al.}}]{Jaskot2024b}
---. 2024{\natexlab{b}}, \apj, 973, 111, \dodoi{10.3847/1538-4357/ad5557}

\bibitem[{{Jennings} {et~al.}(2025){Jennings}, {Henry}, {Mauerhofer}, {et~al.}}]{Jennings_2025}
{Jennings}, R.~M., {Henry}, A., {Mauerhofer}, V., {et~al.} 2025, \apj, 979, 64, \dodoi{10.3847/1538-4357/ad9b13}

\bibitem[{{Ji} {et~al.}(2020){Ji}, {Giavalisco}, {Vanzella}, {et~al.}}]{Ji_2020}
{Ji}, Z., {Giavalisco}, M., {Vanzella}, E., {et~al.} 2020, \apj, 888, 109, \dodoi{10.3847/1538-4357/ab5fdc}

\bibitem[{{Kakiichi} \& {Gronke}(2021)}]{Kakiichi_2021}
{Kakiichi}, K., \& {Gronke}, M. 2021, \apj, 908, 30, \dodoi{10.3847/1538-4357/abc2d9}

\bibitem[{{Kannan} {et~al.}(2022){Kannan}, {Garaldi}, {Smith}, {et~al.}}]{Kannan_2022}
{Kannan}, R., {Garaldi}, E., {Smith}, A., {et~al.} 2022, \mnras, 511, 4005, \dodoi{10.1093/mnras/stab3710}

\bibitem[{{Katz} {et~al.}(2020){Katz}, {{\v{D}}urov{\v{c}}{\'\i}kov{\'a}}, {Kimm}, {et~al.}}]{Katz_2020}
{Katz}, H., {{\v{D}}urov{\v{c}}{\'\i}kov{\'a}}, D., {Kimm}, T., {et~al.} 2020, \mnras, 498, 164, \dodoi{10.1093/mnras/staa2355}

\bibitem[{{Kerutt} {et~al.}(2024){Kerutt}, {Oesch}, {Wisotzki}, {et~al.}}]{Kerutt_2024}
{Kerutt}, J., {Oesch}, P.~A., {Wisotzki}, L., {et~al.} 2024, \aap, 684, A42, \dodoi{10.1051/0004-6361/202346656}

\bibitem[{{Komarova} {et~al.}(2024){Komarova}, {Oey}, {Hernandez}, {et~al.}}]{Komarova_2024}
{Komarova}, L., {Oey}, M.~S., {Hernandez}, S., {et~al.} 2024, \apj, 967, 117, \dodoi{10.3847/1538-4357/ad3962}

\bibitem[{{Leitet} {et~al.}(2013){Leitet}, {Bergvall}, {Hayes}, {et~al.}}]{Leitet_2013}
{Leitet}, E., {Bergvall}, N., {Hayes}, M., {et~al.} 2013, \aap, 553, A106, \dodoi{10.1051/0004-6361/201118370}

\bibitem[{{Leitet} {et~al.}(2011){Leitet}, {Bergvall}, {Piskunov}, {et~al.}}]{Leitet_2011}
{Leitet}, E., {Bergvall}, N., {Piskunov}, N., {et~al.} 2011, \aap, 532, A107, \dodoi{10.1051/0004-6361/201015654}

\bibitem[{{Leitherer} {et~al.}(2018){Leitherer}, {Byler}, {Lee}, \& {Levesque}}]{Leitherer_2018}
{Leitherer}, C., {Byler}, N., {Lee}, J.~C., \& {Levesque}, E.~M. 2018, \apj, 865, 55, \dodoi{10.3847/1538-4357/aada84}

\bibitem[{{Leitherer} {et~al.}(2016){Leitherer}, {Hernandez}, {Lee}, {et~al.}}]{Leitherer_2016}
{Leitherer}, C., {Hernandez}, S., {Lee}, J.~C., {et~al.} 2016, \apj, 823, 64, \dodoi{10.3847/0004-637X/823/1/64}

\bibitem[{{Leitherer} {et~al.}(1999){Leitherer}, {Schaerer}, {Goldader}, {et~al.}}]{Leitherer_1999}
{Leitherer}, C., {Schaerer}, D., {Goldader}, J.~D., {et~al.} 1999, \apjs, 123, 3, \dodoi{10.1086/313233}

\bibitem[{{Ma} {et~al.}(2016){Ma}, {Hopkins}, {Kasen}, {et~al.}}]{Ma_2016}
{Ma}, X., {Hopkins}, P.~F., {Kasen}, D., {et~al.} 2016, \mnras, 459, 3614, \dodoi{10.1093/mnras/stw941}

\bibitem[{{Ma} {et~al.}(2020){Ma}, {Quataert}, {Wetzel}, {et~al.}}]{Ma_2020}
{Ma}, X., {Quataert}, E., {Wetzel}, A., {et~al.} 2020, \mnras, 498, 2001, \dodoi{10.1093/mnras/staa2404}

\bibitem[{{Madau} \& {Haardt}(2015)}]{Madau_2015}
{Madau}, P., \& {Haardt}, F. 2015, \apjl, 813, L8, \dodoi{10.1088/2041-8205/813/1/L8}

\bibitem[{{Markwardt}(2009)}]{Markwardt_2009}
{Markwardt}, C.~B. 2009, in Astronomical Society of the Pacific Conference Series, Vol. 411, Astronomical Data Analysis Software and Systems XVIII, ed. D.~A. {Bohlender}, D.~{Durand}, \& P.~{Dowler}, 251, \dodoi{10.48550/arXiv.0902.2850}

\bibitem[{{Marques-Chaves} {et~al.}(2021){Marques-Chaves}, {Schaerer}, {{\'A}lvarez-M{\'a}rquez}, {et~al.}}]{Marques-Chaves_2021}
{Marques-Chaves}, R., {Schaerer}, D., {{\'A}lvarez-M{\'a}rquez}, J., {et~al.} 2021, \mnras, 507, 524, \dodoi{10.1093/mnras/stab2187}

\bibitem[{{Marques-Chaves} {et~al.}(2024){Marques-Chaves}, {Schaerer}, {Kuruvanthodi}, {et~al.}}]{Marques-Chaves_2024}
{Marques-Chaves}, R., {Schaerer}, D., {Kuruvanthodi}, A., {et~al.} 2024, \aap, 681, A30, \dodoi{10.1051/0004-6361/202347411}

\bibitem[{{Mascia} {et~al.}(2023){Mascia}, {Pentericci}, {Calabr{\`o}}, {et~al.}}]{Mascia_2023}
{Mascia}, S., {Pentericci}, L., {Calabr{\`o}}, A., {et~al.} 2023, \aap, 672, A155, \dodoi{10.1051/0004-6361/202345866}

\bibitem[{{Mascia} {et~al.}(2024){Mascia}, {Pentericci}, {Calabr{\`o}}, {et~al.}}]{Mascia_2024}
---. 2024, \aap, 685, A3, \dodoi{10.1051/0004-6361/202347884}

\bibitem[{{Mauerhofer} {et~al.}(2021){Mauerhofer}, {Verhamme}, {Blaizot}, {et~al.}}]{Mauerhofer_2021}
{Mauerhofer}, V., {Verhamme}, A., {Blaizot}, J., {et~al.} 2021, \aap, 646, A80, \dodoi{10.1051/0004-6361/202039449}

\bibitem[{{Meynet} {et~al.}(1994){Meynet}, {Maeder}, {Schaller}, {Schaerer}, \& {Charbonnel}}]{Meynet_1994}
{Meynet}, G., {Maeder}, A., {Schaller}, G., {Schaerer}, D., \& {Charbonnel}, C. 1994, \aaps, 103, 97

\bibitem[{{Mu{\~n}oz} {et~al.}(2024){Mu{\~n}oz}, {Mirocha}, {Chisholm}, {Furlanetto}, \& {Mason}}]{Munoz_2024}
{Mu{\~n}oz}, J.~B., {Mirocha}, J., {Chisholm}, J., {Furlanetto}, S.~R., \& {Mason}, C. 2024, \mnras, 535, L37, \dodoi{10.1093/mnrasl/slae086}

\bibitem[{{Naidu} {et~al.}(2018){Naidu}, {Forrest}, {Oesch}, {et~al.}}]{Naidu_2018}
{Naidu}, R.~P., {Forrest}, B., {Oesch}, P.~A., {et~al.} 2018, \mnras, 478, 791, \dodoi{10.1093/mnras/sty961}

\bibitem[{{Naidu} {et~al.}(2022){Naidu}, {Matthee}, {Oesch}, {et~al.}}]{Naidu_2022}
{Naidu}, R.~P., {Matthee}, J., {Oesch}, P.~A., {et~al.} 2022, \mnras, 510, 4582, \dodoi{10.1093/mnras/stab3601}

\bibitem[{{Naidu} {et~al.}(2020){Naidu}, {Tacchella}, {Mason}, {et~al.}}]{Naidu_2020}
{Naidu}, R.~P., {Tacchella}, S., {Mason}, C.~A., {et~al.} 2020, \apj, 892, 109, \dodoi{10.3847/1538-4357/ab7cc9}

\bibitem[{{Nakajima} {et~al.}(2020){Nakajima}, {Ellis}, {Robertson}, {et~al.}}]{Nakajima_2020}
{Nakajima}, K., {Ellis}, R.~S., {Robertson}, B.~E., {et~al.} 2020, \apj, 889, 161, \dodoi{10.3847/1538-4357/ab6604}

\bibitem[{{Neufeld}(1990)}]{Neufeld_1990}
{Neufeld}, D.~A. 1990, \apj, 350, 216, \dodoi{10.1086/168375}

\bibitem[{{Newville} {et~al.}(2014){Newville}, {Stensitzki}, {Allen}, \& {Ingargiola}}]{Newville_2014}
{Newville}, M., {Stensitzki}, T., {Allen}, D.~B., \& {Ingargiola}, A. 2014, {LMFIT: Non-Linear Least-Square Minimization and Curve-Fitting for Python}, 0.8.0,  Zenodo, \dodoi{10.5281/zenodo.11813}

\bibitem[{{Orlitov{\'a}} {et~al.}(2018){Orlitov{\'a}}, {Verhamme}, {Henry}, {et~al.}}]{Orlitova_2018}
{Orlitov{\'a}}, I., {Verhamme}, A., {Henry}, A., {et~al.} 2018, \aap, 616, A60, \dodoi{10.1051/0004-6361/201732478}

\bibitem[{{Pahl} {et~al.}(2025){Pahl}, {Topping}, {Shapley}, {et~al.}}]{Pahl_2025}
{Pahl}, A., {Topping}, M.~W., {Shapley}, A., {et~al.} 2025, \apj, 981, 134, \dodoi{10.3847/1538-4357/adb1ab}

\bibitem[{{Pahl} {et~al.}(2021){Pahl}, {Shapley}, {Steidel}, {et~al.}}]{Pahl_2021}
{Pahl}, A.~J., {Shapley}, A., {Steidel}, C.~C., {et~al.} 2021, \mnras, 505, 2447, \dodoi{10.1093/mnras/stab1374}

\bibitem[{{Pahl} {et~al.}(2023){Pahl}, {Shapley}, {Steidel}, {et~al.}}]{Pahl_2023}
---. 2023, \mnras, 521, 3247, \dodoi{10.1093/mnras/stad774}

\bibitem[{{Parker} {et~al.}(2024){Parker}, {Berg}, {Gazagnes}, {et~al.}}]{Parker2024}
{Parker}, K.~S., {Berg}, D.~A., {Gazagnes}, S., {et~al.} 2024, \apj, 977, 104, \dodoi{10.3847/1538-4357/ad87cd}

\bibitem[{{Reddy} {et~al.}(2016{\natexlab{a}}){Reddy}, {Steidel}, {Pettini}, \& {Bogosavljevi{\'c}}}]{Reddy_2016b}
{Reddy}, N.~A., {Steidel}, C.~C., {Pettini}, M., \& {Bogosavljevi{\'c}}, M. 2016{\natexlab{a}}, \apj, 828, 107, \dodoi{10.3847/0004-637X/828/2/107}

\bibitem[{{Reddy} {et~al.}(2016{\natexlab{b}}){Reddy}, {Steidel}, {Pettini}, {et~al.}}]{Reddy_2016}
{Reddy}, N.~A., {Steidel}, C.~C., {Pettini}, M., {et~al.} 2016{\natexlab{b}}, \apj, 828, 108, \dodoi{10.3847/0004-637X/828/2/108}

\bibitem[{{Rivera-Thorsen} {et~al.}(2019){Rivera-Thorsen}, {Dahle}, {Chisholm}, {et~al.}}]{Rivera-Thorsen_2019}
{Rivera-Thorsen}, T.~E., {Dahle}, H., {Chisholm}, J., {et~al.} 2019, Science, 366, 738, \dodoi{10.1126/science.aaw0978}

\bibitem[{{Rivera-Thorsen} {et~al.}(2017){Rivera-Thorsen}, {{\"O}stlin}, {Hayes}, {et~al.}}]{Rivera-Thorsen_2017}
{Rivera-Thorsen}, T.~E., {{\"O}stlin}, G., {Hayes}, M., {et~al.} 2017, \apj, 837, 29, \dodoi{10.3847/1538-4357/aa5d0a}

\bibitem[{{Robertson} {et~al.}(2015){Robertson}, {Ellis}, {Furlanetto}, {et~al.}}]{Robertson_2015}
{Robertson}, B.~E., {Ellis}, R.~S., {Furlanetto}, S.~R., {et~al.} 2015, \apjl, 802, L19, \dodoi{10.1088/2041-8205/802/2/L19}

\bibitem[{{Rosdahl} {et~al.}(2022){Rosdahl}, {Blaizot}, {Katz}, {et~al.}}]{Rosdahl_2022}
{Rosdahl}, J., {Blaizot}, J., {Katz}, H., {et~al.} 2022, \mnras, 515, 2386, \dodoi{10.1093/mnras/stac1942}

\bibitem[{{Rosdahl} \& {Teyssier}(2015)}]{Rosdahl_2015}
{Rosdahl}, J., \& {Teyssier}, R. 2015, \mnras, 449, 4380, \dodoi{10.1093/mnras/stv567}

\bibitem[{{Roy} {et~al.}(2024){Roy}, {Heckman}, {Henry}, {et~al.}}]{Roy_2024}
{Roy}, N., {Heckman}, T., {Henry}, A., {et~al.} 2024, arXiv e-prints, arXiv:2410.13254, \dodoi{10.48550/arXiv.2410.13254}

\bibitem[{{Saldana-Lopez} {et~al.}(2022){Saldana-Lopez}, {Schaerer}, {Chisholm}, {et~al.}}]{Saldana-Lopez_2022}
{Saldana-Lopez}, A., {Schaerer}, D., {Chisholm}, J., {et~al.} 2022, \aap, 663, A59, \dodoi{10.1051/0004-6361/202141864}

\bibitem[{{Schaerer} {et~al.}(2011){Schaerer}, {Hayes}, {Verhamme}, \& {Teyssier}}]{Schaerer_2011}
{Schaerer}, D., {Hayes}, M., {Verhamme}, A., \& {Teyssier}, R. 2011, \aap, 531, A12, \dodoi{10.1051/0004-6361/201116709}

\bibitem[{{Shapley} {et~al.}(2016){Shapley}, {Steidel}, {Strom}, {et~al.}}]{Shapley_2016}
{Shapley}, A.~E., {Steidel}, C.~C., {Strom}, A.~L., {et~al.} 2016, \apjl, 826, L24, \dodoi{10.3847/2041-8205/826/2/L24}

\bibitem[{{Sharma} {et~al.}(2016){Sharma}, {Theuns}, {Frenk}, {et~al.}}]{Sharma_2016}
{Sharma}, M., {Theuns}, T., {Frenk}, C., {et~al.} 2016, \mnras, 458, L94, \dodoi{10.1093/mnrasl/slw021}

\bibitem[{{Simmonds} {et~al.}(2024){Simmonds}, {Tacchella}, {Hainline}, {et~al.}}]{Simmonds_2024}
{Simmonds}, C., {Tacchella}, S., {Hainline}, K., {et~al.} 2024, \mnras, 527, 6139, \dodoi{10.1093/mnras/stad3605}

\bibitem[{{Soderblom}(2021)}]{Soderblom_2021}
{Soderblom}, D.~R. 2021, in COS Data Handbook v. 5.0, Vol.~5, 5

\bibitem[{{Stanway} {et~al.}(2016){Stanway}, {Eldridge}, \& {Becker}}]{Stanway_2016}
{Stanway}, E.~R., {Eldridge}, J.~J., \& {Becker}, G.~D. 2016, \mnras, 456, 485, \dodoi{10.1093/mnras/stv2661}

\bibitem[{{Steidel} {et~al.}(2018){Steidel}, {Bogosavljevi{\'c}}, {Shapley}, {et~al.}}]{Steidel_2018}
{Steidel}, C.~C., {Bogosavljevi{\'c}}, M., {Shapley}, A.~E., {et~al.} 2018, \apj, 869, 123, \dodoi{10.3847/1538-4357/aaed28}

\bibitem[{{Vanzella} {et~al.}(2016){Vanzella}, {de Barros}, {Vasei}, {et~al.}}]{Vanzella_2016}
{Vanzella}, E., {de Barros}, S., {Vasei}, K., {et~al.} 2016, \apj, 825, 41, \dodoi{10.3847/0004-637X/825/1/41}

\bibitem[{{Vasei} {et~al.}(2016){Vasei}, {Siana}, {Shapley}, {et~al.}}]{Vasei_2016}
{Vasei}, K., {Siana}, B., {Shapley}, A.~E., {et~al.} 2016, \apj, 831, 38, \dodoi{10.3847/0004-637X/831/1/38}

\bibitem[{{Verhamme} {et~al.}(2015){Verhamme}, {Orlitov{\'a}}, {Schaerer}, {et~al.}}]{Verhamme_2015}
{Verhamme}, A., {Orlitov{\'a}}, I., {Schaerer}, D., {et~al.} 2015, \aap, 578, A7, \dodoi{10.1051/0004-6361/201423978}

\bibitem[{{Verhamme} {et~al.}(2017){Verhamme}, {Orlitov{\'a}}, {Schaerer}, {et~al.}}]{Verhamme_2017}
---. 2017, \aap, 597, A13, \dodoi{10.1051/0004-6361/201629264}

\bibitem[{{Verhamme} {et~al.}(2008){Verhamme}, {Schaerer}, {Atek}, \& {Tapken}}]{Verhamme_2008}
{Verhamme}, A., {Schaerer}, D., {Atek}, H., \& {Tapken}, C. 2008, \aap, 491, 89, \dodoi{10.1051/0004-6361:200809648}

\bibitem[{{Verhamme} {et~al.}(2006){Verhamme}, {Schaerer}, \& {Maselli}}]{Verhamme_2006}
{Verhamme}, A., {Schaerer}, D., \& {Maselli}, A. 2006, \aap, 460, 397, \dodoi{10.1051/0004-6361:20065554}

\bibitem[{{Vidal-Garc{\'\i}a} {et~al.}(2017){Vidal-Garc{\'\i}a}, {Charlot}, {Bruzual}, \& {Hubeny}}]{Vidal_Garcia_2017}
{Vidal-Garc{\'\i}a}, A., {Charlot}, S., {Bruzual}, G., \& {Hubeny}, I. 2017, \mnras, 470, 3532, \dodoi{10.1093/mnras/stx1324}

\bibitem[{{Virtanen} {et~al.}(2020){Virtanen}, {Gommers}, {Oliphant}, {et~al.}}]{Virtanen_2020}
{Virtanen}, P., {Gommers}, R., {Oliphant}, T.~E., {et~al.} 2020, Nature Methods, 17, 261, \dodoi{10.1038/s41592-019-0686-2}

\bibitem[{{Wang} {et~al.}(2019){Wang}, {Heckman}, {Leitherer}, {et~al.}}]{Wang_2019}
{Wang}, B., {Heckman}, T.~M., {Leitherer}, C., {et~al.} 2019, \apj, 885, 57, \dodoi{10.3847/1538-4357/ab418f}

\bibitem[{{Wise} {et~al.}(2014){Wise}, {Demchenko}, {Halicek}, {et~al.}}]{Wise_2014}
{Wise}, J.~H., {Demchenko}, V.~G., {Halicek}, M.~T., {et~al.} 2014, \mnras, 442, 2560, \dodoi{10.1093/mnras/stu979}

\bibitem[{{Wu} \& {Kravtsov}(2024)}]{Wu_2024}
{Wu}, Z., \& {Kravtsov}, A. 2024, The Open Journal of Astrophysics, 7, 56, \dodoi{10.33232/001c.121193}

\bibitem[{{Xu} {et~al.}(2022){Xu}, {Henry}, {Heckman}, {et~al.}}]{Xu_2022}
{Xu}, X., {Henry}, A., {Heckman}, T., {et~al.} 2022, \apj, 933, 202, \dodoi{10.3847/1538-4357/ac7225}

\bibitem[{{Yang} {et~al.}(2020){Yang}, {Wang}, {Fan}, {et~al.}}]{Yang_2020}
{Yang}, J., {Wang}, F., {Fan}, X., {et~al.} 2020, \apj, 904, 26, \dodoi{10.3847/1538-4357/abbc1b}

\bibitem[{{Yung} {et~al.}(2020){Yung}, {Somerville}, {Finkelstein}, {et~al.}}]{Yung_2020}
{Yung}, L.~Y.~A., {Somerville}, R.~S., {Finkelstein}, S.~L., {et~al.} 2020, \mnras, 496, 4574, \dodoi{10.1093/mnras/staa1800}

\end{thebibliography}


\startlongtable
\begin{longrotatetable}
\begin{deluxetable*}{l|ccccccccccc|cccccc|c}
	 \tablewidth{0pt}
	 \setlength{\tabcolsep}{4pt}
	 \tabletypesize{\scriptsize}
	 \tablecaption{\textbf{Galaxy properties and $f_{\mathrm{esc}}^{\: \mathrm{LyC}}$ estimates.} \textit{Column 1:} Galaxy. \textit{Columns 2-9:} measurements used to estimate $f_{\mathrm{esc}}^{\: \mathrm{AFT}}$: $\log(M_\star)$, $E(B-V)_{\mathrm{neb}}$ (measured from the Balmer decrement), $\log$(SFR), star formation rate surface density $\log(\Sigma_{\mathrm{SFR}})$, absolute magnitude at 1500$\AA$ M$_{1500}$, $\beta_{\mathrm{UV}}$-slope, ionization parameter O$_{32}$, and the H$\beta$ equivalent width. The $\beta$-slopes were measured in this work while the other measurements are from \cite{Berg_2022}. In addition to their use in estimating $f_{\mathrm{esc}}^{\: \mathrm{AFT}}$, the $\beta_{\mathrm{UV}}$-slopes were also used to estimate $f_{\mathrm{esc}}^{\: \beta}$ and the O$_{32}$ ratios were also used to estimate $f_{\mathrm{esc}}^{\: \mathrm{O}_{32}}$. \textit{Column 10:} our measurements of the Ly$\alpha$ peak separations (used to estimate $f_{\mathrm{esc}}^{\: v_{\mathrm{sep}}}$). \textit{Columns 11-12:} the LIS covering fractions from \citep{Parker2024} and the resulting \ion{H}{1} covering fractions (estimated using Equation \ref{eq:cf}) which were used to predict $f_{\mathrm{esc}}^{\: C_f}$. The measurements of $E(B-V)_\star$ (the UV dust attenuation) used in the $C_f$-method and AFT-method are listed in Table \ref{tab:fit_props}, along with the other stellar continuum fit properties. \textit{Columns 13-18:} The resulting individual i$f_{\mathrm{esc}}^{\: \mathrm{LyC}}$ predictions. \textit{Column 19:} The median escape fraction based on these indirect estimates.}
     \tablehead{\colhead{Galaxy} \vline & \colhead{log(M$_\star$)} & \colhead{E(B-V)$_{neb}$} & 
    \colhead{log(SFR)} & \colhead{log($\Sigma_{SFR}$)} & \colhead{M$_{1500}$} & 
    \colhead{$\beta_{UV}$} & \colhead{O$_{32}$} & \colhead{W$_{\lambda}$(H$\beta$)} & \colhead{$v_{sep}^{Ly\alpha}$} & 
    \colhead{$C_f$(LIS)} & \colhead{$C_f$(\ion{H}{1})} \vline & \colhead{$f_{\mathrm{esc}}^{\: C_f}$} & 
    \colhead{$f_{\mathrm{esc}}^{\: \beta}$} & \colhead{$f_{\mathrm{esc}}^{\: v_{\mathrm{sep}}}$} & \colhead{$f_{\mathrm{esc}}^{\: \mathrm{O}_{32}}$} &
    \colhead{$f_{\mathrm{esc}}^{\: \mathrm{sim}}$} & \colhead{$f_{\mathrm{esc}}^{\: \mathrm{AFT}}$} \vline & \colhead{$\left<f_{\mathrm{esc}}^{\: \mathrm{LyC}}\right>$} \\ 
    \colhead{} \vline & \colhead{(M$_\odot$)} & \colhead{(mag)} & \colhead{(M$_\odot$/yr)} & 
    \colhead{(M$_\odot$/yr/kpc$^2$)} & \colhead{(mag)} & \colhead{} & \colhead{} & \colhead{($\AA$)} & \colhead{(km/s)} & 
    \colhead{} & \colhead{} \vline & \colhead{(\%)} & \colhead{(\%)} & \colhead{(\%)} & \colhead{(\%)} & 
    \colhead{(\%)} & \colhead{(\%)} \vline & \colhead{(\%)}}
	 \startdata
	 J0021+0052 & 9.09 & 0.11 & 1.07 & -0.19 & -20.6 & -1.93 & 2.0 & 91.4 & 561 & 0.94 & 0.71 & 6.1$_{-1.3}^{+1.3}$ & 3.0$_{-2.0}^{+2.0}$ & 0.2$_{-0.2}^{+0.7}$ & 1.2$_{-0.8}^{+0.7}$ & 6.1$_{-8.1}^{+8.1}$ & 2.3$_{-1.8}^{+4.6}$ & 2.6$_{-1.3}^{+1.2}$\\ 
	 J0036-3333 & 9.14 & 0.12 & 1.01 & 0.03 & -18.6 & -2.12 & 1.1 & - & 426 & 0.86 & 0.86 & 3.2$_{-3.1}^{+3.1}$ & 5.1$_{-3.4}^{+3.4}$ & 3.4$_{-1.0}^{+1.0}$ & 0.7$_{-0.6}^{+0.7}$ & 0.6$_{-0.6}^{+0.6}$ & 1.4$_{-1.1}^{+2.9}$ & 2.3$_{-1.2}^{+0.6}$\\ 
	 J0127-0619 & 8.74 & 0.28 & -0.75 & -0.39 & -14.1 & -0.79 & 1.1 & - & - & 0.51 & 1.00 & 0.0$_{-0.0}^{+0.00}$ & 0.12$_{-0.07}^{+0.07}$ & - & 0.7$_{-0.6}^{+0.7}$ & 0.1$_{-0.27}^{+0.27}$ & 0.1$_{-0.10}^{+0.25}$ & 0.1$_{-0.05}^{+0.17}$\\ 
	 J0144+0453 & 7.65 & 0.09 & -0.81 & -0.61 & -13.3 & -1.96 & 2.1 & - & - & 0.74 & 1.00 & 0.0$_{-1.45}^{+1.45}$ & 3.2$_{-2.3}^{+2.3}$ & - & 1.3$_{-0.7}^{+0.8}$ & 0.3$_{-4.5}^{+4.5}$ & 2.1$_{-1.7}^{+4.4}$ & 1.3$_{-0.6}^{+1.0}$\\ 
	 J0337-0502 & 7.06 & 0.00 & -0.32 & -0.33 & -16.9 & -2.40 & 6.2 & - & - & 0.79 & 1.00 & 0.0$_{-0.64}^{+0.64}$ & 10.9$_{-7.9}^{+7.9}$ & - & 7.0$_{-1.6}^{+1.7}$ & 0.3$_{-1.1}^{+1.1}$ & 18.8$_{-15.0}^{+39.4}$ & 7.0$_{-6.1}^{+0.8}$\\ 
	 J0405-3648 & 6.61 & 0.02 & -1.81 & -1.17 & -11.0 & -1.93 & 0.6 & - & - & 0.38 & 0.78 & 9.6$_{-1.4}^{+1.4}$ & 3.0$_{-2.2}^{+2.2}$ & - & 0.6$_{-0.6}^{+0.7}$ & 7.4$_{-6.7}^{+6.7}$ & 0.6$_{-0.5}^{+1.3}$ & 3.0$_{-2.0}^{+1.5}$\\ 
	 J0808+3948 & 9.12 & 0.15 & 1.26 & -0.24 & -20.2 & -2.55 & 0.8 & 9.7 & 448 & 0.83 & 1.00 & 0.0$_{-1.97}^{+1.97}$ & 16.8$_{-12.9}^{+12.8}$ & 2.6$_{-0.7}^{+0.7}$ & 0.6$_{-0.6}^{+0.7}$ & 5.8$_{-10.3}^{+10.3}$ & 1.5$_{-1.2}^{+3.0}$ & 2.0$_{-1.0}^{+0.9}$\\ 
	 J0823+2806 & 9.38 & 0.23 & 1.48 & 0.02 & -18.9 & -0.86 & 2.0 & 82.7 & - & 0.99 & 1.00 & 0.0$_{-0.0}^{+0.00}$ & 0.15$_{-0.08}^{+0.08}$ & - & 1.2$_{-0.7}^{+0.6}$ & 0.00$_{-0.01}^{+0.01}$ & 1.2$_{-0.9}^{+2.4}$ & 0.1$_{-0.02}^{+0.62}$\\ 
	 J0926+4427 & 8.76 & 0.11 & 1.03 & -0.92 & -20.6 & -2.10 & 3.1 & 108.2 & 445 & 0.34 & 0.76 & 7.8$_{-0.8}^{+0.8}$ & 4.7$_{-3.5}^{+3.5}$ & 2.7$_{-1.8}^{+1.8}$ & 2.2$_{-0.8}^{+0.8}$ & 3.6$_{-7.4}^{+7.4}$ & 5.0$_{-3.9}^{+10.1}$ & 4.2$_{-1.6}^{+1.7}$\\ 
	 J0934+5514 & 6.27 & 0.00 & -1.52 & -0.68 & -14.1 & -2.65 & 8.7 & - & - & 0.50 & 0.86 & 14.0$_{-5.5}^{+5.5}$ & 22.5$_{-17.3}^{+17.3}$ & - & 13.2$_{-3.0}^{+3.0}$ & 2.7$_{-3.5}^{+3.5}$ & 7.7$_{-6.1}^{+16.1}$ & 13.2$_{-5.6}^{+2.0}$\\ 
	 J0938+5428 & 9.15 & 0.09 & 1.05 & -0.55 & -20.5 & -2.24 & 1.9 & 76.2 & 657 & 0.83 & 1.00 & 0.0$_{-0.18}^{+0.18}$ & 7.0$_{-5.1}^{+5.1}$ & 0.0$_{-0.00}^{+0.42}$ & 1.0$_{-0.6}^{+0.7}$ & 0.0$_{-0.22}^{+0.22}$ & 5.6$_{-4.5}^{+11.4}$ & 0.5$_{-0.3}^{+0.4}$\\ 
	 J0940+2935 & 6.71 & 0.08 & -2.01 & -1.41 & -10.7 & -2.12 & 0.7 & 37.0 & - & 0.50 & 0.85 & 4.1$_{-1.3}^{+1.3}$ & 5.0$_{-3.6}^{+3.6}$ & - & 0.7$_{-0.6}^{+0.7}$ & 3.3$_{-3.5}^{+3.5}$ & 0.1$_{-0.10}^{+0.25}$ & 3.3$_{-2.7}^{+0.6}$\\ 
	 J0942+3547 & 7.56 & 0.04 & -0.76 & -0.80 & -16.3 & -2.17 & 2.6 & 82.6 & 268 & 0.35 & 0.76 & 10.2$_{-1.4}^{+1.4}$ & 5.7$_{-4.0}^{+4.0}$ & 17.7$_{-2.0}^{+2.0}$ & 1.7$_{-0.7}^{+0.7}$ & 4.1$_{-7.3}^{+7.3}$ & 1.2$_{-0.9}^{+2.4}$ & 4.9$_{-1.7}^{+3.1}$\\ 
	 J0944-0038 & 6.83 & 0.22 & -0.78 & 0.42 & -12.8 & -0.88 & 2.9 & 281.2 & 397 & 0.87 & 1.00 & 0.0$_{-0.0}^{+0.05}$ & 0.15$_{-0.09}^{+0.10}$ & 4.7$_{-3.1}^{+3.1}$ & 1.9$_{-0.9}^{+0.8}$ & 0.03$_{-0.04}^{+0.04}$ & 0.01$_{-0.01}^{+0.03}$ & 0.09$_{-0.04}^{+0.06}$\\ 
	 J0944+3442 & 8.19 & 0.20 & -0.01 & -0.84 & -15.1 & -1.23 & 1.4 & 16.5 & 551 & 0.75 & 1.00 & 0.0$_{-0.37}^{+0.37}$ & 0.4$_{-0.3}^{+0.3}$ & 0.4$_{-0.4}^{+1.4}$ & 0.8$_{-0.7}^{+0.8}$ & 0.2$_{-0.6}^{+0.6}$ & 1.0$_{-0.8}^{+2.0}$ & 0.4$_{-0.2}^{+0.3}$\\ 
	 J1016+3754 & 6.72 & 0.07 & -1.17 & -0.38 & -14.2 & -2.21 & 4.6 & 101.6 & 361 & 0.45 & 0.77 & 12.3$_{-4.8}^{+4.8}$ & 6.5$_{-4.5}^{+4.5}$ & 6.8$_{-3.3}^{+3.3}$ & 4.1$_{-1.2}^{+1.1}$ & 7.5$_{-13.4}^{+13.4}$ & 2.0$_{-1.6}^{+4.1}$ & 6.7$_{-2.6}^{+1.4}$\\ 
	 J1024+0524 & 7.89 & 0.05 & 0.21 & -0.53 & -18.2 & -2.16 & 2.1 & 93.2 & 438 & 0.51 & 0.87 & 5.6$_{-3.7}^{+3.7}$ & 5.6$_{-4.0}^{+3.9}$ & 3.0$_{-1.1}^{+1.1}$ & 1.3$_{-0.7}^{+0.7}$ & 0.1$_{-0.3}^{+0.3}$ & 2.4$_{-1.9}^{+4.8}$ & 2.7$_{-1.3}^{+0.8}$\\ 
	 J1025+3622 & 8.87 & 0.13 & 1.04 & -0.52 & -20.3 & -2.06 & 2.4 & 67.6 & 467 & 0.66 & 0.93 & 1.6$_{-1.8}^{+1.8}$ & 4.2$_{-3.0}^{+3.0}$ & 2.1$_{-1.4}^{+1.4}$ & 1.6$_{-0.7}^{+0.7}$ & 0.1$_{-0.9}^{+0.9}$ & 4.2$_{-3.4}^{+8.6}$ & 1.8$_{-0.7}^{+0.8}$\\ 
	 J1044+0353 & 6.80 & 0.27 & -0.59 & -0.40 & -15.1 & -0.79 & 19.0 & 280.6 & 447 & 0.95 & 1.00 & 0.0$_{-0.0}^{+0.00}$ & 0.12$_{-0.06}^{+0.06}$ & 2.7$_{-2.7}^{+2.9}$ & 62.0$_{-12.4}^{+12.3}$ & 0.4$_{-0.7}^{+0.7}$ & 34.9$_{-27.7}^{+65.1}$ & 1.5$_{-1.1}^{+3.6}$\\ 
	 J1105+4444 & 8.98 & 0.33 & 0.69 & -0.26 & -17.3 & -0.73 & 2.0 & 99.5 & - & 0.95 & 1.00 & 0.0$_{-0.0}^{+0.00}$ & 0.10$_{-0.05}^{+0.05}$ & - & 1.2$_{-0.7}^{+0.8}$ & 0.03$_{-0.02}^{+0.02}$ & 0.2$_{-0.2}^{+0.4}$ & 0.1$_{-0.03}^{+0.23}$\\ 
	 J1112+5503 & 9.59 & 0.27 & 1.60 & -0.08 & -20.4 & -1.28 & 0.9 & 34.6 & 609 & 0.76 & 0.97 & 0.1$_{-0.6}^{+0.6}$ & 0.5$_{-0.3}^{+0.3}$ & 0.0$_{-0.00}^{+0.52}$ & 0.6$_{-0.6}^{+0.8}$ & 0.1$_{-0.29}^{+0.29}$ & 1.1$_{-0.9}^{+2.3}$ & 0.3$_{-0.2}^{+0.3}$\\ 
	 J1119+5130 & 6.77 & 0.08 & -1.58 & -0.90 & -13.3 & -2.05 & 2.0 & 47.4 & 614 & 0.60 & 0.92 & 3.1$_{-0.9}^{+0.9}$ & 4.1$_{-2.9}^{+2.9}$ & 0.0$_{-0.00}^{+0.62}$ & 1.1$_{-0.6}^{+0.9}$ & 0.7$_{-3.3}^{+3.3}$ & 0.6$_{-0.5}^{+1.3}$ & 0.9$_{-0.2}^{+1.3}$\\ 
	 J1129+2034 & 8.09 & 0.22 & -0.37 & -0.17 & -13.1 & -1.42 & 1.8 & 196.2 & - & 0.87 & 1.00 & 0.0$_{-0.1}^{+0.09}$ & 0.7$_{-0.4}^{+0.4}$ & - & 1.0$_{-0.7}^{+0.7}$ & 0.2$_{-0.5}^{+0.5}$ & 0.1$_{-0.05}^{+0.13}$ & 0.2$_{-0.1}^{+0.3}$\\ 
	 J1132+1411 & 7.31 & 0.17 & -1.07 & -1.68 & -13.6 & -1.39 & 0.8 & 144.1 & - & 0.99 & 1.00 & 0.0$_{-0.0}^{+0.00}$ & 0.6$_{-0.4}^{+0.4}$ & - & 1.8$_{-0.8}^{+0.7}$ & 0.05$_{-0.02}^{+0.02}$ & 0.04$_{-0.03}^{+0.08}$ & 0.05$_{-0.01}^{+0.06}$\\ 
	 J1132+5722 & 8.68 & 0.00 & 0.44 & -0.20 & -15.8 & -2.55 & 2.7 & 73.1 & - & 0.63 & 0.93 & 6.9$_{-4.2}^{+4.2}$ & 16.7$_{-14.3}^{+14.4}$ & - & 0.6$_{-0.6}^{+0.7}$ & 1.0$_{-2.6}^{+2.6}$ & 9.2$_{-7.3}^{+18.8}$ & 6.9$_{-5.6}^{+1.2}$\\ 
	 J1144+4012 & 9.89 & 0.22 & 1.51 & -0.34 & -19.9 & -1.26 & 0.6 & 13.4 & - & 0.86 & 1.00 & 0.0$_{-0.0}^{+0.05}$ & 0.4$_{-0.3}^{+0.3}$ & - & 0.5$_{-0.5}^{+0.7}$ & 0.1$_{-0.25}^{+0.25}$ & 2.4$_{-1.9}^{+4.8}$ & 0.4$_{-0.4}^{+0.1}$\\ 
	 J1148+2546 & 8.14 & 0.20 & 0.53 & -0.12 & -18.1 & -0.87 & 3.7 & 152.7 & 624 & 0.87 & 1.00 & 0.0$_{-0.1}^{+0.09}$ & 0.15$_{-0.10}^{+0.09}$ & 0.0$_{-0.00}^{+0.77}$ & 2.8$_{-0.9}^{+1.0}$ & 0.04$_{-0.05}^{+0.05}$ & 0.6$_{-0.5}^{+1.3}$ & 0.1$_{-0.03}^{+0.23}$\\ 
	 J1150+1501 & 6.84 & 0.21 & -1.33 & -0.07 & -13.7 & -1.55 & 2.3 & 131.3 & - & 0.79 & 1.00 & 0.0$_{-0.21}^{+0.21}$ & 1.0$_{-0.6}^{+0.6}$ & - & 1.4$_{-0.7}^{+0.8}$ & 0.2$_{-0.4}^{+0.4}$ & 0.1$_{-0.1}^{+0.3}$ & 0.2$_{-0.05}^{+0.40}$\\ 
	 J1157+3220 & 9.04 & 0.14 & 0.97 & 0.52 & -17.1 & -2.07 & 1.2 & 49.3 & 735 & 0.50 & 0.86 & 4.1$_{-3.9}^{+3.9}$ & 4.4$_{-2.9}^{+2.9}$ & 0.0$_{-0.00}^{+0.19}$ & 0.8$_{-0.7}^{+0.6}$ & 0.7$_{-3.2}^{+3.2}$ & 3.1$_{-2.5}^{+6.4}$ & 2.0$_{-1.4}^{+1.1}$\\ 
	 J1200+1343 & 8.12 & 0.26 & 0.75 & -0.36 & -18.5 & -1.10 & 5.1 & 210.5 & 510 & 0.84 & 1.00 & 0.0$_{-0.69}^{+0.69}$ & 0.3$_{-0.2}^{+0.2}$ & 1.1$_{-0.4}^{+0.4}$ & 4.9$_{-1.3}^{+1.2}$ & 0.3$_{-1.4}^{+1.4}$ & 1.1$_{-0.9}^{+2.2}$ & 0.7$_{-0.2}^{+0.5}$\\ 
	 J1225+6109 & 7.12 & 0.23 & -1.08 & -0.16 & -13.4 & -1.41 & 4.7 & 88.6 & - & 0.85 & 1.00 & 0.0$_{-0.11}^{+0.11}$ & 0.7$_{-0.4}^{+0.4}$ & - & 4.3$_{-1.2}^{+1.2}$ & 0.1$_{-0.45}^{+0.45}$ & 0.6$_{-0.5}^{+1.2}$ & 0.6$_{-0.4}^{+0.3}$\\ 
	 J1253-0312 & 7.65 & 0.26 & 0.56 & 0.33 & -18.2 & -1.31 & 8.0 & 239.6 & 425 & 0.74 & 1.00 & 0.0$_{-0.30}^{+0.30}$ & 0.5$_{-0.3}^{+0.3}$ & 3.4$_{-0.7}^{+0.7}$ & 11.3$_{-2.7}^{+2.3}$ & 0.5$_{-0.8}^{+0.8}$ & 1.9$_{-1.5}^{+3.9}$ & 1.2$_{-0.4}^{+1.0}$\\ 
	 J1314+3452 & 7.56 & 0.23 & -0.67 & 0.44 & -12.8 & -1.32 & 2.3 & - & - & 0.90 & 1.00 & 0.0$_{-0.0}^{+0.00}$ & 0.5$_{-0.3}^{+0.3}$ & - & 1.3$_{-0.5}^{+0.7}$ & 0.1$_{-0.99}^{+0.99}$ & 0.3$_{-0.2}^{+0.6}$ & 0.3$_{-0.00}^{+0.60}$\\ 
	 J1323-0132 & 6.31 & 0.05 & -0.72 & -0.55 & -16.0 & -2.25 & 37.8 & 250.0 & 177 & 0.00 & 0.54 & 19.6$_{-6.4}^{+6.4}$ & 7.2$_{-5.2}^{+5.2}$ & 52.2$_{-10.0}^{+10.0}$ & 100.0$_{-0.0}^{+0.0}$ & 22.8$_{-11.5}^{+11.5}$ & 100.0$_{-0.0}^{+0.0}$ & 37.5$_{-29.81}^{+0.00}$\\ 
	 J1359+5726 & 8.41 & 0.15 & 0.42 & -0.38 & -18.6 & -1.94 & 2.6 & 60.4 & - & 0.72 & 1.00 & 0.0$_{-0.68}^{+0.68}$ & 3.0$_{-2.0}^{+2.0}$ & - & 1.7$_{-0.8}^{+0.7}$ & 0.1$_{-0.22}^{+0.22}$ & 4.0$_{-3.2}^{+8.2}$ & 1.7$_{-1.3}^{+0.3}$\\ 
	 J1416+1223 & 9.59 & 0.16 & 1.57 & -0.11 & -20.6 & -1.89 & 0.8 & 28.5 & 597 & 0.76 & 1.00 & 0.0$_{-0.57}^{+0.57}$ & 2.6$_{-1.7}^{+1.7}$ & 0.0$_{-0.00}^{+0.75}$ & 0.7$_{-0.7}^{+0.6}$ & 0.0$_{-0.17}^{+0.17}$ & 3.4$_{-2.7}^{+7.0}$ & 0.4$_{-0.2}^{+0.5}$\\ 
	 J1418+2102 & 6.22 & 0.19 & -1.13 & -0.54 & -13.8 & -1.33 & 11.5 & 220.6 & 395 & 0.74 & 1.00 & 0.0$_{-0.41}^{+0.41}$ & 0.5$_{-0.3}^{+0.3}$ & 4.8$_{-1.2}^{+1.2}$ & 22.5$_{-4.8}^{+6.7}$ & 0.5$_{-1.0}^{+1.0}$ & 2.3$_{-1.8}^{+4.6}$ & 1.4$_{-0.6}^{+1.4}$\\ 
	 J1428+1653 & 9.56 & 0.14 & 1.22 & -0.77 & -20.8 & -1.74 & 1.2 & 45.7 & 521 & 0.60 & 0.91 & 3.5$_{-3.2}^{+3.2}$ & 1.7$_{-1.4}^{+1.3}$ & 0.8$_{-0.8}^{+0.8}$ & 0.7$_{-0.7}^{+0.7}$ & 0.4$_{-0.4}^{+0.4}$ & 11.1$_{-8.8}^{+22.7}$ & 1.3$_{-0.6}^{+0.6}$\\ 
	 J1429+0643 & 8.80 & 0.12 & 1.42 & -0.46 & -20.9 & -1.36 & 4.2 & 169.3 & 525 & 0.66 & 0.94 & 1.3$_{-0.2}^{+0.2}$ & 0.6$_{-0.4}^{+0.4}$ & 0.8$_{-0.7}^{+0.7}$ & 3.4$_{-1.1}^{+1.0}$ & 0.1$_{-0.3}^{+0.3}$ & 6.0$_{-4.8}^{+12.3}$ & 1.0$_{-0.2}^{+0.4}$\\ 
	 J1444+4237 & 6.48 & 0.11 & -1.94 & -1.08 & -11.8 & -2.27 & 4.1 & - & - & 0.69 & 0.98 & 0.9$_{-4.5}^{+4.5}$ & 7.8$_{-5.6}^{+5.6}$ & - & 3.4$_{-1.0}^{+0.9}$ & 1.3$_{-3.7}^{+3.7}$ & 1.2$_{-0.9}^{+2.5}$ & 1.3$_{-0.1}^{+2.7}$\\ 
	 J1448-0110 & 7.61 & 0.23 & 0.39 & 0.01 & -17.7 & -0.99 & 8.0 & 144.4 & - & 0.94 & 1.00 & 0.0$_{-0.33}^{+0.33}$ & 0.2$_{-0.1}^{+0.1}$ & - & 11.4$_{-2.5}^{+2.4}$ & 0.0$_{-0.17}^{+0.17}$ & 5.1$_{-4.0}^{+10.3}$ & 0.2$_{-0.08}^{+0.20}$\\ 
	 J1521+0759 & 9.00 & 0.15 & 0.95 & -0.47 & -20.3 & -2.42 & 1.5 & 27.4 & 430 & 0.52 & 0.95 & 3.9$_{-10.0}^{+10.0}$ & 11.8$_{-9.0}^{+9.0}$ & 3.2$_{-1.7}^{+1.7}$ & 1.0$_{-0.7}^{+0.6}$ & 0.1$_{-0.2}^{+0.2}$ & 21.9$_{-17.4}^{+44.8}$ & 3.6$_{-2.4}^{+2.5}$\\ 
	 J1525+0757 & 10.10 & 0.21 & 1.00 & -0.48 & -19.8 & -1.83 & 0.5 & 19.3 & - & 0.88 & 1.00 & 0.0$_{-0.0}^{+0.02}$ & 2.2$_{-1.8}^{+1.8}$ & - & 0.5$_{-0.5}^{+0.7}$ & 0.1$_{-0.3}^{+0.3}$ & 0.6$_{-0.5}^{+1.2}$ & 0.5$_{-0.5}^{+0.2}$\\ 
	 J1545+0858 & 7.52 & 0.14 & 0.37 & -0.31 & -18.5 & -1.43 & 8.6 & 194.7 & 279 & 0.51 & 0.90 & 2.4$_{-1.7}^{+1.7}$ & 0.7$_{-0.4}^{+0.4}$ & 15.7$_{-2.0}^{+2.0}$ & 13.2$_{-3.6}^{+3.4}$ & 0.4$_{-0.8}^{+0.8}$ & 11.2$_{-8.9}^{+22.9}$ & 6.8$_{-4.3}^{+1.4}$\\ 
	 J1612+0817 & 9.78 & 0.20 & 1.58 & -0.18 & -21.1 & -1.29 & 0.7 & 27.9 & 583 & 0.86 & 1.00 & 0.0$_{-0.0}^{+0.00}$ & 0.5$_{-0.3}^{+0.3}$ & 0.0$_{-0.00}^{+2.33}$ & 0.6$_{-0.6}^{+0.7}$ & 0.2$_{-0.3}^{+0.3}$ & 2.8$_{-2.2}^{+5.7}$ & 0.4$_{-0.1}^{+0.5}$\\ 
	 \enddata
\tablecomments{The measurements of E(B-V)$_\star$ used in the $C_f$-method are listed in Table \ref{tab:fit_props}, along with the other stellar continuum fit properties.}
	 \label{tab:results}
\end{deluxetable*}
\end{longrotatetable}



\end{document}